\documentstyle[prl,aps,amsfonts,graphicx,twoside]{revtex}

\input{macro.lib}
\input{epsf}

\begin{document}

\draft

\title{Thermodynamics and excitations of the one-dimensional Hubbard
model}

\author{
T.~Deguchi$^{(1)}$\footnote[2]{Address after March 1, 1999: Department
of Physics, Ochanomizu University, 2-1-1 Ohtsuka, Bunkyo-ku, Tokyo
112-8610, Japan}, F.~H.~L.~Essler$^{(2)}$, F.~G\"ohmann$^{(1)}$,
A.~Kl\"umper$^{(3)}$, V.~E.~Korepin$^{(1)}$, and K.~Kusakabe$^{(4)}$.
}

\address{
 $^{(1)}$Institute for Theoretical Physics, State University of New
York at Stony Brook, Stony Brook, NY 11794-3840, USA\\
 $^{(2)}$Department of Physics, Theoretical Physics, Oxford University,
1 Keble Road, Oxford OX1 NP, UK\\
 $^{(3)}$Institut f\"ur Theoretische Physik, Universit\"at zu K\"oln,
Z\"ulpicher Str.\ 77, 50937 K\"oln, Germany\\
 $^{(4)}$Graduate School of Science and Technology, Niigata University,
Ikarashi, Niigata, 950-2181, Japan
}

\maketitle

\begin{abstract}
We review fundamental issues arising in the exact solution of the
one-dimensional Hubbard model. We perform a careful analysis of the
Lieb-Wu equations, paying particular attention to so-called `string
solutions'. Two kinds of string solutions occur: $\La$ strings,
related to spin degrees of freedom and $k$-$\La$ strings, describing
spinless bound states of electrons. Whereas $\La$ strings were
thoroughly studied in the literature, less is known about $k$-$\La$
strings. We carry out a thorough analytical and numerical analysis of
$k$-$\La$ strings. We further review two different approaches to the
thermodynamics of the Hubbard model, the Yang-Yang approach and the
quantum transfer matrix approach, respectively. The Yang-Yang approach
is based on strings, the quantum transfer matrix approach is not. We
compare the results of both methods and show that they agree.
Finally, we obtain the dispersion curves of all elementary excitations
at zero magnetic field for the less than half-filled band by considering
the zero temperature limit of the Yang-Yang approach.\\[2ex]
PACS numbers: 71.10.Fd, 71.10.Pm, 71.27.+a\\[1ex]
{\it Key words: Hubbard model, strongly correlated electrons,
thermodynamics, excitations}
\end{abstract}

\tableofcontents

\clearpage

\section{Introduction}
The Hubbard model was introduced as a simple effective model for the
study of correlation effects of d-electrons in transition metals
\cite{Gutzwiller63,Hubbard63} (see also \cite{Kanamori63}). It is
believed to provide a qualitative description of the magnetic
properties of these materials and the Mott metal-insulator
transition \cite{Gebhard97}. Despite of its appealing conceptual
simplicity rigorous results for the Hubbard model are rare. 
The dimension of the underlying lattice is a crucial parameter.
Two of the most important theorems valid for arbitrary lattice
dimension are due to Nagaoka \cite{Nagaoka66} and to Lieb \cite{Lieb89}.
Nagaoka's theorem states, that creating a single hole in a half-filled
connected lattice for infinitely repulsive interaction renders the ground
state ferromagnetic. Lieb's theorem is valid for arbitrary finite
repulsion. It states that on a bipartite lattice at half-filling the
ground state has spin $S = \2 ||B| - |A||$, where $|B|$ ($|A|$) is the
number of sites in the $B$ ($A$) sublattice. For reviews of rigorous
results about the Hubbard model in arbitrary dimensions see
\cite{Lieb95,Tasaki98}. Some simplifications occur in the limit of
infinite lattice dimension \cite{MeVo89,Vollhardt89,Mueller-Hartmann89}.
However, most exact results have been obtained for the one-dimensional
lattice. This is because a complete set of eigenfunctions of the
Hubbard Hamiltonian is only known for this case. The one-dimensional
Hubbard model was solved by Lieb and Wu \cite{LiWu68}. They used the
nested Bethe ansatz discovered in \cite{Yang67,Gaudin67},

There exists a vast literature on the Hubbard model. Some of the most
significant results have been collected in two reprint volumes. The
volume \cite{Montorsi92} gives a general overview for the years
1963-1990\footnote{See also \cite{Rasetti91}}. For a review of the
history of the exact solution in one dimension including a rather
exhaustive list of references until about 1992 we refer the reader to
\cite{KoEsBo}.

Most of the references listed in \cite{KoEsBo} are based on the seminal
1968 paper \cite{LiWu68} by Lieb and Wu. In this paper the problem of
diagonalizing the Hamiltonian was reduced to solving a set of coupled
nonlinear equations known as the Bethe ansatz or Lieb-Wu equations. Lieb
and Wu calculated the ground state energy of the system. They showed
that the model at half-filling is an insulator for arbitrary positive
value of the coupling $U$. In other words, they showed that the
half-filled model undergoes a Mott transition at critical coupling
$U = 0$.

In 1972 Takahashi proposed a classification of the solutions to the
Lieb-Wu equations \cite{Takahashi72}, which is commonly referred to as
`Takahashi's string hypothesis'. Analogous classifications are used
in all models solvable by the Bethe Ansatz method (see
e.g. \cite{Takahashi71,FaTa84} for the case of the Heisenberg model and
\cite{TsWi83} for the case of the Anderson model, which bears certain
similarities with the Hubbard model). The string hypothesis is the
basis of many subsequent publications. In the paper \cite{Takahashi72},
Takahashi used it to obtain a set of nonlinear integral equations that
determines the thermodynamics of the Hubbard model. By solving these
equations in some limiting cases, he was able to calculate the
low temperature specific heat in \cite{Takahashi74}.

At the beginning of the 80's Woynarovich resumed the study of the
excitation spectrum of the Hubbard model \cite{Woynarovich82a,%
Woynarovich82b,Woynarovich83a,Woynarovich83b} which was started ten
years earlier \cite{Ovchinnikov70,Coll74}. He gave a detailed analysis
of the charge excitations at half-filling \cite{Woynarovich82a} and was
the first to study gapped, spin-singlet charge excitations at
half-filling. These involve the first examples of excitations which in
the sequel will be called $k$-$\La$-string excitations
\cite{Woynarovich82b}. In his article \cite{Woynarovich82a}
Woynarovich presented the explicit form of the Bethe ansatz wave
function (see equations (\ref{sectorq})-(\ref{wwfsa}) below).

Since the publication of the reprint volume \cite{KoEsBo} there have
been several interesting developments. Two of the present authors
\cite{EsKo94a,EsKo94b} showed that the excitation spectrum at 
half-filling in the absence of a magnetic field is given by the
scattering states of only four elementary excitations. Two of them carry
charge but no spin, the other two carry spin but no charge. These
elementary excitations are called holon, antiholon and spinon with spin
up or down, respectively. They form the fundamental representation of
SO(4). In the same articles \cite{EsKo94a,EsKo94b} the $S$-matrix of
the four quasiparticles was obtained. Thus there is now a complete and
satisfactory picture on the level of elementary excitations of
spin-charge separation in the one-dimensional Hubbard model at
half-filling. Spin and charge degrees of freedom can be excited
separately, but the corresponding quasiparticles do interact, albeit
weakly. The interaction is seen in the non-triviality of the
$S$-matrix. Spin-charge separation is one of the most interesting
properties of the one-dimensional Hubbard model. Recently experimental
evidence was found for the existence of spin-charge separation in quasi
one-dimensional materials \cite{KimEtal96,KimEtal97}. 

Another interesting recent development is the calculation of the bulk
thermodynamic properties of the Hubbard model within the quantum
transfer matrix approach \cite{KlBa96,JKS98}. In contrast to the
traditional approach \cite{Takahashi72,Takahashi97,Takahashi99}, the
quantum transfer matrix approach leads to a finite number of non-linear
integral equations that determine the Gibbs free energy. This enables
high-precision numerical calculation of thermodynamic quantities such
as the charge and spin susceptibilities over the entire range of
doping, temperature and magnetic field. It further opens the
interesting perspective of calculating the correlation length at
arbitrary finite temperature.

It was shown in the papers \cite{CNC94a,CNC94b,PCCS97} how to use a
pseudo particle approach in order to obtain transport properties
(optical conductivity) of the one-dimensional Hubbard model.

There was also progress in the understanding of the algebraic structure
of the Hubbard model. Shiroishi and Wadati showed \cite{ShWa95}, that
the $R$-matrix, which was constructed earlier by Shastry
\cite{Shastry86b,Shastry88b,OWA87}, and which underlies the 
integrability of the Hubbard model, satisfies the Yang-Baxter equation.
Martins and Ramos \cite{RaMa97,MaRa98} were able to construct a variant
of the algebraic Bethe ansatz for the Hubbard model. They obtained the
eigenvalue of the transfer matrix of the two-dimensional statistical
covering model (see also \cite{YuDe97}). This result was later used in
the quantum transfer matrix approach to the thermodynamics
\cite{JKS98}. Another interesting algebraic result was the discovery of
a quantum group symmetry of the Hubbard model on the infinite line.
The Hamiltonian is invariant under the direct sum of two Y(su(2))
Yangians \cite{UgKo94}. The relation of these Yangians to Shastry's
$R$-matrix was clarified in \cite{MuGo97a,MuGo98a}, where it was also
shown that the eigenstates of the Hubbard Hamiltonian on the infinite
interval, at zero density transform like irreducible representations of
one of the Yangians.

The purpose of this article is to give a pedagogical introduction
to the Bethe ansatz solution of the one-dimensional Hubbard model and
at the same time to fill some gaps in the previous literature. We
present a detailed account of the Bethe ansatz solution for periodic
boundary conditions and of the thermodynamics of the model. There are
two approaches to the thermodynamics. The approach of Takahashi
\cite{Takahashi72,Takahashi97,Takahashi99} relies on a string
hypothesis for the Hubbard model and is a natural generalization of
Yang and Yang's thermodynamic Bethe ansatz for the delta interacting
Bose gas \cite{YaYa69}. The second approach \cite{JKS98} is built on a
lattice path integral formulation of the partition function. We compare
both approaches and discuss their specific advantages. Special
attention is given to an aspect which, although fundamental, was
largely ignored in the previous literature, namely $k$-$\La$ strings.
$k$-$\La$ strings are spin-singlet bound states of electrons.

The Bethe ansatz for the one-dimensional Hubbard model \cite{LiWu68}
gives the eigenfunctions and eigenvalues of the Hubbard Hamiltonian
parametrized by two sets of quantum numbers $\{k_j\}$ and
$\{\la_l\}$, which are solutions of the Lieb-Wu equations (see
formulae (\ref{bak}), (\ref{bas}) below). The $k_j$ and $\la_l$ are
called charge momenta and spin rapidities, respectively. The Lieb-Wu
equations have finite and infinite solutions $k_j$, $\la_l$. They
should be considered separately, because they have different occupation
numbers. Every finite solution $k_j$ (or $\la_l$) can be occupied only
once. If two finite $k_j$ (or $\la_l$) coincide, the wave function
vanishes (see formulae (\ref{perm2}), (\ref{perm3}) and below). By
contrast, infinite $k_j$ (or $\la_l$) can be generally occupied more
than once. Later we shall explain that the multiplicities of occupation
of the infinite $k$'s and $\la$'s are given by the dimensions of the
representations of corresponding su(2) symmetry algebras. In order to
study this carefully the `regular' Bethe ansatz was defined in
\cite{EKS92a}. `Regular' means that all $k$'s and $\la$'s are finite.
They may be real or complex.

Takahashi's string hypothesis \cite{Takahashi72} is a statement about
the structure of the regular solutions of the Lieb-Wu equations in the
thermodynamic limit. Except for the real solutions (all $k_j$ and all
$\la_l$ real) there are solutions involving complex $k_j$ and $\la_l$.
The complex momenta and rapidities occur in two kinds of configurations,
which are symmetric with respect to the real axis. These configurations
are called strings. There are $\La$ strings involving only spin
rapidities and $k$-$\La$ strings, which involve charge momenta as
well. The $\La$ strings can be interpreted as bound states of magnons,
whereas the $k$-$\La$ strings describe spin singlet bound states of
electrons.

The $\La$ strings in the Hubbard model are similar to the $\La$ strings
in the isotropic Heisenberg spin chain, which have been extensively
studied in the literature \cite{Bethe31,Takahashi71,Gaudin71,EKS92c}.
Much less attention has been given to the $k$-$\La$ strings, which are
peculiar to the Hubbard model. They play an important role at half-%
filling, where the $k$-$\La$ two string enters the calculation of the
phase shift in the holon-holon scattering \cite{EsKo94a,EsKo94b}.
Holons are the lowest lying charge excitations of the Hubbard model.
At half-filling they have a gap. Below half-filling they are gapless,
however, whereas all $k$-$\Lambda$ strings lead to gapped excitations
in the thermodynamic limit (see section VI). Hence, $k$-$\La$ strings
below half-filling do not contribute to the low energy properties of
the model. What is their physical significance then? They do
contribute to the high temperature thermodynamic properties of the
Hubbard model. This is, in fact, the context in which they were first
introduced \cite{Takahashi72}. On the other hand, $k$-$\La$ strings
are interesting as a curious kind of excitations, which does not exist
in more simple integrable models.

According to the string hypothesis the $k$-$\La$ strings approach
certain ideal configurations as the number of lattice sites becomes
large. We call these configurations `ideal $k$-$\La$ strings'.
They are characterized by $m$ complex $\la$'s and $2m$ complex
$k$'s. The $\la$'s involved in an ideal $k$-$\La$ string have common
real part $\La'$, and their imaginary parts are $(m - 1) \frac{\i U}{4},
(m - 3) \frac{\i U}{4}, \dots, - (m - 1) \frac{\i U}{4}$. In the
repulsive case ($U > 0$) the $k$'s are given as
\beann
     k^1 & = & \p - \arcsin(\La' + m \tst{\frac{\i U}{4}}) \qd, \\
     k^2 & = & \arcsin(\La' + (m - 2) \tst{\frac{\i U}{4}}) \qd, \\
     k^3 & = & \p - k^2 \qd, \\
	 & \vdots & \\
     k^{2m - 2} & = & \arcsin(\La' - (m - 2) \tst{\frac{\i U}{4}})
		      \qd, \\
     k^{2m - 1} & = & \p - k^{2m - 2} \qd, \\
     k^{2m} & = & \p - \arcsin(\La' - m \tst{\frac{\i U}{4}}) \qd.
\eeann
This means that $\Re \sin (k^\a) = \La'$. Hence, the $\sin (k^\a)'s$
and the $\la$'s form a string in the complex plane.

In the thermodynamic limit, when the strings become ideal, the
variables describing their width can be eliminated from the Lieb-Wu
equations. We call the resulting set of equations discrete Takahashi
equations.

Some reformulation of the string hypothesis may be necessary before it
will be possible to achieve a rigorous mathematical proof. Yet, the
string hypothesis has passed many tests, and there is no doubt by now
that it describes the physics of the Hubbard model correctly. We
summarize our understanding of the issue and present several important
tests and consequences of the string hypothesis. We shall mostly
concentrate on the Hubbard model below half-filling, since the case of
half-filling was treated elsewhere \cite{EsKo94a,EsKo94b}.

Let us outline the plan of this article.

In section II we summarize known basic results about the Hubbard model.
This section contains a review of the Bethe ansatz solution of Lieb
and Wu \cite{LiWu68}. We present the wave function in the form given
by Woynarovich \cite{Woynarovich82a} and discuss its discrete
symmetries, namely the symmetries under permutations of electrons and
quantum numbers and the particle-hole and spin-reversal symmetries. This
leads to the notion of regular Bethe ansatz states. We proceed with
explaining the SO(4) symmetry \cite{HeLi71,Yang89,Pernici90,YaZh90,%
GoMu97b}, which is characteristic of the Hubbard model. Then we
introduce the discrete Takahashi equations. SO(4) symmetry and discrete
Takahashi equations are the prerequisites for the proof of completeness
of the Bethe ansatz \cite{EKS92b}, which is reviewed at the end of the
section.

Section III comprises a rigorous analytical study of the Lieb-Wu 
equations for one down spin. For the sake of pedagogical clarity we
mostly focus on the cases of two and three electrons. This is our first
test of the string hypothesis. The result is positive: $k$-$\La$
strings do exist as solutions of the Lieb-Wu equations. They become
ideal in the thermodynamic limit. Furthermore, the counting of solutions
implied by the discrete Takahashi equations agrees in this case with
the counting obtained directly from the Lieb-Wu equations.

Section IV is devoted to the self-consistent solution of the Lieb-Wu
equations for three electrons and one down spin. Self consistency
arguments underly the derivation of the discrete Takahashi equations.
In the simple case considered in this section we can be more explicit.
We calculate the deviations of the rapidities and momenta (solutions
of the Lieb-Wu equations) from their ideal positions (solutions of the
discrete Takahashi equations). It turns out that these deviations
vanish exponentially in the thermodynamic limit. This is our second
positive test of the string hypothesis.

In section V we complement the analytical considerations of the
previous sections with numerical results. Numerical data based on the
Lieb-Wu equations are compared with data, which were obtained
independently of the Bethe ansatz by direct numerical diagonalization
of the Hamiltonian. We find perfect agreement between the two numerical
methods. The energy levels obtained by the two methods agree within a
numerical error of $\CO(10^{- 15})$. Our numerical study confirms the
completeness of the Bethe ansatz and the correctness of the counting of
the solutions implied by the string hypothesis. This is our third
successful test of the string hypothesis.

In section VI we review Takahashi's approach to the thermodynamics
of the Hubbard model \cite{Takahashi72}. We show that the dressed
energies of all $k$-$\La$ strings (bound states of electrons) follow
from Takahashi's integral equations (thermodynamic Bethe ansatz
equations) in the zero temperature limit. We can actually do better.
Starting from the thermodynamic Bethe ansatz equations and passing to
the zero temperature limit we obtain a complete classification of all
elementary excitations at zero magnetic field, below half-filling.

Takahashi derived his equations in order to calculate thermodynamic
quantities such as the specific heat or charge and spin
susceptibilities for the Hubbard model \cite{Takahashi72,%
Takahashi74,KUO89,UKO90}. Nowadays there is an independent method to
calculate these quantities, which does not rely on strings. It is
called the quantum transfer matrix method \cite{SuIn87,SAW90,JKS98}.
In section VII we compare the results of both methods and find that
they agree well. The string hypothesis also passes this significant
test. If the string hypothesis would miss one of the elementary
excitations, it should be visible in the thermodynamics.

Section VIII contains a brief conclusion and a list of interesting
open problems.

In appendix A we present a derivation of the Bethe ansatz wave function
for the Hubbard model. It can be seen from the derivation that charge
momenta and spin rapidities do not have to be real. To every solution
of the Lieb-Wu equations there corresponds a well defined periodic wave
function. This is particularly the case for the $k$-$\Lambda$ string
solutions.

In appendix B we derive the algebraic Bethe ansatz solution of the
inhomogeneous isotropic Heisenberg model, which is needed to construct
the Bethe ansatz wave function in appendix A.

Appendix C contains the tables of our numerical data for three
electrons and one down spin.

%%%%%%%%%%%%%%%%%%%%%%%%%%%%%%%%%%%%%%%%%%%%%%%%%%%%%%%%%%%%%%%%%%%%%%%%
%   scd1.tex                                                           %
%%%%%%%%%%%%%%%%%%%%%%%%%%%%%%%%%%%%%%%%%%%%%%%%%%%%%%%%%%%%%%%%%%%%%%%%
\section{Bethe ansatz for the Hubbard model}
\subsection{Eigenfunctions and eigenvalues}
The Hamiltonian of the one-dimensional Hubbard model on a periodic
$L$-site chain may be written as
\beq \label{ham}
     H = - \sum_{j=1}^L \sum_{\s = \auf, \ab}
	   (c_{j, \s}^+ c_{j+1, \s} + c_{j+1, \s}^+ c_{j, \s})
	   + U \sum_{j=1}^L
	   (n_{j \auf} - \tst{\2})(n_{j \ab} - \tst{\2}) \qd.
\eeq
$c_{j, \s}^+$ and $c_{j, \s}$ are creation and annihilation operators
of electrons in Wannier states, and periodicity is guaranteed by
setting $c_{L+1, \s} = c_{1, \s}$. $n_{j,\s} = c_{j, \s}^+ c_{j, \s}$
is the particle number operator for electrons of spin $\sigma$ at site
$j$, $U$ is the coupling constant. The eigenvalue problem for the
Hubbard Hamiltonian (\ref{ham}) was solved by Lieb and Wu \cite{LiWu68}
using the nested Bethe ansatz \cite{Yang67}. The Hubbard Hamiltonian
conserves the number of electrons $N$ and the number of down spins $M$.
The corresponding Schr\"odinger equation can therefore be solved for
fixed $N$ and $M$. Since the Hamiltonian is invariant under
particle-hole transformations and under reversal of spins
\cite{LiWu68}, we may set $2M \le N \le L$. We shall denote the
positions and spins of the electrons by $x_j$ and $\s_j$, respectively.
The Bethe ansatz eigenfunctions of the Hubbard Hamiltonian (\ref{ham})
depend on the relative ordering of the $x_j$. There are $N!$ possible
orderings of the coordinates of $N$ electrons. Any ordering may be
related to a permutation $Q$ of the numbers $1, \dots, N$ through the
inequality
\beq \label{sectorq}
     1 \le x_{Q1} \le x_{Q2} \le \dots \le x_{QN} \le L \qd.
\eeq
This inequality divides the configuration space of $N$ electrons into
$N!$ sectors, which can be labeled by the permutations~$Q$. The Bethe
ansatz eigenfunctions of the Hubbard Hamiltonian (\ref{ham}) in the
sector $Q$ are given as
\beq \label{wwf}
     \ps (x_1, \dots, x_N; \s_1, \dots, \s_N) =
	\sum_{P \in S_N} \sign(PQ) \, \ph_P (\s_{Q1}, \dots, \s_{QN})
	\exp \left( \i \sum_{j=1}^N k_{Pj} x_{Qj} \right) \qd.
\eeq
Here the $P$-summation extends over all permutations of the numbers
$1, \dots, N$. These permutations form the symmetric group $S_N$.
The function $\sign(Q)$ is the sign function on the symmetric group,
which is $- 1$ for odd permutations and $+ 1$ for even permutations.
The spin dependent amplitudes $\ph_P (\s_{Q1}, \dots, \s_{QN})$ can be
found in Woynarovich's paper \cite{Woynarovich82a}. They are of the
form of the Bethe ansatz wave functions of an inhomogeneous XXX spin
chain,
\beq \label{wswf}
     \ph_P (\s_{Q1}, \dots, \s_{QN}) = \sum_{\p \in S_M}
	A(\la_{\p 1}, \dots, \la_{\p M})
	\prod_{l=1}^M F_P (\la_{\p l}; y_l) \qd.
\eeq
Here $F_P (\la; y)$ is defined as
\beq
     F_P (\la; y) = \frac{1}{\la - \sin k_{Py} + \i U/4}
		    \prod_{j=1}^{y-1} \frac{\la - \sin k_{Pj} - \i U/4}
		    {\la - \sin k_{Pj} + \i U/4} \qd,
\eeq
and the amplitudes $A(\la_1, \dots, \la_M)$ are given by
\beq \label{wwfsa}
     A(\la_1, \dots, \la_M) = \prod_{1 \le m < n \le M}
	\frac{\la_m - \la_n - \i U/2}{\la_m - \la_n} \qd.
\eeq
$y_j$ in the above equations denotes the position of the $j$th down
spin in the sequence $\s_{Q1}, \dots, \s_{QN}$. The $y$'s are thus
`coordinates of down spins on electrons'. Below we shall illustrate
the notation through an explicit example.

The wave functions (\ref{wwf}) are characterized by two sets of
quantum numbers $\{k_j\}$ and $\{\la_l\}$. These quantum numbers
may be generally complex. The $k_j$ and $\la_l$ are called charge
momenta and spin rapidities, respectively. The charge momenta and spin
rapidities satisfy the Lieb-Wu equations
\bea \label{bak}
     e^{\i k_j L} & = & \prod_{l=1}^M \frac{\la_l - \sin k_j - \i U/4}
                                      {\la_l - \sin k_j + \i U/4} \qd,
		                      \qd j = 1, \dots, N \qd, \\
				      \label{bas}
     \prod_{j=1}^N \frac{\la_l - \sin k_j - \i U/4}
                        {\la_l - \sin k_j + \i U/4} & = &
     \prod_{m=1 \atop m \ne l}^M \frac{\la_l - \la_m - \i U/2}
                        {\la_l - \la_m + \i U/2} \qd,
			\qd l = 1, \dots, M \qd.
\eea
A derivation of the wave function (\ref{wwf}) and the Lieb-Wu equations
(\ref{bak}), (\ref{bas}) is presented in appendices A and B.

The wave functions (\ref{wwf}) are joint eigenfunctions of the
Hubbard Hamiltonian (\ref{ham}) and the momentum operator%
\footnote{For a proper definition of the momentum operator see
appendix B of \cite{GoMu97b}} with eigenvalues
\beq \label{enmom}
     E = - 2 \sum_{j=1}^N \cos k_j + \frac{U}{4}(L - 2N) \qd, \qd
     P = \left( \sum_{j=1}^N k_j \right) \mod \, 2\p \qd.
\eeq

The `coordinates of down spins' $y_j$ which enter (\ref{wswf}) depend
on $(\s_1, \dots, \s_N)$ {\it and} on $(x_1, \dots, x_N)$. The
following example should help to understand the notation. Let $L = 12$,
$N = 5$, $M = 2$, and let, for example, $(x_1, \dots, x_5) =
(7,3,5,1,8)$, $(\s_1, \dots, \s_5) = (\auf \auf \auf \ab \ab)$. Then
$x_4 \le x_2 \le x_3 \le x_1 \le x_5$, i.e.\ $Q = (4,2,3,1,5)$. It
follows that $(x_{Q1}, \dots, x_{Q5}) = (1,3,5,7,8)$ and $(\s_{Q1},
\dots, \s_{Q5}) = (\ab \auf \auf \auf \ab)$. Thus $y_1 = 1$, $y_2 = 5$.

Whenever it will be necessary, we shall indicate the dependence of the
wave functions (\ref{wwf}) on the charge momenta and spin rapidities
by subscripts, $\ps = \ps_{k_1, \dots, k_N; \la_1, \dots, \la_M}$.
Let us consider the symmetries of the eigenfunctions under permutations,
\bea \label{perm1}
     \ps (x_{P1}, \dots, x_{PN}; \s_{P1}, \dots, \s_{PN}) & = &
	 \sign(P) \ps (x_1, \dots, x_N; \s_1, \dots, \s_N) \qd, \qd
	 P \in S_N \qd, \\ \label{perm2}
     \ps_{k_{P1}, \dots, k_{PN}; \la_1, \dots, \la_M} & = &
	 \sign(P) \ps_{k_1, \dots, k_N; \la_1, \dots, \la_M} \qd, \qd
	 P \in S_N \qd, \\ \label{perm3}
     \ps_{k_1, \dots, k_N; \la_{P1}, \dots, \la_{PM}} & = &
	 \ps_{k_1, \dots, k_N; \la_1, \dots, \la_M} \qd, \qd
	 P \in S_M \qd.
\eea
Equation (\ref{perm1}) means that the eigenfunctions respect the Pauli
principle. (\ref{perm2}) and (\ref{perm3}) describe their properties
with respect to permutations of the quantum numbers. They are totally
antisymmetric with respect to interchange of the charge momenta $k_j$,
and they are totally symmetric with respect to interchange of the spin
rapidities $\la_l$. Hence, in order to find all Bethe ansatz wave
functions we have to solve the Lieb-Wu equations (\ref{bak}),
(\ref{bas}) modulo permutations of the sets $\{k_j\}$ and $\{\la_l\}$.
The $k_j$'s have to be mutually distinct, since otherwise the wave
function vanishes due to (\ref{perm2}). In fact, the $\la_l$'s have to
be mutually distinct, too. This is called the `Pauli principle for
interacting Bosons' (see \cite{KBIBo}). We would like to emphasize that
there are no further restrictions on the solutions of (\ref{bak}),
(\ref{bas}). In particular, the spin and charge rapidities do {\it not}
have to be real.

Bethe ansatz states on a finite lattice of length $L$ that have finite
momenta $k_j$ and rapidities $\la_l$, a non-negative value of the
total spin ($N - 2M > 0$), and a total number of electrons not larger
than the length of the lattice ($N \le L$) are called {\it regular}
(cf.\ \cite{EKS92a}, page 562).

There exist two discrete symmetries of the model which can be used to
obtain additional eigenstates from the regular ones \cite{LiWu68}.
The Hamiltonian is invariant under exchange of up and down spins.
This symmetry allows for obtaining eigenstates with negative value
$N- 2M$ of the total spin from eigenstates with positive value of the 
total spin. This symmetry does not affect the number of electrons.
Thus, its action on regular states does not lead above half filling.
States above half filling ($N > L$) can be obtained by employing the
transformation $c_{j \s} \rightarrow (- 1)^j c_{j \s}^+$, $c_{j \s}^+
\rightarrow (- 1)^j c_{j \s}$, $\s = \auf, \ab$, which leaves the
Hamiltonian (\ref{ham}) invariant, but maps the empty Fock state
$|0\>$ to the completely filled Fock state $|\auf \ab\>$.
\subsection{SO(4) symmetry}
The Hubbard Hamiltonian (\ref{ham}) is invariant under rotations in
spin space. The corresponding su(2) Lie algebra is generated by the
operators
\beq \label{rot}
     \begin{array}{r@{\qd, \qd}c@{\qd, \qd}l}
        \dst{\z = \sum_{j=1}^L c_{j \auf}^+ c_{j \ab}} &
        \dst{\z^\dagger = \sum_{j=1}^L c_{j \ab}^+ c_{j \auf}} &
        \dst{\z^z = \tst{\2} \sum_{j=1}^L (n_{j \ab} - n_{j \auf}) \qd.}
	   \\[2ex]
        [\z,\z^\dagger] = - 2 \z^z &
        [\z,\z^z] = \z &
        [\z^\dagger,\z^z] = - \z^\dagger \qd.
     \end{array}
\eeq
For lattices of even length $L$ there is another representation of
su(2), which commutes with the Hubbard Hamiltonian \cite{HeLi71,%
Yang89,Pernici90}.
This representation generates the so-called $\h$-pairing symmetry,
\beq \label{eta}
     \begin{array}{r@{\qd, \qd}c@{\qd, \qd}l}
        \dst{\h = \sum_{j=1}^L (- 1)^j c_{j \auf} c_{j \ab}} &
        \dst{\h^\dagger = \sum_{j=1}^L (- 1)^j
	   c_{j \ab}^+ c_{j \auf}^+} &
        \dst{\h^z = \tst{\2} \sum_{j=1}^L (n_{j \ab} + n_{j \auf}) 
	   - \tst{\2 L} \qd.}
	   \\[2ex]
        [\h,\h^\dagger] = - 2 \h^z &
        [\h,\h^z] = \h &
        [\h^\dagger,\h^z] = - \h^\dagger \qd.
     \end{array}
\eeq
The generators of both algebras commute with one-another. They combine
into a representation of su(2)$\oplus$su(2).

The $\h$-pairing symmetry connects sectors of the Hilbert space with
different numbers of electrons. The operator $\h^\dagger$, for instance,
creates a local pair of electrons of opposite spin and momentum $\p$.
Hence, in order to consider the action of the $\h$-symmetry on
eigenstates we write them in second quantized form.
\beq \label{sq}
     |k_1, \dots, k_N; \la_1, \dots, \la_M \> =
	\sum_{x_1, \dots, x_N = 1}^L
        \ps_{k_1, \dots, k_N; \la_1, \dots, \la_M} 
	(x_1, \dots, x_N; \s_1, \dots, \s_N)
	c_{x_1, \s_1}^+ \dots c_{x_N, \s_N}^+ |0\> \qd,
\eeq
where $\s_1 = \dots = \s_M = \ab$ and $\s_{M + 1} = \dots = \s_N =
\auf$. It is easily seen that
\beq
     (\z^z + \h^z) |k_1, \dots, k_N; \la_1, \dots, \la_M \> =
	(M - \tst{\frac{L}{2}})
        |k_1, \dots, k_N; \la_1, \dots, \la_M \> \qd.
\eeq
Here $M - \frac{L}{2}$ is integer, since $L$ is even. Therefore the
symmetry group generated by the representations (\ref{rot}), (\ref{eta})
is SO(4) rather than SU(2)$\times$SU(2) \cite{YaZh90}.

It was shown in \cite{EKS92a} that the {\it regular} Bethe ansatz
states are lowest weight vectors of both su(2) symmetries (\ref{rot})
and
(\ref{eta}),
\beq \label{lw}
     \z |k_1, \dots, k_N; \la_1, \dots, \la_M \> = 0 \qd, \qd
     \h |k_1, \dots, k_N; \la_1, \dots, \la_M \> = 0 \qd.
\eeq
This is an important theorem. It was the prerequsite for the proof
of completeness (see section D) of the Bethe ansatz for the Hubbard
model in \cite{EKS92b}. The proof of (\ref{lw}) is direct but lengthy
\cite{EKS92a}. $\z$ and $\h$ are applied to the states (\ref{sq}), and
the Lieb-Wu equations (\ref{bak}), (\ref{bas}) are used to reduce the
resulting expressions to zero. We would like to emphasize that the
proof of (\ref{lw}) is {\it not} restricted to real solutions of the
Lieb-Wu equations. It goes through for all solutions corresponding to
regular Bethe ansatz states including the strings.

Since the two su(2) symmetries (\ref{rot}), (\ref{eta}) leave the
Hubbard Hamiltonian (\ref{ham}) invariant, additional eigenstates which
do not belong to the regular Bethe ansatz can be obtained by applying
$\z^\dagger$ and $\h^\dagger$ to regular Bethe ansatz eigenstates. Since
\bea
     \z^z |k_1, \dots, k_N; \la_1, \dots, \la_M \> & = &
        (M - \tst{\frac{N}{2}}) |k_1, \dots, k_N; \la_1, \dots, \la_M \>
	\qd, \\
     \h^z |k_1, \dots, k_N; \la_1, \dots, \la_M \> & = &
        \tst{\2}(N - L) |k_1, \dots, k_N; \la_1, \dots, \la_M \> \qd,
\eea
a state $|k_1, \dots, k_N; \la_1, \dots, \la_M \>$ has spin
$\2 (N - 2M)$ and $\h$-spin $\2 (L - N)$. The dimension of the
corresponding multiplet is thus given by
\beq \label{so4mult}
     \dim_{M, N} = (N - 2M + 1)(L - N + 1) \qd.
\eeq
The states in this multiplet are of the form
\beq \label{so4ext}
     |k_1, \dots, k_N; \la_1, \dots, \la_M; \a; \be \> =
        (\z^\dagger)^\a (\h^\dagger)^\be
	|k_1, \dots, k_N; \la_1, \dots, \la_M \> \qd,
\eeq
where $\a = 0, \dots, N - 2M$ and $\be = 0, \dots, L - N$. Note that
states of the form (\ref{so4ext}) can be obtained from regular Bethe
ansatz states with $\tilde N \ge N$, $\tilde M \ge M$ by formally
setting some of the charge momenta and spin rapidities equal to
infinity \cite{Woynarovich82a,Gaudin83,FaTa84}.
\subsection{Discrete Takahashi equations}
Let us now formulate Takahashi's string hypothesis \cite{Takahashi72}
more precisely: All regular solutions $\{k_j\}$, $\{\la_l\}$ of the
Lieb-Wu equations (\ref{bak}), (\ref{bas}) consist of three different
kinds of configurations.
\begin{enumerate}
\item
A single real momentum $k_j$.
\item
$m$ $\la$'s combining into a $\La$ string. This includes the case
$m = 1$, which is just a single $\La_\a$.
\item
$2m$ $k$'s and $m$ $\la$'s combining into a $k$-$\La$ string.
\end{enumerate}
For large lattices ($L \gg 1$) and a large number of electrons
($N \gg 1$), almost all strings are close to ideal, i.e.\ the imaginary
parts of the $k$'s and $\la$'s are almost equally spaced.

For ideal $\La$ strings of length $m$ the rapidities involved are
\beq \label{ideal1}
     \La_\a^{m, j} = \La_\a^m + (m - 2j +1) \tst{\frac{\i U}{4}} \qd.
\eeq
Here $\a$ enumerates the strings of the same length $m$, and $j = 1,
\dots, m$ counts the $\la$'s involved in the $\a$th $\La$ string of
length $m$. $\La_\a^m$ is the real center of the string.

The $k$'s and the $\la$'s involved in an ideal $k$-$\La$ string are
(for $U > 0$)
\bea
     k_\a^1 & = & \p - \arcsin({\La'}_\a^m + m \tst{\frac{\i U}{4}})
		  \qd, \nn \\
     k_\a^2 & = & \arcsin({\La'}_\a^m + (m - 2) \tst{\frac{\i U}{4}})
		  \qd, \nn \\
     k_\a^3 & = & \p - k_\a^2 \qd, \nn \\
	 & \vdots & \label{idealk} \\
     k_\a^{2m - 2} & = & \arcsin({\La'}_\a^m - (m - 2)
			 \tst{\frac{\i U}{4}}) \qd, \nn \\
     k_\a^{2m - 1} & = & \p - k_\a^{2m - 2} \qd, \nn \\
     k_\a^{2m} & = & \p - \arcsin({\La'}_\a^m - m \tst{\frac{\i U}{4}})
		     \qd, \nn
\eea
and
\beq \label{ideal3}
     {\La'}_\a^{m, j} = {\La'}_\a^m + (m - 2j + 1) \tst{\frac{\i U}{4}}
			\qd.
\eeq
Again $m$ denotes the `length' of the string, $\a$ enumerates strings
of length $m$, and $j$ counts the $\la$'s involved in a given string.
${\La'}_\a^m$ is the real center of the $k$-$\La$ string. The branch
of $\arcsin(x)$ in (\ref{idealk}) is fixed as $- \pi/2 \le \text{Re}
(\arcsin (x)) \le \pi/2$.

The string hypothesis assumes that almost all solutions of the Lieb-Wu
equations (\ref{bak}), (\ref{bas}) are approximately given by
(\ref{ideal1})-(\ref{ideal3}) with exponentially small corrections of
order $\CO (\exp( - \de L))$, where $\de$ is real and positive and
depends on the specific string under consideration.

Using the string hypothesis inside the Lieb-Wu equations (\ref{bak}),
(\ref{bas}) and taking logarithms afterwards, we arrive at the
following form of Bethe ansatz equations for strings, which we call
discrete Takahashi equations
\begin{eqnarray} \label{t1}
     k_j L & = & 2 \pi I_j - \sum_{n=1}^\infty \sum_{\alpha = 1}^{M_n}
                 \theta \left(
		 \frac{\sin k_j - \Lambda_\alpha^n}{nU/4} \right)
                 - \sum_{n=1}^\infty \sum_{\alpha = 1}^{M_n'}
                 \theta \left(
		 \frac{\sin k_j - {\Lambda'}_\alpha^n}{nU/4} \right),
		 \\ \label{t2}
     \sum_{j=1}^{N - 2M'} \theta \left(
		 \frac{\Lambda_\alpha^n - \sin k_j}{nU/4} \right) & = &
		 2 \pi J_\alpha^n +
		 \sum_{m=1}^\infty \sum_{\beta = 1}^{M_m}
		 \Theta_{nm} \left(
		 \frac{\Lambda_\alpha^n - \Lambda_\beta^m}{U/4} \right),
		 \\ \label{t3}
     L [\arcsin({\Lambda'}_\alpha^n + n \tst{\frac{\i U}{4}})
        + \arcsin({\Lambda'}_\alpha^n - n \tst{\frac{\i U}{4}})] & = &
	         2 \pi {J'}_\alpha^n +
		 \sum_{j=1}^{N - 2M'} \theta \left(
		 \frac{{\Lambda'}_\alpha^n - \sin k_j}{nU/4} \right) +
		 \sum_{m=1}^\infty \sum_{\beta = 1}^{M_m'}
		 \Theta_{nm} \left(
		 \frac{{\Lambda'}_\alpha^n - {\Lambda'}_\beta^m}{U/4}
		 \right).
\end{eqnarray}
Here we assumed $L$ to be even. $I_j$, $J_\alpha^n$, and
${J'}_\alpha^n$ are integer or half-odd integer numbers, according to
the following prescriptions: $I_j$ is integer (half odd integer), if
$\sum_m (M_m + M_m')$ is even (odd); the $J_\a^n$ are integer (half odd
integer), if $N - M_n$ is odd (even); the ${J'}_\a^n$ are integer (half
odd integer), if $L - (N - M_n')$ is odd (even). $M_n$ and $M_m'$ are
the numbers of $\Lambda$ strings of length $n$, and $k$-$\Lambda$
strings of length $m$ in a specific solution of the system (\ref{t1})-%
(\ref{t3}). $M' = \sum_{n=1}^\infty n M_n'$, is the total number of
$\la$'s involved in $k$-$\La$ strings. The integer (half-odd integer)
numbers in (\ref{t1})-(\ref{t3}) have ranges
\begin{eqnarray} \label{r1}
     && - \frac{L}{2} < I_j \le \frac{L}{2}, \\ \label{r2}
     && |J_\alpha^n| \le \frac{1}{2}
        \left(N - 2M' - \sum_{m=1}^\infty t_{nm} M_m - 1 \right), \\
	\label{r3}
     && |{J'}_\alpha^n| \le \frac{1}{2}
        \left(L - N + 2M' - \sum_{m=1}^\infty t_{nm} M_m' - 1 \right),
\end{eqnarray}
where $t_{mn} = 2 \min (m,n) - \delta_{mn}$. The functions $\th$ and
$\Theta_{nm}$ in (\ref{t1})-(\ref{t3}) are defined as $\theta(x) =
2 \arctan(x)$, and
\begin{equation} \label{defthetas}
     \Theta_{nm} (x) = \left\{ \begin{array}{l}
	{\displaystyle
        \theta \left( \frac{x}{|n - m|} \right) +
        2 \theta \left( \frac{x}{|n - m| + 2} \right) + \cdots +
        2 \theta \left( \frac{x}{n + m - 2} \right) +
        \theta \left( \frac{x}{n + m} \right), \: \text{if} \quad
	n \ne m,} \\[3ex]
	{\displaystyle
        2 \theta \left( \frac{x}{2} \right) +
        2 \theta \left( \frac{x}{4} \right) + \cdots +
        2 \theta \left( \frac{x}{2n - 2} \right) +
        \theta \left( \frac{x}{2n} \right), \: \text{if} \quad n = m.}
	\end{array} \right.
\end{equation}

In terms of the parameters of the ideal strings total energy and
momentum (\ref{enmom}) are expressed as
\bea \label{mom}
     P & = & \left[ \sum_{j=1}^{N - 2M'} k_j -
		    \sum_{n=1}^\infty \sum_{\a = 1}^{M_n'}
		    \left(2 \, \Re \arcsin \left( {\La'}_\a^n +
		     n \tst{\frac{\i U}{4}} \right) -
		     (n + 1) \p \right) \right] \mod 2 \p \qd, \\
		     \label{en}
     E & = & - 2 \sum_{j=1}^{N - 2M'} \cos(k_j) +
		 4 \sum_{n=1}^\infty \sum_{\a = 1}^{M_n'}
		 \Re \sqrt{1 - \left( {\La'}_\a^n +
		 n \tst{\frac{\i U}{4}} \right)^2}
		 + \frac{U}{4} (L - 2N) \qd.
\eea

Equations (\ref{t1})-(\ref{r3}) can be used to study all excitations
of the Hubbard model in the thermodynamic limit. They are the basis
for the derivation of Takahashi's integral equations \cite{Takahashi72},
which determine the thermodynamics of the Hubbard model (see sections
VI and VII). Applications of (\ref{t1})-(\ref{r3}) are usually based
on the following assumptions.
\begin{enumerate}
\item
Any set of non-repeating (half odd) integers $I_j$, $J_\a^n$,
${J'}_\a^n$ subject to the constraints (\ref{r1})-(\ref{r3}) specifies
one and only one solution $\{k_j\}$, $\{\La_\a^n\}$, $\{{\La'}_\a^n\}$
of equations (\ref{t1})-(\ref{t3}).
\item
The solutions $\{k_j\}$, $\{\La_\a^n\}$, $\{{\La'}_\a^n\}$ of
(\ref{t1})-(\ref{t3}) specified by a set of non-repeating (half odd)
integers $I_j$, $J_\a^n$, ${J'}_\a^n$ subject to (\ref{r1})-(\ref{r3})
are in one-to-one correspondence to solutions of the Lieb-Wu equations
(\ref{bak}), (\ref{bas}).
\item
For large $L$ and $N$ almost every solution $\{k_j\}$, $\{\la_l\}$ of
the Lieb-Wu equations (\ref{bak}), (\ref{bas}) is exponentially close to
the corresponding solution $\{k_j\}$, $\{\La_\a^n\}$, $\{{\La'}_\a^n\}$
of the discrete Takahashi equations, which means that the strings
contained in $\{k_j\}$, $\{\la_l\}$ are well approximated by the ideal
strings determined by $\{k_j\}$, $\{\La_\a^n\}$, $\{{\La'}_\a^n\}$.
\end{enumerate}
\subsection{Completeness of the Bethe ansatz}
The proof of completeness of the Bethe ansatz given in \cite{EKS92b}
is  based on assumptions (i) and (ii) above. Similar assumptions were
proved for other Bethe ansatz solvable models \cite{KBIBo}. Note that
assumption (ii) does {\it not} mean that the classification of the
solutions of the Lieb-Wu equations (\ref{bak}), (\ref{bas}) into strings
is actually given by (\ref{t1})-(\ref{r3}). There may be a
redistribution between different kinds of strings. This phenomenon was
observed in a number of Bethe ansatz solvable models and was carefully
studied by examples \cite{Bethe31,EKS92c,EKS92b} (see also section
V.B). It turned out that the redistribution did in no case affect the
total number of solutions of the Bethe ansatz equations.

Using (i) and (ii) above, the proof of completeness reduces to a
combinatorial problem based on (\ref{r1})-(\ref{r3}) \cite{EKS92b}.
From (\ref{r1})-(\ref{r3}) we read off the numbers of allowed values
of the (half odd) integers $I_j$, $J_\a^n$, ${J'}_\a^n$ in a given
configuration $\{M_n\}$, $\{M_n'\}$ of strings. These numbers are
\begin{enumerate}
\item
$L$ for a free $k_j$ (not involved in a $k$-$\La$ string),
\item
$N - 2M' - \sum_{m=1}^\infty t_{nm} M_m$ for a $\La$ string of length
$n$,
\item
$L - N + 2M' - \sum_{m=1}^\infty t_{nm} M_m'$ for a $k$-$\La$ string
of length $n$.
\end{enumerate}
The total number of ways to select the $I_j$, $J_\a^n$, ${J'}_\a^n$
(recall that they are assumed to be non-repeating) for a given
configuration $\{M_n\}$, $\{M_n'\}$ is thus
\beq
     n(\{M_n\},\{{M'}_n\}) = {L \choose N - 2M'}
	\prod_{n=1}^\infty
	{N - 2M' - \sum_{m=1}^\infty t_{nm} M_m \choose M_n}
	\prod_{n=1}^\infty
        {L - N + 2M' - \sum_{m=1}^\infty t_{nm} M_m' \choose M_n'} \qd.
\eeq
Hence, the number of regular Bethe ansatz states for given numbers $N$
of electrons and $M$ of down spins is
\beq \label{nreg}
     n_{\rm reg} (M,N) = \sum_{\{M_n\}, \{M_n'\}} n(\{M_n\},\{M_n'\})
			 \qd,
\eeq
where the summation is over all configurations of strings which satisfy
the constraints $N - 2M' \ge 0$ and $M = \sum_{m=1}^\infty m(M_m +
M_m')$. Finally, the total number of states (\ref{so4ext}) in the SO(4)
extended Bethe ansatz is
\beq \label{ntot}
     n_{\rm tot} (L) = \sum_{M, N} n_{\rm reg} (M, N) \dim_{M, N}
                     = \sum_{M, N} n_{\rm reg} (M, N)
		       (N - 2M + 1)(L - N +1) \qd,
\eeq
where the sum is over all $M$, $N$ with $0 \le 2M \le N \le L$. The
sums (\ref{nreg}) and (\ref{ntot}) were calculated in \cite{EKS92b}.
It turns out that
\beq
     n_{\rm tot} (L) = 4^L \qd,
\eeq
which is the dimension of the Hilbert space of the Hubbard model on an
$L$-site chain.

Let us list again the essential steps that led to the above proof of
completeness:
\begin{enumerate}
\item
Impose periodic boundary conditions.
\item
Take Woynarovich's wave function (\ref{wwf})-(\ref{wwfsa}).
\item
Define the Bethe ansatz in the narrow sense of regular Bethe ansatz
(see below (\ref{perm3})). This eliminates infinite $k$'s and $\la$'s
whose multiplicities are not under control.
\item
Prove the lowest weight theorem (\ref{lw}). Then gluing back solutions
with infinite $k$'s and $\la$'s is equivalent to considering the
multiplets (\ref{so4ext}).
\item
The multiplicities of occupation of infinite $k$'s and $\la$'s are given
by the dimensions (\ref{so4mult}) of the multiplets (\ref{so4ext}).
\item
Use Takahashi's integers (\ref{r1})-(\ref{r3}) for counting of the
regular Bethe ansatz states.
\end{enumerate}

%%%%%%%%%%%%%%%%%%%%%%%%%%%%%%%%%%%%%%%%%%%%%%%%%%%%%%%%%%%%%%%%%%%%%%%%
%   scd2.tex                                                           %
%%%%%%%%%%%%%%%%%%%%%%%%%%%%%%%%%%%%%%%%%%%%%%%%%%%%%%%%%%%%%%%%%%%%%%%%
\section{Lieb-Wu equations for a single down spin (I) -- Graphical
solution}
In this section we study the Lieb-Wu equations (\ref{bak}), (\ref{bas})
in the most simple non-trivial case, when there is only one down spin,
$M = 1$. For pedagogical clarity some emphasis will be on the most
instructive cases $N = 2$ and $N = 3$. These are the cases which we
also studied numerically (cf.\ section V). Some of the analytical
calculations, however, are presented for general $N$, simply because
the general arguments are simple enough and enable treating the
cases $N = 2$ and $N = 3$ to some extent simultaneously.

The Lieb-Wu equations for $N = 2$ and $M = 1$ were studied before in
appendix B of \cite{EKS92b}. There the emphasis was on the
redistribution phenomenon mentioned in section II.D. For $U = 0$ the
Hubbard Hamiltonian turns into a free tight binding Hamiltonian,
and there is no bound state of electrons ($k$-$\La$ string) left.
It is therefore clear that bound states decay as the coupling becomes
weaker. In appendix B of \cite{EKS92b} it was shown that each time
a $k$-$\La$ string disappears from the spectrum at a certain critical
value of the coupling $U > 0$, a new real solution emerges. Here we take
a slightly different point of view. We fix $U$ and study the solutions
for large finite $L$. It turns out that there is no redistribution
phenomenon for the most simple $k$-$\La$ strings consisting of two
complex conjugated $k$'s and one real $\la$ as $L \rightarrow \infty$.
These strings always exist for large enough finite $L$, and their
number is in accordance with the counting implied by Takahashi's
discrete equations (\ref{t1})-(\ref{r3}). In this respect the $k$-$\La$
strings of the Hubbard model are different from the $\La$ strings in
the XXX spin chain \cite{EKS92c}.

For $M = 1$ the Lieb-Wu equations (\ref{bak}), (\ref{bas}) read
\bea \label{bak1}
     && e^{\i k_j L} = \frac{\La - \sin k_j - \i U/4}
                         {\La - \sin k_j + \i U/4} \qd, \qd
                          j = 1, \dots, N \qd,\\ \label{bas1}
     && \prod_{j = 1}^N \frac{\La - \sin k_j - \i U/4}
			  {\La - \sin k_j + \i U/4} = 1 \qd.
\eea
Equation (\ref{bas1}) can be replaced by the equation for the
conservation of momentum,
\beq \label{emom}
     e^{\i (k_1 + \dots + k_N)L} = 1 \qd.
\eeq
(\ref{bak1}) and (\ref{emom}) follow from (\ref{bak1}) and (\ref{bas1})
and vice versa. Let us take the logarithm of (\ref{emom}) and solve
(\ref{bak1}) for $\sin k_j - \La$. We obtain the equations
\bea \label{momlog}
     (k_1 + \dots + k_N) \mod \, 2 \p & = & \frac{m2\p}{L} \qd, \qd
        m = 0, \dots, L - 1 \qd, \\ \label{laofk}
     \sin k_j - \La & = & \frac{U}{4} \ctg \left( \frac{k_j L}{2}
	\right)
     \qd, \qd j = 1, \dots, N \qd,
\eea
which are equivalent to (\ref{bak1}) and (\ref{bas1}) but more
convenient for the further discussion.
\subsection{\boldmath All charge momenta $k_j$ real}
The equation
\beq \label{laofq}
     \sin q - \La = \frac{U}{4} \ctg \left( \frac{q L}{2}
	\right)
\eeq
is easily solved graphically for $q$ as a function of $\La$ (see
figure 1). It has at least $L$ branches of solutions $q_\ell (\La)$
belonging to the interval $0 \le q < 2 \p$. There is at least one
branch with $(\ell - 1) \frac{2\p}{L} < q < \ell \frac{2\p}{L}$. Yet,
for $\frac{\p}{2} \le q \le \frac{3 \p}{2}$ there may be more than
one solution in the interval $[(\ell - 1) \frac{2\p}{L}, \ell
\frac{2\p}{L}]$, if $U$ or $L$ is too small. We have the following
uniqueness condition,
\beq
     - 1 = \min_{0 \le q < 2 \p} \6_q \sin q >
	   \max_{0 \le q < 2 \p} \6_q \frac{U}{4}
	   \ctg \left( \frac{qL}{2} \right) =
	   \max_{0 \le q < 2 \p} - \frac{UL}{8}
	   \sin^{-2} \left( \frac{qL}{2} \right) = - \frac{UL}{8} \qd,
\eeq
which is equivalent to
\beq \label{tcd}
     L > \frac{8}{U} \qd.
\eeq
We call this condition Takahashi condition. In the following we will
assume the Takahashi condition to hold.

Equation (\ref{laofq}) defines $\La$ as a function of $q$. We have
\beq
     \frac{d \La}{dq} = \cos q + \frac{UL}{8} \sin^{-2} \left(
		        \frac{qL}{2} \right) \qd.
\eeq
Then, using (\ref{tcd}), $\frac{d \La}{dq} > 0$. Hence, all branches
$q_\ell (\La)$ of solutions of (\ref{laofq}) are monotonically
increasing, $\frac{d q_\ell (\La)}{d \La} > 0$. We can summarize the
properties of the solutions of equation (\ref{laofq}) in the following
lemma.
\begin{lemma}
If $L > \frac{8}{U}$, equation (\ref{laofq}) has exactly $L$ branches
of solutions $q_\ell (\La)$, which have the properties
\beq \begin{array}{r@{\qqd}l}
     (i) & \dst{(\ell - 1) \frac{2\p}{L} \le q_\ell (\La) \le \ell
			   \frac{2\p}{L}} \qd, \\[2ex]
     (ii) & \dst{\frac{d q_\ell (\La)}{d \La} > 0} \qd, \\[2ex]
     (iii) & \dst{\lim_{\La \rightarrow - \infty} q_\ell (\La) =
	     (\ell - 1) \frac{2\p}{L} \qd, \qd
             \lim_{\La \rightarrow \infty} q_\ell (\La) =
	     \ell \frac{2\p}{L}} \qd.
     \end{array}
\eeq
\end{lemma}
(iii) can be seen from figure 1.

\begin{figure}

\begin{center}

\fbox{
\epsfxsize 8cm
\epsffile{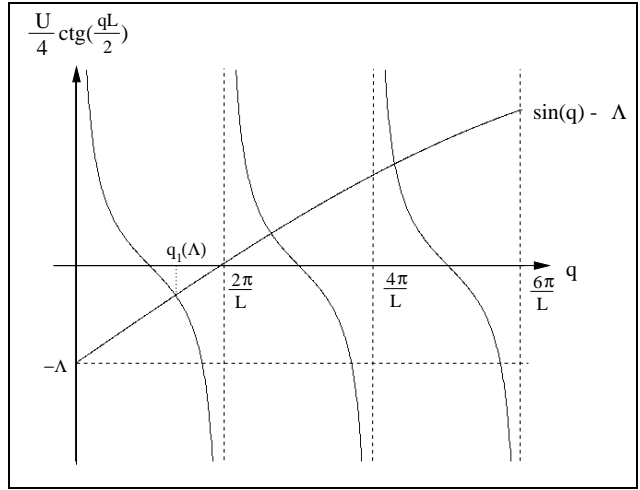}
}

\caption{Sketch of equation (\ref{laofq}).}
\end{center}

\end{figure}

Lemma III.1 is sufficient to characterize and count the real
solutions of (\ref{bak1}) and (\ref{bas1}) (or equivalently
(\ref{momlog}) and (\ref{laofk})). We are seeking for solutions of
(\ref{bak1}) and (\ref{bas1}) modulo permutations, where all $k_j$ are
mutually distinct.  Choose $N$ branches $q_{\ell_1} (\La) < \dots
< q_{\ell_N} (\La)$ of solutions of (\ref{laofq}), and define
\beq
     Q(\La) = q_{\ell_1} (\La) + \dots + q_{\ell_N} (\La) \qd.
\eeq
Then $k_j = q_{\ell_j} (\La)$, $j = 1, \dots, N$, and $\La$ solve
(\ref{bak1}) and (\ref{bas1}), if and only if
\beq \label{count}
     Q(\La) \mod \, 2 \p = m \frac{2\p}{L} \qd.
\eeq
Now, using Lemma III.1,
\beq
     \lim_{\La \rightarrow - \infty} Q(\La) = (\ell_1 + \dots +
					       \ell_N - N)
			        	      \frac{2\p}{L} \qd, \qd
     \lim_{\La \rightarrow \infty} Q(\La) = (\ell_1 + \dots + \ell_N)
					    \frac{2\p}{L} \qd.
\eeq
Furthermore, $\frac{d Q(\La)}{d \La} > 0$. Thus there are precisely
$N - 1$ values $\La_\a$, $\a = 1, \dots, N - 1$, which satisfy
(\ref{count}) for a given choice $\ell_1 < \dots < \ell_N$ of branches
of equation (\ref{laofq}) (recall that we exclude $\La = \pm \infty$).
We summarize our result in the following
\begin{lemma}
Let $L > \frac{8}{U}$. Then there are precisely ${L \choose N} (N - 1)$
solutions with all $k_j$ real to equations (\ref{bak1}), (\ref{bas1}).
To every choice of $N$ mutually distinct intervals ${\cal I}_j =
[(\ell_j - 1) \frac{2 \p}{L}, \ell_j \frac{2 \p}{L}]$, $j = 1, \dots,
N$, $\ell_j \ne \ell_k$, there correspond $N - 1$ solutions with
$k_j \in {\cal I}_j$. These solutions are characterized by $N - 1$
different values of $\La$.
\end{lemma}
Let us show that our result coincides with Takahashi's counting
(\ref{r1})-(\ref{r3}). We have $N$ real $k$'s and one $\La$ string of
length 1. Since there is no $k$-$\La$ string, there is no ${J'}_\a^n$
to specify. $M' = 0$, $M_1 = 1$, $M_j = 0$ for $j > 1$. Thus $|J^1| \le
\2 (N - t_{11} - 1) = \2 (N - 2)$, and there are $N - 1$ possible
values of $J^1$. The number of different sets $\{I_1, \dots, I_N\}$
follows from (\ref{r1}) as $L \choose N$. This means that Takahashi's
counting predicts a total number of ${L \choose N} (N - 1)$ real
solutions, which is in accordance with Lemma III.2 as long as the
Takahashi condition (\ref{tcd}) is satisfied. In the special cases
$N = 2$ and $N = 3$ we find $L \choose 2$ and $2 {L \choose 3}$
solutions, respectively.
\subsection{\boldmath $k$-$\La$ two string}
Let us consider equation (\ref{laofk}) in the case that two of the
$k_j$'s are complex conjugated and the others are real. We may set
$k_- = k_1 = q - \i \x$, $k_+ = k_2 = q + \i \x$ with real $q$ and
real, positive $\x$. It follows from (\ref{laofk}) that
\beq
     \sin(q + \i \x) = \sin(q) \ch(\x) + \i \cos(q) \sh(\x)
        = \La + \frac{U}{4} \frac{\sin(qL) - \i \, \sh(\x L)}
	                         {\ch(\x L) - \cos(qL)} \qd,
\eeq
or, if we separate real and imaginary part of this equation,
\bea \label{resin}
     \sin(q) \ch(\x) & = & \La + \frac{U}{4}
			   \frac{\sin(qL)}{\ch(\x L) - \cos(qL)} \qd,
			   \\ \label{imsin}
     \cos(q) \sh(\x) & = & - \frac{U}{4}
			   \frac{\sh(\x L)}{\ch(\x L) - \cos(qL)} \qd.
\eea
Note that for $\xi = 0$ equation (\ref{imsin}) is satisfied identically
and (\ref{resin}) turns into (\ref{laofq}).

Let us consider the two-electron case $N = 2$ first. Then, by equation
(\ref{momlog}),
\beq \label{momqn}
     q = m \frac{\p}{L} \qd, \qd m = 0, \dots, 2L - 1 \qd.
\eeq
Equation (\ref{imsin}) determines $\x$ as a function of $q = m
\frac{\p}{L}$, and equation (\ref{resin}) determines $\La$. We have
$qL = m \p$. Hence, $\sin(qL) = 0$, $\cos(qL) = (- 1)^{m}$, and
(\ref{resin}) and (\ref{imsin}) decouple into
\bea \label{laq}
     \La & = & \sin(q) \ch(\x) \\ \label{xiq}
     \sh(\x) & = & - \, \frac{U}{4 \cos(q)} \,
		     \frac{\sh(\x L)}{\ch(\x L) - (- 1)^m} \qd.
\eea
This decoupling is a peculiarity of the two particle case, which makes
it more simple than the general case. We can discuss equation
(\ref{xiq}) graphically (see figure 2).

\begin{figure}

\begin{center}

\fbox{
\epsfxsize 8cm
\epsffile{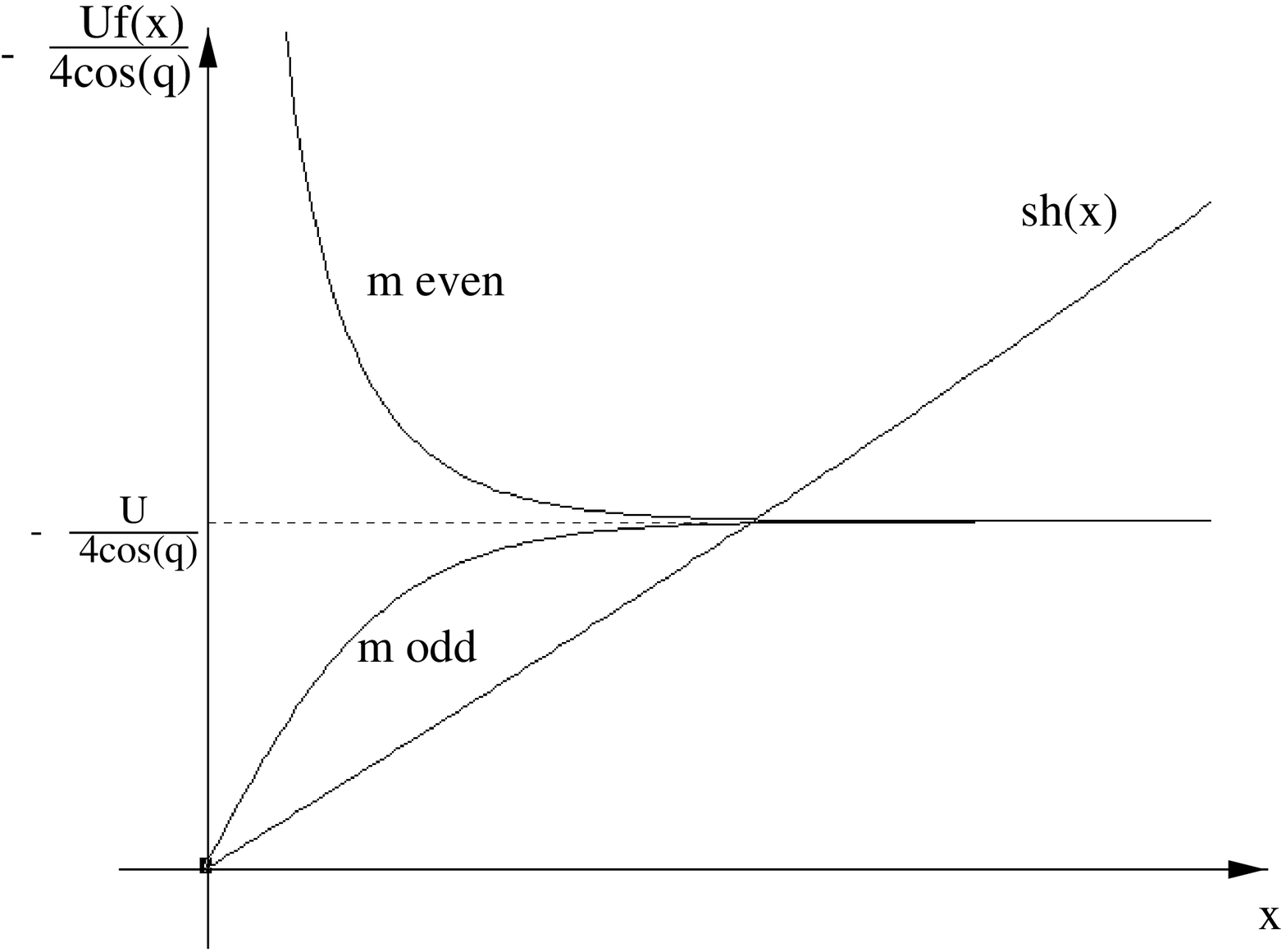}
}

\caption{Sketch of equation (\ref{xiq}).}

\end{center}

\end{figure}

Let us define
\beq \label{fofxi}
     f(\xi) = \frac{\sh(\xi L)}{\ch(\xi L) - (- 1)^m} = \left\{
	\begin{array}{l@{\qd, \qd \mbox{for $m$ }}l}
	   \tanh \left( \frac{\xi L}{2} \right) & \mbox{odd} \\[2ex]
	   \cth \left( \frac{\xi L}{2} \right) & \mbox{even} \qd.
        \end{array} \right.
\eeq
$f(\xi) > 0$ for all $\xi > 0$. Hence, (\ref{xiq}) can have solutions
for positive $\xi$ only if
\beq \label{rangem}
     \frac{\p}{2} < q < \frac{3 \p}{2} \qd, \qd \mbox{for\ } U > 0
     \qd, \qd \frac{\p}{2} < |q - \p| < \p \qd, \qd \mbox{for\ }
     U < 0 \qd.
\eeq
For the sake of simplicity let us concentrate on the repulsive case
$U > 0$. It follows from (\ref{fofxi}) that (\ref{xiq}) always has
exactly one solution $\xi_m$ for every even $m$ which satisfies
(\ref{rangem}). The condition for a solution with odd $m$ to exist is
that the derivative of the right hand side of equation (\ref{xiq}) is
larger than the derivative of the left hand side of equation (\ref{xiq})
as $\xi$ approaches zero from the right, i.e.
\beq
     - \, \frac{UL}{8 \cos(q)} > 1 \qd.
\eeq
This is satisfied for all $q$ with $\frac{\p}{2} < q < \frac{3 \p}{2}$,
if and only if
\beq
     L > \frac{8}{U} \qd.
\eeq
Again we have found the Takahashi condition (\ref{tcd}). If the
Takahashi condition is satisfied, then there is one and only one
$k$-$\La$ two string solution for every $m$ satisfying (\ref{momqn})
and (\ref{rangem}), and we can easily count these solutions. Their
number as a function of $L$ is different for odd and even $L$,
respectively.
\begin{lemma}
Let $L > \frac{8}{U}$, $U > 0$, $N = 2$. Equations (\ref{bak1}) and
(\ref{bas1}) have $k$-$\La$ two-string solutions only if the momentum
$q$ of the center of the string is in the range $\frac{\p}{2} < q
< \frac{3 \p}{2}$. The allowed values of $q$ in that range are
quantized as $q = m \frac{\p}{L}$. For every $q = m \frac{\p}{L}$
there is one and only one $k$-$\La$ two string. The total number of
$k$-$\La$ two strings is $L$ for $L$ odd and $L - 1$ for $L$ even.
\end{lemma}
Comparing our result with the prediction of Takahashi's counting
(\ref{r1})-(\ref{r3}) we find again agreement. Now there is no free
$k_j$ and no $\La$ string, thus no $I_j$ and no $J_\a^n$. Furthermore,
$M_1' = M' = 1$, and $M_j = 0$ for $j > 1$. Thus $|{J'}^1| \le
\2 (L - t_{11} - 1) = \2 (L - 2)$, which means that there are $L - 1$
possible values of ${J'}^1$. This agrees with our Lemma since $L$ was
assumed to be even in (\ref{r1})-(\ref{r3}).

Note that $\lim_{L \rightarrow \infty} f(\xi) = 1$, pointwise for all
$\xi > 0$. This suggests the notion of an ideal string determined by
the equations
\beq \label{ideal2}
     \La = \sin(q) \ch(\x) \qd, \qd
     \sh(\x) = - \, \frac{U}{4 \cos(q)} \qd, \qd
     q = m \frac{\p}{L} \qd.
\eeq
Replacing (\ref{laq}) and (\ref{xiq}) by (\ref{ideal2}) means to
replace the curves denoted by `$m$ odd' and `$m$ even' in figure 2
by the dashed line. Clearly, solutions of (\ref{laq}), (\ref{xiq})
are in one-to-one correspondence to solutions of (\ref{ideal2}), if
the Takahashi condition is satisfied. We can formulate the following
corrolary of Lemma III.3.
\begin{lemma}
Let $L > \frac{8}{U}$, $N = 2$. Let $m_L$ be a sequence of integers
($\frac{L}{2} < m_L < \frac{3L}{2}$), such that $\lim_{L \rightarrow
\infty} m_L \frac{\p}{L} = q$, $\frac{\p}{2} \le q \le \frac{3 \p}{2}$.
According to Lemma III.3 this defines a sequence $\xi_{m_L}$ of
solutions of (\ref{xiq}). This sequence has the limit
\beq \label{limes2}
     \lim_{L \rightarrow \infty} \xi_{m_L} = 
        - \, \arsh \left( \frac{U}{4 \cos(q)} \right) \qd,
\eeq
i.e.\ in the thermodynamic limit all $k$-$\La$ two strings are driven
to the ideal string positions.
\end{lemma}
Proof: (\ref{limes2}) follows from (\ref{xiq}), since the sequence
$\xi_{m_L}$ is bounded from below. Let us prove the latter statement.
Assume the contrary. Then there is a subsequence $\xi_{m_{L_j}}$
of $\xi_{m_L}$ which goes to zero, and it follows from (\ref{xiq})
that $\lim_{j \rightarrow \infty} f(\xi_{n_{L_j}}) = 0$. This means
(i) $m_{L_j}$ is odd for all sufficiently large $j$, and (ii)
$\lim_{j \rightarrow \infty} \sh(\xi_{m_{L_j}} L_j) = 
\lim_{j \rightarrow \infty} \xi_{m_{L_j}} L_j = 0$. We conclude that
\beq
     \lim_{j \rightarrow \infty}
        \frac{\sh(\xi_{m_{L_j}})}{\sh(\xi_{m_{L_j}} L_j)} = 
     \lim_{j \rightarrow \infty}
	\frac{1}{L_j} = 0 =
     \lim_{j \rightarrow \infty}
	- \frac{U}{8 \cos(m_{L_j} \frac{2 \p}{L_j})} = 
	- \frac{U}{8 \cos(q)} \qd,
\eeq
which is a contradiction. Thus, the lemma is proved.

Let us now proceed with the case $N = 3$. We have to solve the following
system of equations (cf.\ (\ref{momlog}), (\ref{laofk}), (\ref{resin}),
(\ref{imsin})),
\bea \label{n31}
     2 q + k_3 & = & m \frac{2 \p}{L} \qd, \qd m = 0, \dots, 3L -1
		     \qd, \\ \label{n32}
     \sin(k_3) - \La & = & \frac{U}{4}
                           \ctg \left( \frac{k_3 L}{2} \right)
			   \qd, \\ \label{n33}
     \La & = & \sin(q) \ch(\xi) - \frac{U}{4} \frac{\sin(qL)}
	       {\ch(\xi L) - \cos(qL)} \qd, \\ \label{n34}
     \cos(q) \sh(\xi) & = & - \frac{U}{4} \frac{\sh(\xi L)}
			      {\ch(\xi L) - \cos(qL)} \qd.
\eea
We choose a branch of solution $q_\ell (\La)$ of (\ref{n32}) and insert
it into (\ref{n31}). This yields
\beq \label{cmm}
     q = m \frac{\p}{L} - \frac{q_\ell (\La)}{2} \qd.
\eeq
Using this result, (\ref{n33}) and (\ref{n34}) turn into
\bea \label{la3}
     \La & = & \sin \left(\tst{m \frac{\p}{L} - \frac{q_\ell (\La)}{2}}
		    \right) \ch(\xi) - \frac{U}{4}
	       \frac{\sin(q_\ell (\La) L/2)}
		    {\cos(q_\ell (\La) L/2) - (- 1)^m \ch(\xi L)}
	       \qd, \\ \label{xi3}
     \cos \left(\tst{m \frac{\p}{L} - \frac{q_\ell (\La)}{2}} \right)
	\sh(\xi) & = & - \frac{U}{4} \frac{\sh(\xi L)}
                         {\ch(\xi L) - (- 1)^m \cos(q_\ell (\La) L/2)}
			  \qd.
\eea
This is a system of two equations in two unknowns $\xi$ and $\La$. In
contrast to the case $N = 2$, which was considered above, these
equations do not decouple.

We shall first consider the solutions $\xi_{\ell, m} (\La)$ of equation
(\ref{xi3}). For this purpose we define
\beq
     f_a (\xi) = \frac{\sh(\xi L)}{\ch(\xi L) - a} \qd, \qd
     a = (- 1)^m \cos \left( \frac{q_\ell (\La) L}{2} \right) \qd, \qd
     b = - \, \frac{U}{4 \cos \left( \tst{m \frac{\p}{L} -
		 \frac{q_\ell (\La)}{2}} \right)} \qd.
\eeq
With these definitions equation (\ref{xi3}) turns into
\beq \label{xi3ofla}
     \sh(\xi) = b f_a (\xi) \qd,
\eeq
Note that $a$ is real and $|a| \le 1$. Hence, $f_a (\xi) > 0$ for all
positive $\xi$, and a necessary condition for (\ref{xi3ofla}) to have
a solution is $b > 0$. This means that
\beq \label{rangeq}
     \frac{\p}{2} \le m \frac{\p}{L} - \frac{q_\ell (\La)}{2} \le
	\frac{3 \p}{2} \qd, \qd \mbox{for\ } U > 0
     \qd, \qd \frac{\p}{2} \le \left| m \frac{\p}{L} -
	\frac{q_\ell (\La)}{2} - \p \right| \le \p
     \qd, \qd \mbox{for\ } U < 0 \qd.
\eeq
Let us concentrate again on the case $U > 0$. The inequality for the
case $U > 0$ in (\ref{rangeq}) holds for all real $\La$, if and only
if $\frac{L}{2} \le m - \ell \le \frac{3L}{2} - 1$ (cf.\ Lemma III.1).
Equation (\ref{xi3ofla}) can be easily solved graphically for fixed $a$
($|a| < 1$) and $b > 0$ (see figure 3). The function $f_a (\xi)$ has
the following properties. (i) $a < 0$: $f_a (\xi)$ is monotonically
increasing ($f_a' (\xi) > 0$) and concave ($f_a'' (\xi) < 0$),
$f_a (0) = 0$, $f_a' (0) = \frac{L}{1 - a} \ge \frac{L}{2}$ and
$\lim_{\xi \rightarrow \infty} f_a (\xi) = 1$. (ii) $a > 0$: $f_a (\xi)$
has a single positive maximum $\xi^{(0)}$ and a single positive
turning point $\xi^{(1)}$, $\xi^{(0)} < \xi^{(1)}$, $f_a (0) = 0$,
$f_a' (0) = \frac{L}{1 - a} \ge \frac{L}{2}$ and $\lim_{\xi \rightarrow
\infty} f_a (\xi) = 1$. These properties are sufficient to conclude
that (\ref{xi3ofla}) has a unique solution $\xi_{\ell, m} (\La)$ for
all real $\La$, if and only if $\frac{L}{2} \le m - \ell \le
\frac{3L}{2} - 1$ (recall that $U > 0$) and the Takahashi condition
$L > \frac{8}{U}$ is satisfied. Note that $f_a (\xi)$ as a function of
$a$ interpolates between the two branches of the function $f(\xi)$,
equation (\ref{fofxi}). These two branches are the two dashed lines
envelopping the functions $f_a (\xi)$ in figure 3.
\begin{figure}

\begin{center}

\fbox{
\epsfxsize 8cm
\epsffile{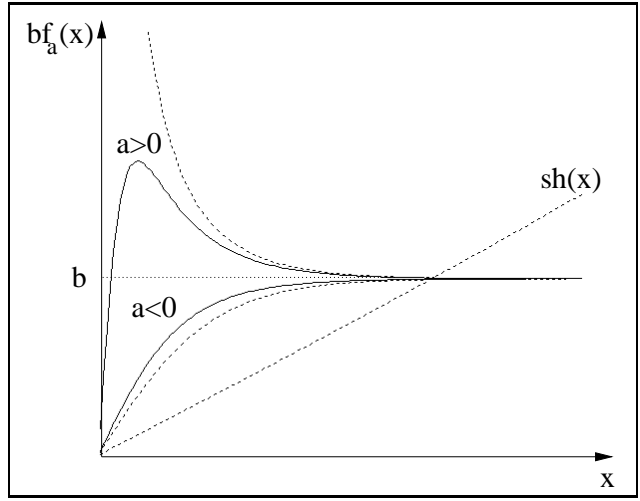}
}

\caption{Sketch of equation (\ref{xi3ofla}).}

\end{center}

\end{figure}

Lemma III.1 and Lemma III.3 allow us to understand the behaviour of the
solutions $\xi_{\ell, m} (\La)$ of (\ref{xi3}) as $\La \rightarrow
\pm \infty$. Lemma III.1 applied to (\ref{xi3}) implies
\bea
     \cos \left((m - \ell + 1) \tst{\frac{\p}{L}} \right) \sh(\xi) & = &
     - \frac{U}{4} \frac{\sh(\xi L)}{\ch(\xi L) - (- 1)^{m - \ell + 1}}
	\qd, \qd \mbox{for\ } \La \rightarrow - \infty \qd, \\
     \cos \left((m - \ell) \tst{\frac{\p}{L}} \right) \sh(\xi) & = &
     - \frac{U}{4} \frac{\sh(\xi L)}{\ch(\xi L) - (- 1)^{m - \ell}}
	\qd, \qd \mbox{for\ } \La \rightarrow \infty \qd.
\eea
Using Lemma III. 3 we conclude that
\beq \label{limxila}
     \lim_{\La \rightarrow - \infty} \xi_{\ell, m} (\La) =
	   \xi_{m - \ell + 1} \qd, \qd
     \lim_{\La \rightarrow \infty} \xi_{\ell, m} (\La) =
	   \xi_{m - \ell} \qd,
\eeq
where $\xi_m$ is the unique solution of equation (\ref{xiq}). This
solution exists (cf.\ (\ref{rangem}) and recall that $U > 0$), if and
only if $\frac{L}{2} < m < \frac{3L}{2}$. Hence, the range of validity
of (\ref{limxila}) is restricted to
\beq \label{rangexila}
     \frac{L}{2} < m - \ell < \frac{3L}{2} - 1 \qd.
\eeq

Let us insert $\xi_{\ell, m} (\La)$ into (\ref{la3}). Then (\ref{la3})
turns into
\beq \label{lag}
     \La = g_{\ell, m} (\La) \qd,
\eeq
where $g_{\ell, m} (\La)$ is defined by
\beq \label{defg}
     g_{\ell, m} (\La) = \sin \left(\tst{m \frac{\p}{L} -
				  \frac{q_\ell (\La)}{2}} \right)
		  \ch(\xi_{\ell, m} (\La)) -
		     \frac{U}{4} \frac{\sin(q_\ell (\La) L/2)}
			{\cos(q_\ell (\La) L/2) -
			   (- 1)^m \ch(\xi_{\ell, m} (\La) L)} \qd.
\eeq
Using (\ref{limxila}) we obtain the asymptotics of $g_{\ell, m} (\La)$,
\beq \label{limgla}
     \lim_{\La \rightarrow - \infty} g_{\ell, m} (\La) =
	\sin \left( (m - \ell + 1) \tst{\frac{\p}{L}} \right)
	   \ch(\xi_{m - \ell + 1}) \qd, \qd
     \lim_{\La \rightarrow \infty} g_{\ell, m} (\La) =
	\sin \left( (m - \ell) \tst{\frac{\p}{L}} \right)
	   \ch(\xi_{m - \ell}) \qd.
\eeq
We see that $g_{\ell, m} (\La)$ has finite asymptotics for $\La
\rightarrow \pm \infty$. Since $g_{\ell, m} (\La)$ is continuous in
$\La$, we arrive at the conclusion that there exists a solution
$\La_{\ell, m}$ of (\ref{lag}) for every pair $(\ell, m)$, which
satisfies (\ref{rangexila}). Hence, we have shown the following
\begin{lemma}
Let $L > \frac{8}{U}$, $U > 0$, $N = 3$. Equations (\ref{bak1}) and
(\ref{bas1}) have solutions consisting of one $k$-$\Lambda$ two string
and a single real $k$ only if the momentum $q$ of the center of the
string is in the range $\frac{\p}{2} < q < \frac{3 \p}{2}$. For every
choice of branch of the real momentum $k_3$ (cf.\ figure 1), there
exist $L - 1$ $k$-$\La$ two strings for odd $L$ and $L - 2$ $k$-$\La$
two strings for even $L$. This gives a total number of $L(L - 1)$
solutions of this type for $L$ odd and of $L(L - 2)$ for $L$ even,
respectively.
\end{lemma}
Let us compare with Takahashi's counting (\ref{r1})-(\ref{r3}). We
have a single free $k_j$ and no $\La$ string, which means that there
is one $I_j = I$ and no $J_\a^n$. It follows from (\ref{r1}) that
$I$ may take $L$ different values. Furthermore, $M_1' = M' = 1$, and
$M_j = 0$ for $j > 1$. Thus $|{J'}^1| \le \2 (L - 3)$, which leads to
$L - 2$ possible values of ${J'}^1$. Takahashi's counting therefore
gives $L(L - 2)$ solutions with one $k$-$\Lambda$ string and one real
$k$. This is in aggreement with our Lemma.

We are now ready to state the following generalization of Lemma III.4,
\begin{lemma}
Let $L > \frac{8}{U}$, $N = 3$. Choose two sequences of integers
$\ell_L$ and $m_L$, such that $\lim_{L \rightarrow \infty}
(m_L - \ell_L) \frac{\p}{L} = q$, $\frac{\p}{2} \le q \le
\frac{3 \p}{2}$. This defines a sequence of solutions
$\xi_{\ell_L, m_L} (\La)$ of equation (\ref{xi3}), which has the limit
\beq
     \lim_{L \rightarrow \infty} \xi_{\ell_L, m_L} (\La) =
	- \, \arsh \left( \frac{U}{4 \cos(q)} \right) \qd,
\eeq
uniformly in $\La$. Thus all strings corresponding to the sequence
$\xi_{\ell_L, m_L} (\La)$ are driven to their ideal positions.
\end{lemma}
Proof: There is no $\La$  and no subsequence $\xi_{\ell_{L_j},
m_{L_j}} (\La)$, such that $\lim_{j \rightarrow \infty}
\xi_{\ell_{L_j}, m_{L_j}} (\La) = 0$. This can be seen in similar way
as in the proof of Lemma III.4.

Let us finally note that our considerations for the case $N = 3$
readily generalize to arbitrary $N$. For arbitrary $N$ we have to
consider $N - 2$ copies of equation (\ref{n32}) in the system of
equations (\ref{n31})-(\ref{n34}). We further have to replace
$q_\ell (\La)$ in (\ref{cmm}) by $Q(\La) = q_{\ell_1} (\La) + \dots
+ q_{\ell_{N - 2}} (\La)$ with $\ell_1 < \dots < \ell_{N - 2}$. The
properties of $Q(\La)$ (monotonicity and asymptotics) then follow from
Lemma III.1, and all considerations go through as in the case $N = 3$.
\subsection{Summary}
In this section we have studied $k$-$\La$ string solutions of the
Lieb-Wu equations (\ref{bak}), (\ref{bas}). We have shown that such
solutions exist and that they are driven to certain ideal string
positions in the limit of a large lattice. We have further shown that
for a large enough lattice of finite length their number is in
accordance with the number of corresponding solutions of the discrete
Takahashi equations (\ref{t1})-(\ref{t3}).

%%%%%%%%%%%%%%%%%%%%%%%%%%%%%%%%%%%%%%%%%%%%%%%%%%%%%%%%%%%%%%%%%%%%%%%%
%   scd3.tex                                                           %
%%%%%%%%%%%%%%%%%%%%%%%%%%%%%%%%%%%%%%%%%%%%%%%%%%%%%%%%%%%%%%%%%%%%%%%%
\section{Lieb-Wu equations for a single down spin (II) --
Self-consistent solution}
\subsection{Zeroth order and the discrete Takahashi equations}
In this section we present the self-consistent solution of the Lieb-Wu
equations (\ref{bak}), (\ref{bas}) for the case of three electrons
and one $k$-$\La$ string. The Lieb-Wu equations (\ref{bak}), (\ref{bas})
provide a self-consistent way of calculation of the deviation of the
strings from their ideal positions. We show that every solution of the
discrete Takahashi equations gives an approximate solution to the
Lieb-Wu equations (\ref{bak}), (\ref{bas}), and we calculate the leading
order corrections. These corrections vanish exponentially fast as the
number of lattice sites $L$ becomes large.

The Lieb-Wu equations for three electrons and one down spin are
\bea \label{bak3}
     && e^{\i k_j L} = \frac{\La - \sin k_j - \i U/4}
                         {\La - \sin k_j + \i U/4} \qd, \qd
                          j = 1, 2, 3 \qd,\\ \label{bas3}
     && \prod_{j = 1}^3 \frac{\La - \sin k_j - \i U/4}
			  {\La - \sin k_j + \i U/4} = 1 \qd.
\eea
As in section III.A we may replace equation (\ref{bas3}) by the
equation for the conservation of momentum,
\beq \label{emom3}
     e^{\i (k_1 + k_2 + k_3)L} = 1 \qd.
\eeq
(\ref{bak3}) and (\ref{emom3}) follow from (\ref{bak3}) and (\ref{bas3})
and vice versa.

Let us follow the usual self-consistent strategy for obtaining a
$k$-$\La$ string solution. As in section III.C we shall use the
notation $k_- = k_1 = q - \i \xi$, $k_+ = q + \i \xi$, $\xi > 0$,
i.e.\ we assume that $k_1$ and $k_2$ are part of a $k$-$\La$ string.
In order to facilitate comparison with the previous literature
we shall also use the abbreviations $\Ph = \Re \sin(k_+)
= \Re \sin(k_-)$ and $\chi = \Im \sin(k_-) = - \Im \sin(k_+)$. We
introduce $\de$ as a measure of the deviation of the string from its
ideal position. Then
\beq \label{dev}
     \sin(k_-) = \Ph + \i \chi = \La + \frac{\i U}{4} + \de \qd, \qd
     \sin(k_+) = \Ph - \i \chi = \La - \frac{\i U}{4} + \bar \de \qd.
\eeq
Inserting (\ref{dev}) into (\ref{bak3}) gives
\beq \label{epm}
     e^{\i k_- L} = 1 + \frac{\i U}{2 \de} \qd, \qd
     e^{- \i k_+ L} = 1 - \frac{\i U}{2 \bar \de} \qd.
\eeq
We may consider the first equation in (\ref{dev}) as the equation, that
defines $\de$. The second equation in (\ref{dev}) is not independent.
It is the complex conjugated of the first one and may be dropped for
that reason. Similarly, we may also drop the second equation in
(\ref{epm}). Then we are left with six independent equations,
(\ref{bak3}) for $j = 3$, (\ref{emom3}), and the real and imaginary
parts of the first equations in (\ref{dev}) and (\ref{epm}). Note that
$k_1 + k_2 = 2q$. Therefore our six equations are equivalent to
\bea \label{basc1}
     e^{2 \i q L} & = & \frac{\La - \sin k_3 + \i U/4}
                         {\La - \sin k_3 - \i U/4} \qd, \\ \label{basc2}
     e^{\i k_3 L} & = & \frac{\La - \sin k_3 - \i U/4}
                         {\La - \sin k_3 + \i U/4} \qd, \\ \label{basc3}
     \sin(k_-) & = & \La + \frac{\i U}{4} + \de \qd, \\ \label{basc4}
     \de & = & \frac{\i U}{2} \, \frac{1}{e^{\i k_- L} - 1} \qd.
\eea
Every $k$-$\La$ string solution of the Lieb-Wu equations (\ref{bak3}),
(\ref{bas3}) gives a solution of equations (\ref{basc1})-(\ref{basc4})
(with real $q$, $k_3$, $\La$ and real positive $\xi$) and vice versa.

If $L$ is large and $\xi = \Im k_+ = \CO (1)$, then $\de$ is very
small and may be neglected in equation (\ref{basc3}). Then (\ref{basc4})
decouples from the other equations, which become
\bea \label{pt1}
     e^{2 \i q^{(0)} L} & = & \frac{\La_0 - \sin k_3^{(0)} + \i U/4}
                         {\La_0 - \sin k_3^{(0)} - \i U/4} \qd, \\
			    \label{pt2}
     e^{\i k_3^{(0)} L} & = & \frac{\La_0 - \sin k_3^{(0)} - \i U/4}
                         {\La_0 - \sin k_3^{(0)} + \i U/4} \qd, \\
			    \label{pt3}
     \sin(k_-^{(0)}) & = & \La_0 + \frac{\i U}{4} \qd.
\eea
If, on the other hand, $\de$ is very small in (\ref{basc3}), we may
neglect it in first approximation and solve (\ref{pt3}) instead. Now
(\ref{pt3}) implies that $\xi^{(0)} = \CO (1)$ (see below). Then, using
(\ref{basc4}), we see that $\de$ is indeed small for large $L$. This
means the assumption $\de$ be small for large $L$ is self-consistent.

Let us be more precise. We set $k_-^{(0)} = q^{(0)} - \i \xi^{(0)}$ with
real $q^{(0)}$ and real, positive $\xi^{(0)}$. Separating (\ref{pt3})
into real and imaginary part we obtain two equations which relate the
three unknowns $q^{(0)}$, $\xi^{(0)}$ and $\La_0$,
\bea \label{idla}
     \La_0 & = & \sin(q^{(0)}) \ch(\xi^{(0)}) \qd, \\ \label{idxi}
     \xi^{(0)} & = & - \, \arsh \left( \frac{U}{4 \cos(q^{(0)})} \right)
		       \qd.
\eea
Let us concentrate on positive coupling $U > 0$ for simplicity. Since
we are assuming that $\xi^{(0)} > 0$, the range of $q^{(0)}$ is then
restricted to, say, $\frac{\p}{2} < q^{(0)} < \frac{3 \p}{2}$ by
equation (\ref{idxi}). We obtain the uniform estimate
\beq \label{xiest}
     \xi^{(0)} > \arsh(U/4) \qd.
\eeq
In order to test, if a solution $q^{(0)}$, $\xi^{(0)}$, $k_3^{(0)}$,
$\La_0$ of (\ref{pt1})-(\ref{pt3}) is a good approximate solution of
(\ref{basc1})-(\ref{basc4}), we use it to estimate the modulus of
$\de$ for large $L$,
\beq \label{deest}
     |\de| \approx \frac{U}{2} \, e^{- \xi^{(0)} L} <
	\frac{U}{2} \, e^{- \arsh(U/4) L} \qd.
\eeq
The inequality follows from (\ref{xiest}). We conclude that $|\de|$
becomes very small for large $L$. For $U = 4$ and $L = 24$, for
instance, the estimate (\ref{deest}) gives $|\de| < 1.3 \cdot 10^{- 9}$,
whereas the difference between two real $k_3$'s is of the order of
$\frac{2 \p}{L} = 0.26$ (cf.\ section III.B). For large $L$ it becomes
impossible to numerically distinguish between solutions of
(\ref{basc1})-(\ref{basc4}) and (\ref{pt1})-(\ref{pt3}), respectively.

If we fix the branch of the arcsin as $- \frac{\p}{2} \le \arcsin (z)
\le \frac{\p}{2}$, it follows from the inequality $\frac{\p}{2} <
q^{(0)} < \frac{3 \p}{2}$ that
\beq \label{kmla}
     k_-^{(0)} = \p - \arcsin \left( \La_0 + \tst{\frac{\i U}{4}} \right)
	     \qd.
\eeq
Inserting (\ref{kmla}) into (\ref{pt1}) leads to
\beq \label{ptp}
     e^{2 \i \, \Re \arcsin(\La_0 + \i U/4) L} =
	\frac{\La_0 - \sin k_3^{(0)} - \i U/4}
	     {\La_0 - \sin k_3^{(0)} + \i U/4}
	\qd.
\eeq
We thus have eliminated $k_-^{(0)}$ from the system of equations
(\ref{pt1})-(\ref{pt3}), and we are left with two independent equations
(\ref{pt2}) and (\ref{ptp}). These two equations determine the two
real unknowns $\La_0$ and $k_3^{(0)}$. Taking logarithms of (\ref{pt2})
and (\ref{ptp}) we arrive at Takahashi's discrete equations (\ref{t1}),
(\ref{t3}) for one $k$-$\La$ string and one real $k$.

We have seen that the equations (\ref{pt1})-(\ref{pt3}) determine
Takahashi's {\it ideal strings}. Equations (\ref{basc1})-(\ref{basc2})
on the other hand, are equations for {\it non-ideal strings}, which
solve the Lieb-Wu equations (\ref{bak3}), (\ref{bas3}). Thus, $\de$ is
a measure for the deviation of the strings from their ideal positions.
We have further seen that the assumption that $\de$ be small is
self-consistent. In particular, every solution of the equations
(\ref{pt1})-(\ref{pt3}), which are equivalent to Takahashi's discrete
equations, is an approximate solution of equations (\ref{basc1})-%
(\ref{basc4}). The approximation becomes extremely accurate for large
$L$.
\subsection{First order corrections}
Inserting any solution of (\ref{pt1})-(\ref{pt3}) into (\ref{basc4})
we have
\beq \label{deexp}
     \de = \frac{\i U}{2} \, e^{- \i q^{(0)} L} e^{- \xi^{(0)} L} +
	   \CO \left(e^{- 2 \xi^{(0)} L} \right) =
           \frac{U}{2} \, \sin(q^{(0)} L) e^{- \xi^{(0)} L} +
           \frac{\i U}{2} \, \cos(q^{(0)} L) e^{- \xi^{(0)} L} +
	   \CO \left(e^{- 2 \xi^{(0)} L} \right) \qd.
\eeq
So the relevant parameter, which controls the deviation of the strings
from their ideal positions is $\eps = e^{- \xi^{(0)} L}$. Every
solution of (\ref{pt1})-(\ref{pt3}) is an approximate solution of
(\ref{basc1})-(\ref{basc4}). Let us calculate the leading order
corrections. We expect them to be proportional to $\eps$,
\bea \label{qeps}
     q & = & q^{(0)} + q^{(1)} \eps + \CO (\eps^2) \qd, \\ \label{xieps}
     \xi & = & \xi^{(0)} + \xi^{(1)} \eps + \CO (\eps^2) \qd, \\
	       \label{k3eps}
     k_3 & = & k_3^{(0)} + k_3^{(1)} \eps + \CO (\eps^2) \qd, \\
	       \label{laeps}
     \La & = & \La_0 + \La_1 \eps + \CO (\eps^2) \qd.
\eea
The $\eps$-expansion of $\de$ is given in equation (\ref{deexp}) above.
We also introduce
\bea \label{phieps}
     \Ph & = & \sin(q^{(0)}) \ch(\xi^{(0)}) + \Ph^{(1)} \eps
	       + \CO (\eps^2) \qd, \\
     \chi& = & \frac{U}{4} + \chi^{(1)} \eps + \CO (\eps^2) \qd,
	       \label{chieps}
\eea
since it is the quantities $\Ph = \Re \sin(k_-)$ and $\chi =
\Im \sin(k_-)$ which actually form strings in the complex plane.

The two sets of variables $q$, $\xi$ and $\Ph$, $\chi$ are not
independent. Inserting (\ref{qeps}) and (\ref{xieps}) into the left
hand side of (\ref{phieps}) and (\ref{chieps}) and comparing leading
orders in $\eps$ we find
\beq \label{li1}
     \left( \begin{array}{c} \Ph^{(1)} \\ \chi^{(1)} \end{array}
	\right) = 
     \left( \begin{array}{cc} \cos(q^{(0)}) \ch(\xi^{(0)}) &
	\sin(q^{(0)}) \sh(\xi^{(0)})\\ \sin(q^{(0)}) \sh(\xi^{(0)}) &
	- \cos(q^{(0)}) \ch(\xi^{(0)}) \end{array} \right)
     \left( \begin{array}{c} q^{(1)} \\ \xi^{(1)} \end{array} \right)
	\qd.
\eeq
Let us insert (\ref{deexp}), (\ref{phieps}) and (\ref{chieps}) into
(\ref{basc3}). We obtain to leading order
\bea \label{li2}
     \Ph^{(1)} & = & \La_1 + \frac{U}{2} \sin(q^{(0)} L) \qd, \\
     \chi^{(1)} & = & \frac{U}{2} \cos(q^{(0)} L) \qd, \label{li3}
\eea
i.e.\ we have already found the leading order correction $\chi^{(1)}$,
equation (\ref{li3}). Next we insert (\ref{qeps}), (\ref{k3eps}) and
(\ref{laeps}) into (\ref{basc1}) and (\ref{basc2}) and linearize in
$\eps$. The resulting equations are
\bea \label{li4}
     2 q^{(1)} & = & - \frac{U}{2L} \,
		       \frac{\La_1 - \cos(k_3^{(0)}) \, k_3^{(1)}}
			    {\left(\La_0 - \sin(k_3^{(0)})\right)^2
			     + \frac{U^2}{16}} \qd, \\ \label{li5}
     k_3^{(1)} & = & \frac{U}{2L} \,
		     \frac{\La_1 - \cos(k_3^{(0)}) \, k_3^{(1)}}
		          {\left(\La_0 - \sin(k_3^{(0)})\right)^2 +
			   \frac{U^2}{16}} = - 2 q^{(1)} \qd.
\eea
The equation $2 q^{(1)} + k_3^{(1)} = 0$ following from (\ref{li4}),
(\ref{li5}) is, of course, a consequence of momentum conservation.

Equations (\ref{li1})-(\ref{li5}) are a system of six linear equations
for six unknowns, $q^{(1)}$, $\xi^{(1)}$, $\Ph^{(1)}$, $\chi^{(1)}$,
$k_3^{(1)}$ and $\La_1$. Note that in the derivation of these equations
we have used all of the equations (\ref{basc1})-(\ref{basc4}).
Equations (\ref{li2}) and (\ref{li3}) came out of (\ref{basc3}),
(\ref{basc4}) was used to obtain the expansion (\ref{deexp}) for $\de$
in terms of $\eps$, and (\ref{li4}), (\ref{li5}) follow from
(\ref{basc1}) and (\ref{basc2}). Equation (\ref{li1}) is a consequence
of the definition of $\Ph$ and $\chi$. Being a set of linear equations,
(\ref{li1})-(\ref{li5}) are readily solved,
\bea \label{pert1}
     \La_1 & = & - \frac{U}{2} \left( \sin(q^{(0)} L) + \tan (q^{(0)})
		      \tanh(\xi^{(0)}) \cos(q^{(0)} L) \right) C(L) \\
		      \label{pert2}
           & = & - \frac{U}{2} \left( \sin(q^{(0)} L) + \tan (q^{(0)})
		      \tanh(\xi^{(0)}) \cos(q^{(0)} L) \right)
		      + \CO(1/L) \qd, \\ \label{pert3}
     k_3^{(1)} & = & \frac{\La_1}{\frac{2L}{U} \left[ \left( \La_0
			- \sin(k_3^{(0)}) \right)^2 + \frac{U^2}{16}
			\right] + \cos(k_3^{(0)})} \\ \label{pert4}
           & = & - \, \frac{U^2 \left( \sin(q^{(0)} L) + \tan (q^{(0)})
		      \tanh(\xi^{(0)}) \cos(q^{(0)} L) \right)} 
                      {4L \left[ \left( \La_0 - \sin(k_3^{(0)})
				    \right)^2 + \frac{U^2}{16}
			\right]} + \CO(1/L^2) \qd, \\ \label{pert5}
     \Ph^{(1)} & = & \La_1 + \frac{U}{2} \sin(q^{(0)} L) =
                        - \frac{U}{2} \tan (q^{(0)}) \tanh(\xi^{(0)})
			\cos(q^{(0)} L) + \CO(1/L) \qd, \\ \label{pert6}
     \chi^{(1)} & = & \frac{U}{2} \cos(q^{(0)} L) \qd.
\eea
The function $C(L)$ in equation (\ref{pert1}) which gives the explicit
form of the $\CO(1/L)$ and $\CO(1/L^2)$ corrections in the remaining
equations is
\beq
     C(L) = \left[ 1 + \frac{\cos(q^{(0)}) \ch(\xi^{(0)})
		  (1 + \tan^2 (q^{(0)}) \tanh^2 (\xi^{(0)}))}
                  {\frac{4L}{U} \left[ \left( \La_0
			- \sin(k_3^{(0)}) \right)^2 + \frac{U^2}{16}
			\right] + 2 \cos(k_3^{(0)})} \right]^{- 1}
          = 1 + \CO(1/L) \qd.
\eeq
Equations (\ref{pert1})-(\ref{pert6}) give a complete description of
the leading order deviation of a non-ideal string from its ideal
position in the presence of one real $k$. The deviations of $q$ and
$\xi$, which are determined by $q^{(1)}$ and $\xi^{(1)}$, follow from
the equations (\ref{li1}) and (\ref{pert5}), (\ref{pert6}). In order to
see that $C(L) - 1$ is indeed of order $\CO (1/L)$ on has to use
(\ref{idla}) and (\ref{idxi}).
\subsection{Summary}
In this section we have presented a self-consistent solution of the
Lieb-Wu equations for the case of three electrons and one $k$-$\La$
string. Recall that the existence of these solutions was shown in the
previous section. Here we showed that a self-consistent approach
naturally leads to Takahashi's discrete equations. We showed that
Takahashi's discrete equations provide a highly accurate approximate
solution of the Lieb-Wu equations in the limit of a large lattice.
We also showed that there is a natural parameter $\eps =
e^{- \xi^{(0)} L}$ that measures the deviation of solutions of the
Lieb-Wu equations from the corresponding solutions of the discrete
Takahashi equations. Employing an algebraic perturbation theory we
explicitly calculated the leading order deviation in $\eps$ of a
non-ideal $k$-$\La$ string from its ideal position in the presence of
a real $k$.

%%%%%%%%%%%%%%%%%%%%%%%%%%%%%%%%%%%%%%%%%%%%%%%%%%%%%%%%%%%%%%%%%%%%%%%%
%     scd4.tex                                                         %
%%%%%%%%%%%%%%%%%%%%%%%%%%%%%%%%%%%%%%%%%%%%%%%%%%%%%%%%%%%%%%%%%%%%%%%%
\section{Lieb-Wu equations for a single down spin (III) -- Numerical
solution}
\subsection{Numerical method}

In section III the existence of $k$-$\La$ string solutions was
analytically proven under the Takahashi condition for the simplest
non-trivial cases with one down spin, $M=1$. The deviation of ideal
string solutions given by the discrete Takahashi equations from the
corresponding solutions of the Lieb-Wu equations  was evaluated
analytically in section IV, and it was shown that the corrections
vanish exponentially fast as the lattice size $L$ becomes large.
To confirm these analytical arguments on the existence of $k$-$\La$
strings, we utilize a complementary numerical approach. Although
the tractable system size is limited, we can directly obtain $k$-$\La$
strings and verify the completeness of the Bethe ansatz for arbitrary
$U$. We note that numerical solutions for low-lying particle-hole
and $\Lambda$ string excitations in small finite systems can be found
in the literature (see e.g.\ \cite{Schulz93}). As far as we know,
however, our data are the first example of a numerical study that
confirms the completeness of the Bethe ansatz for a small finite system.

We study again the cases $N=2$ and $N=3$ for both pedagogical clarity
and technical simplicity. We shall employ two numerically exact
methods, (1) the numerical diagonalization of a real-symmetric matrix
using the Householder-QR method (we call it Method 1), (2) a numerical
method to solve coupled nonlinear equations using the Brent
method (we call it Method 2). These techniques themselves are
conventional. They allow us to obtain numerically exact solutions for
the Hubbard model. Here we use the term `numerically exact' to state
that our numerical solutions give exact numbers except for the
inevitable rounding error in the computation.%
\footnote{We will indicate errors by the difference of the left hand
side and the right side of each equation in (\ref{bak}), (\ref{bas}), 
evaluated within our numerical treatment. Then it will become clear that
in all the equations which we tested the relative error is negligible
and of the order of the rounding error expected for the double-precision
calculation.}

Our strategy here is the following. 
\begin{enumerate}
\item Obtain all energy eigenvalues with fixed $N$ and $M$ as a
function of $U$ using Method 1. This step gives a complete list of
energy eigenvalues. 
\item Obtain numerical (real and/or complex) solutions by solving the
Lieb-Wu equations with Method 2. 
\item List up all eigenvalues obtained by the two methods and compare
them with one another. This step gives a confirmation of completeness. 
\end{enumerate}
For our numerical study we used the following form of the Hubbard
Hamiltonian,
\beq
     H = - \sum_{j=1}^L \sum_{\sigma= \uparrow, \downarrow}
         \left( c_{j+1, \sigma}^+ c_{j, \sigma}
         + c_{j-1, \sigma}^+ c_{j, \sigma} \right) 
         + U \sum_{j=1}^{L} n_{j, \uparrow} 
         n_{j, \downarrow} \qd. 
\label{hamiltonian} 
\eeq
This form is different from (\ref{ham}) by a shift of the chemical
potential and by a constant energy shift. For fixed particle number
$N$ this leads to a shift of the spectrum by $\frac{U}{4} (2N - L)$.

In order to perform a numerical diagonalization we used basis vectors
$c_{x_1}^+ \dots c_{x_{N - M}}^+ c_{y_1}^+ \dots 
c_{y_M}^+ |0\>$ to
represent the Hamiltonian as a matrix in the sector of fixed $N$ and
$M$. The number of different configurations, $x_1 < \dots < x_{N - M}$,
$y_1 < \dots < y_M$, in this sector is ${L \choose N - M} \times
{L \choose M}$ and determines the size of the matrix. In order to
employ method 2, we rewrote the Lieb-Wu equations into a proper 
set of real equations, which will be presented in the following 
subsections. Hereafter in this section we assume that $L$ is an even
integer.

%%%%%%%%%%%%%%%% N=2 %%%%%%%%%%%%%%%%

\subsection{Numerical solution for two electrons}
\subsubsection{Equations for the $k$-$\Lambda$ string}
Let us discuss numerical $k$-$\La$ two string solutions for the two
electron system with one down spin ($N = 2$, $M = 1$). We shall use the
same notation as in section III. The variables to be determined are 
$k_+$, $k_-$ and $ \Lambda$, where $k_+$ is the complex conjugate 
of $k_-$ and $\Lambda$ is real. We may write
\beq
     k_+ = q + \i \x \qd, \qd k_- = q - \i \x  \qd,
\eeq
where $0 \le q < 2 \pi $ and $\x > 0$. Then the total momentum is
\beq 
     P = 2q = \frac {2 m \pi} L \mod 2 \pi \qd.
\eeq
This equation restricts the admissible values of $q$. It is easy to
obtain $\La$ as a function of $q$ and $\xi$. We find $\La = \sin(q)
\cosh(\x )$ (cf.\ equation (\ref{laq})). Thus we are left with a single
equation that determines $\xi$. For our numerical calculations we
wrote it in the form
\beq
     \exp( \i m \pi + \x L ) = \frac{-\cos(m \p/L) \sinh(\xi) + U/4}
                                    {-\cos(m \p/L) \sinh(\xi) - U/4}
                               \qd. \label{realeq}
\eeq
This equation is equivalent to (\ref{xiq}). Therefore, for positive $U$,
the allowed values of $m$ are restricted to
\beq 
     m = L/2 + 1, \ldots, 3L/2 - 1
\eeq
(cf.\ equation (\ref{rangem})). Equation (\ref{realeq}) was already
studied in appendix B of \cite{EKS92b}. There it was shown that there
is a redistribution phenomenon as $U$ becomes small. $k$-$\La$ strings
corresponding to odd values of $m$ collapse at critical values of
$U$ given by $U_m = (8/L)|\cos (m\pi/L)|$.
\subsubsection{Numerical solutions for $N=2$}
As a typical example, let us present some numerical $k$-$\La$ two
string solutions for $N=2$ (two electrons). For a $k$-$\La$ two string
we show the dependence of energy eigenvalues, imaginary parts of
charge momenta and the deviation from the ideal string positions on $U$.
We put $m=16$ and $L=16$ (16 sites). Then we have $q=\pi$, i.e. 
$$
     k_+ = \pi + \i \xi \qd, \qd  k_- = \pi - \i \xi \qd,
           \qd \La = 0 \qd. 
$$
In the list below the deviation of the string from its ideal position, 
$k_{\pm}^{(ideal)} = \pi \pm \i \, \arsh (U/4)$ is measured by
$\Im \delta = \sinh (\xi ) - U/4$. 

\vskip 0.6cm 

(1)  $U=10$
\begin{eqnarray}
{\rm Householder} \qquad   
 E & = & 10.7703296143355  \nonumber  \\     
{\rm Bethe \, Ansatz} \qquad
 E & = & 10.7703296143355 \nonumber \\ 
 \x & = &  1.64723114637774 \nonumber \\
 \Im \delta & = &  1.78994596922166 \times 10^{-11} \nonumber  
\end{eqnarray}

(2)  $U=1$
\begin{eqnarray}
{\rm Householder} \qquad      
  E & = & 4.13148449882288 \nonumber \\ 
{\rm Bethe \, Ansatz}  \qquad 
 E & = & 4.13148449882289 \nonumber \\ 
 \x &=&  0.255705305537532 \nonumber \\ 
 \Im \delta & =&  8.50098694369150 \times 10^{-3} \nonumber
\end{eqnarray}

(3) $U=0.1$
\begin{eqnarray}
{\rm Householder} \qquad 
E & = & 4.00668762410970     \nonumber \\ 
{\rm Bethe \, Ansatz} \qquad               
  E & = & 4.00668762410971 \nonumber \\ 
 \x & = & 5.78176505356310 \times 10^{-2} \nonumber \\ 
 \Im \delta &= & 3.28498688384022 \times 10^{-2}  \nonumber  
\end{eqnarray}

(4) $U=0.01$
\begin{eqnarray}
{\rm Householder} \qquad   
E& = & 4.00062917225929   \nonumber \\   
{\rm Bethe \, Ansatz} \qquad               
   E & = & 4.00062917225931 \nonumber \\                       
 \x & = & 1.77363435624506 \times 10^{-2} \nonumber \\ 
 \Im \delta &=&  1.52372734873120 \times 10^{-2} \nonumber 
\end{eqnarray}

The string with $m=16$ does exist for any $U>0$ and is actually the
highest level in the spectrum for $N=2$. We note that the non-ideal
string approaches the ideal $k$-$\La$ string, when $U$ becomes large,
even for such a small system. Yet, in accordance with our expectations,
the ideal string does not provide a good approximation for small $U$ in
finite systems.

In table 1 we show a complete list of eigenstates for the case of one
down-spin ($M=1$, $S_z=0$) and $U=1.5$ for a 6-site system ($L=6$).
Note that the value of $U = 1.5$ is greater than $\max U_m=1.1547$,
or, in the language of section III, the Takahashi condition is
satisfied. Let us explain the table.
\begin{enumerate} 
\item 
The 36 eigenstates are listed in increasing order with respect to 
their energy. 
\item 
$S$ and $P$ denote the spin and momentum of the eigenstate,
respectively.
\item 
The energy eigenvalues obtained by direct diagonalization of the 
Hamiltonian (the Householder method) and by the Bethe ansatz method 
coincide within an error of $\CO(10^{-15})$. 
\item 
The last digit for each numerical value has a rounding error.
\item
There are 5 $k$-$\La$ string solutions among the 36 eigenstates, which
is consistent with the number $L-1 = 5$ obtained by Takahashi's counting
(\ref{r1})-(\ref{r3}). 
\end{enumerate} 

Let us give more explanations on the table. In fact, it confirms the
completeness of the Bethe ansatz as discussed in section II.D. 
We first note that there are 36 eigenstates for the case of two
electrons and one down-spin ($N=2$ and $M=1$) on a 6-site lattice
($L=6$). We recall that the two electrons with one up-spin and one
down-spin can occupy the same site. The ${6 \choose 1} \times
{6 \choose 1} = 36$ eigenstates in table 1 can be classified
into the following types.
\begin{enumerate}
\item 15 eigenstates with two real charge momenta $k_1$, $k_2$ 
and one real spin-rapidity $\Lambda$.  
\item 5 eigenstates with one $k$-$\Lambda$ two string.
\item 15 eigenstates with $S=1$ and $S^z=0$ belonging to spin triplets.
\item 1 $\h$-pairing state.
\end{enumerate} 
Let us consider case (i). In table 1 there are 15 states with real
charge momenta which have total spin zero ($S=0$). This agrees
with our analytical arguments in section III on the number of real,
regular Bethe ansatz solutions (cf.\ Lemma III.2 and below), which
should be ${6 \choose 2} = 15$. Let us also recall that the $I_j$ are
integer (half-integer) when  $\sum_m M_m + M_m^{'}$ is even (odd). 
Thus, for the case $M=1$, the $I_j$ should be half-odd integer,
which is in agreement with the results shown in table 1.

We now consider case (ii). From the analytic discussion in section
III.C, we should have 5 eigenstates with $k$-$\La$ strings. This
is in accordance with our numerical data in table 1. Recall that
we showed in section III.C (below Lemma III.3) that also Takahashi's
counting, using equations (\ref{r1})-(\ref{r3}), leads to the same
number of $k$-$\La$ string solutions.

\begin{center}
\tabcolsep4mm
\begin{tabular}{|c|c|r|r|r|}  
\hline
     No. &  Energy   & $S$ &    $P/(\p/3)$ & type of solution
\nonumber \\  \hline
      1 & $-3.82047006625301$ %-3.82047006625302     
        &  0  &    0     &    real  $I_j = -0.5, 0.5$
\nonumber \\  \hline
      2 & $-3.00000000000000$ &  1  &   1    &     triplet $I_j = 0, 1$
\nonumber \\  \hline
      3 & $-3.00000000000000$ &  1  &  5      &   triplet $I_j = 0, -1$
\nonumber \\  \hline
      4 & $-2.60959865138515$ %-2.60959865138514     
        &  0   &  1    &     real  $I_j = -0.5, 1.5$
\nonumber \\  \hline
      5 & $-2.60959865138515$ %-2.60959865138514     
        &  0 &   5   &      real  $I_j = -1.5, 0.5$
\nonumber \\  \hline
      6 & $-2.00000000000000$ &  1  &   0     &    triplet $I_j = 1, -1$
\nonumber \\  \hline
      7 & $-1.86256622153075$ %-1.86256622153075     
        &  0  &   2      &   real  $I_j = 0.5, 1.5$
\nonumber \\  \hline
      8 & $-1.86256622153075$ %-1.86256622153075     
        &  0 &   4     &    real  $I_j = -0.5, -1.5$
\nonumber \\  \hline
      9 & $-1.53909577258373$ %-1.53909577258374     
        &  0   &  0    &     real  $I_j = -1.5, 1.5$
\nonumber \\  \hline
     10 & $-1.00000000000000$ &  1  &   2    &     triplet $I_j = 0, 2$
\nonumber \\  \hline
     11 & $-1.00000000000000$ &  1  &  4    &     triplet $I_j = 0, -2$
\nonumber \\  \hline
     12 & $-0.594210897273940$ %-0.594210897273940     
        &   0  &   2    &     real  $I_j = -0.5, 2.5$
\nonumber \\  \hline
     13 & $-0.594210897273940$ %-0.594210897273939     
        &   0  &  4    &     real  $I_j = -2.5, 0.5$
\nonumber \\  \hline
     14 &  0.00000000000000 &   0  &   3    &     real  $I_j = 0.5, 2.5$
\nonumber \\  \hline
     15 &  0.00000000000000 &   0  &   3    &     real  $I_j = -2.5, -0.5$
\nonumber \\  \hline
     16 &  0.00000000000000 &   1  &   3     &    triplet $I_j = 1, 2$
\nonumber \\  \hline
     17 &  0.00000000000000 &   1  &   3     &    triplet $I_j = -1, -2$
\nonumber \\  \hline
     18 &  0.00000000000000 &   1  &   3     &    triplet $I_j = 0, 3$
\nonumber \\  \hline
     19 &  0.00000000000000 &   1  &   1    &     triplet $I_j = 2, -1$
\nonumber \\  \hline
     20 &  0.00000000000000 &   1  &   5    &     triplet $I_j = -2, 1$
\nonumber \\  \hline
     21 &  0.474357244982949 % 0.474357244982949     
        &   0  &   1    &     real  $I_j = -1.5, 2.5$
\nonumber \\  \hline
     22 &  0.474357244982949 % 0.474357244982951     
        &   0 &   5    &     real  $I_j = -2.5, 1.5$
\nonumber \\  \hline
     23 &  1.00000000000000 &  1  &   2     &    triplet $I_j = 3, -1$
\nonumber \\  \hline
     24 &  1.00000000000000 &  1  &  4    &     triplet $I_j =3, 1$
\nonumber \\  \hline
     25 &  1.43079477929458 % 1.43079477929458     
        &  0  &  4    &    real  $I_j = 1.5, 2.5$
\nonumber \\  \hline
     26 &  1.43079477929458 % 1.43079477929458     
        &  0  &   2     &   real  $I_j = -2.5, -1.5$
\nonumber \\  \hline
     27 &  1.50000000000000 &  0  &   3       &  $\h$-pair
\nonumber \\  \hline
     28 &  2.00000000000000 &  1  &   0   &      triplet $I_j = 2, -2$
\nonumber \\  \hline
     29 &  2.49213401737360 % 2.49213401737361     
        &  0  &   0    &     real  $I_j =-2.5, 2.5$
\nonumber \\  \hline
     30 &  2.52598233951011 %    2.52598233951011     
        &  0  &   2    &     complex $m=8$, $\xi = 0.710224864788777$
\nonumber \\  \hline
     31 &  2.52598233951011 %    2.52598233951011     
        &  0  &   4    &     complex $m=4$, $\xi = 0.710224864788777$
\nonumber \\  \hline
     32 &  3.00000000000000 &  1  &  5   &     triplet $I_j = 2, 3$
\nonumber \\  \hline
     33 &  3.00000000000000 &  1  &   1   &     triplet $I_j = -2, 3$
\nonumber \\  \hline
     34 &  3.63524140640220  %    3.63524140640219     
        &  0  &   1    &     complex $m=7$, $\xi = 0.313056827256169$
\nonumber \\  \hline
     35 &  3.63524140640220  %    3.63524140640220     
        &  0  &   5    &     complex $m=5$, $\xi = 0.313056827256169$
\nonumber \\  \hline
     36 &  4.36743182146314  % 4.36743182146314     
        &  0  &   0    &     complex $m=6$, $\xi = 0.425405934759021$
\nonumber \\  \hline
\end{tabular}
\vskip 0.6cm 
{Table 1: Classification of all energy levels for $L = 6$, $N = 2$,
$M = 1$ and $U = 1.5$.} 
\end{center}
We consider case (iii). There are 15 states with $S=1$ and $S_z=0$. 
We describe  them  as {\it triplets} in table 1. They are obtained by
multiplying the spin-lowering operator $\zeta^{\dagger}$ to the regular
Bethe states with $S=1$ and $S_z=1$. (For the notation for the $SO(4)$
symmetry see section II.B). The regular Bethe states with $S=1$ and
$S_z=1$ correspond to $N=2$ and $M=0$. We recall again that the $I_j$'s
are integer (half-integer) when  $\sum_m M_m + M_m^{'}$ is even (odd). 
Thus, the $I_j$'s which belong to the $S^z = 0$ states in spin triplets
should be integer-valued. They should take one of the values $-2$,
$-1,$ $0$, $1$, $2$, or $3$, which means that there are ${6 \choose 2}
= 15$ states according to Takahashi's counting (\ref{r1})-(\ref{r3}).

There is one $\h$-pairing state. The energy of this state is equal to
$U$, since the two electrons occupy the same site. 

Now, let us sum up  all the numbers of the different types of
eigenstates: 
\beq 
     15 + 5 + 15 + 1 = 36  \quad . 
\eeq
Thus, we have shown that all the energy eigenstates obtained by Method
1 are  confirmed by Method 2. In particular, we have confirmed
numerically the completeness of the Bethe ansatz.

Let us consider the total momentum. Using equations (\ref{t1})-%
(\ref{t3}) we can express the total momentum $P$ of the eigenstates
with real charge momenta in terms of $I_1$ and $I_2$,
\beq 
     P = {\frac {2 \pi} L} (I_1 + I_2) \; \mod 2 \pi.   
\eeq
This formula is consistent with table 1. 

Let us now discuss the $U$-dependence of the spectrum. In figure 4
we show the spectral flow from strong-coupling to weak-coupling, where
the Takahashi condition does not hold.  
\begin{figure}

\begin{center}

\includegraphics[width = 8cm,angle = 270]{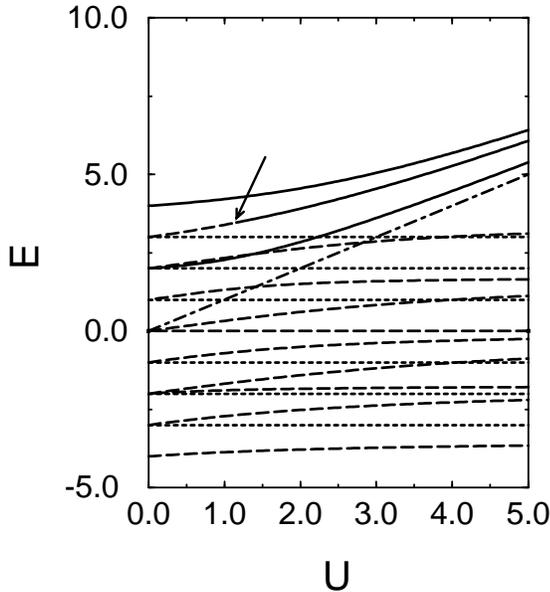}

\caption{The spectral flows for $N = 2$ and $M = 1$ for a 6-site
lattice. Solid lines denote the $k$-$\Lambda$ strings. Dashed lines
denote the energy of real roots ($S=0$). Dotted lines denote the
triplet states ($S=1$). The dash-dotted line denotes the energy of the
$\h$-pair. At $U_5=U_7=1.1547$, as indicated by the arrow,
two $k$-$\Lambda$ two strings with $m=5,7$ collapse into real
solutions.}

\end{center}

\end{figure}

In figure 4 we show the redistribution phenomenon discussed in
sections II.D, III.A and above. There are five $k$-$\La$ two-string
solutions in table 1. The entries 30, 31 and 36 correspond to even $m$.
According to our discussion above, these $k$-$\La$ two strings are
stable as $U$ becomes small. Entry number 36 is the highest energy
level in the figure. Entries number 30 and 31 are degenerate. They
correspond to the third highest level at $U = 5$. The entries number
34 and 35 correspond to odd $m = 5, 7$. The corresponding levels
are degenerate. At $U_5 = U_7 =1.1547$ these $k$-$\La$ two-string
solutions collapse into pairs of real charge momenta. This is indicated
by the arrow in figure 4. On the other hand, all five $k$-$\Lambda$
string solutions do exist as long as the Takahashi condition is
satisfied.

%%%%%%%%%%%%%%%% N=3 %%%%%%%%%%%%%%%%

\subsection{Numerical solution for three electrons} 
\subsubsection{Equations for the $k$-$\Lambda$ string}

We shall now consider the set of solutions to the Lieb-Wu equations  
for $N=3$. The $k$-$\La$ strings to be searched for have two complex
$k_j$, a real $k_j$ and a real $\La$. We express these momenta and
rapidities by
\beq
     k_1 = q - \i \xi \qd, \qd  k_2 = q + \i \xi \qd,
           \qd k_3 \qd, \qd \La \qd. 
\eeq
The total momentum is $2 \pi m/L$, so that 
\beq
     k_1 + k_2 + k_3 = 2 \pi m/L \qd, \qd 2 \pi (\ell - 1)/L < k_3
                         < 2\pi \ell /L \qd. 
\eeq
We define $\Re \delta$ and $\Im \delta$ by 
\begin{eqnarray} 
     \sin k_1 & = & \Lambda + \i U/4 +  \Re \delta + \i \, \Im \delta  
                    \qd, \\ 
     \sin k_2 & = & \Lambda - \i U/4 +  \Re \delta - \i \, \Im \delta  
                    \qd.
\end{eqnarray} 
We recall that $\Re \delta$ and $\Im \delta$ describe the deviations 
from the ideal string solutions. Now we derive equations for four
variables, $k_3$, $q$, $\x$ and $\La$, starting from the Lieb-Wu
equations (\ref{bak3}), (\ref{bas3}). Taking logarithms, we obtain a
set of equations of the form $f_i=0$ ($i=1,\ldots 4$). Here the $f_i$'s
are defined by 
\begin{eqnarray} 
     f_1 & = & L k_3
               - 2 \arctan \left( \frac{\Lambda - \sin k_3}{U/4} \right)
	       - 2 \pi(\ell - 1/2) \qd, \\
     f_2 & = & - 2 L \arcsin
                \bigg(1/2 \Big\{ \sqrt{(\Lambda + \Re \delta + 1)^2 +
                (U/4 + \Im \delta )^2} \nn \\ 
             && \qqqd -  \sqrt{(\Lambda + \Re \delta -1)^2 
                 + (U/4 + \Im \delta)^2} \Big\} \bigg)  + 
                 L k_3 - 2 \pi(\ell - 1/2 - J') \qd, \\
     f_3 & = & \exp(- 2 L \xi)- \frac{(\Re \delta)^2 + (\Im \delta)^2} 
                                      {(\Re \delta)^2 + (U/2 + \Im
                                            \delta)^2} \qd, \\
     f_4 & = & - 2 \arctan \left( \frac{\Re \delta}{\Im \delta}
                                                              \right)
                  + 2 \arctan \left( \frac{\Re \delta}{U/2 +
                                             \Im \delta} \right)
                  - 2 \arctan \left( \frac{\Lambda - \sin k_3}{U/4}
                        \right) - \sigma \pi \qd,
\end{eqnarray}
where 
\beq
     \Re \delta = \sin q \cosh \xi - \Lambda \qd, \qd 
     \Im \delta = - U/4 - \cos q \sinh \xi \qd,
\eeq
and the parameter $\sigma$ is given by 
\beq
     \sigma = 1 \qd {\rm for} \qd k_3 < 
              \frac{2 \pi}{L} (\ell - 1/2) \qd, \qd
     \sigma = - 1 \qd {\rm for} \qd k_3 > 
              \frac {2 \pi}{L} (\ell - 1/2) \qd.
\eeq
Note that the equation $f_1 = 0$ is equivalent to equation (\ref{laofk})
for $j = 3$.

To solve this set of coupled equations by Method 2, we need a
proper initial guess. We employ the ideal strings given by 
the discrete Takahashi equations as initial approximation. 
The fact that the ideal strings provide a good estimate for
the true solution is crucial to reach the correct answer. 
This is because the $f_i$'s are rather singular functions having many
diverging points. Very often, the true solution is very close to a
divergent point. We cannot approach the solution from a point beyond
a branch cut using an iterative way like the Brent method.

\subsubsection{Numerical examples of $k$-$\Lambda$ string solutions}
We present some numerical solutions for one $k$-$\La$ two string
and one real $k$. The parameters in the examples below are $N=3$
(three electrons), $L=10$ (10 sites) and $U=5$. 
 
\vskip 0.6cm 
(1) $m=10$, $\ell=1$ ($I_3=0, J^{'}=1/2$)
\begin{eqnarray}
{\rm Householder} \qquad   
 E & = & 4.46666961980768  \nonumber  \\     
{\rm Bethe \, Ansatz} \qquad
E & = &  4.46666961980768   \nonumber  \\     
  q& = &    2.98893848049280    \nonumber  \\     
  \xi  & =&     1.05674954466496    \nonumber  \\     
  k_3  & =&      0.305308346193987    \nonumber  \\     
  \Lambda & =&   0.245232889885225    \nonumber  \\     
  \Re \delta & =&    -6.42851441900183 \times 10^{-5} \nonumber  \\
  \Im \delta & = &    2.84511548476196 \times 10^{-6} \nonumber     
\end{eqnarray}

(2) $m=7$, $\ell=1$ ($I_3=0, J^{'}=7/2$)
\begin{eqnarray} 
{\rm Householder} \qquad E & = & 3.48250616148668  \nonumber  \\     
  {\rm Bethe \,  ansatz} \qquad 
E & = &  3.48250616148668  \nonumber  \\         
  q& = &     1.92454676428806    \nonumber  \\     
  \xi  & = &     1.99506896270533     \nonumber  \\    
  k_3    & = &    0.549136186449599     \nonumber  \\    
  \Lambda  & = &   3.51250692222577     \nonumber  \\    
  \Re \delta & = &    2.08765360554253 \times 10^{-9}  \nonumber  \\
  \Im \delta  & = &   4.99459629210719 \times 10^{-9}  \nonumber    
\end{eqnarray} 

Let us discuss possible numerical errors for the above solutions. 
Their numerical errors may be evaluated by the residual, $f_i$, 
given for these solutions as follows.  

{\vskip 0.6cm }

(1) $m=10$, $\ell=1$ ($I_3=0, J=1/2$)
 \begin{eqnarray} 
 f_1 & = & -1.24900090270330 \times  10^{-15} \nonumber \\
 f_2 & = & -3.99680288865056 \times  10^{-15} \nonumber \\
 f_3 & = &  5.25259688971173 \times  10^{-22} \nonumber \\
 f_4 & = &  4.53592718940854 \times  10^{-12} \nonumber
\end{eqnarray} 
with $\sigma=+1$. 

(2) $m=7$, $\ell=1$ ($I_3=0, J=7/2$)
\begin{eqnarray} 
  f_1 & =&  -1.33226762955019 \times 10^{-15} \nonumber \\
  f_2 & =&   1.42108547152020 \times 10^{-15} \nonumber \\ 
  f_3 & =&  -4.66698631842825 \times 10^{-25} \nonumber \\
  f_4 & =&  -7.67983898697366 \times 10^{-10} \nonumber
\end{eqnarray} 
with $\sigma=-1$. 

In comparison to other error values $f_i$ the number $f_4$ has a
rather large value. However, in $f_4$, we have an expression like
$\e/\e'$ ($\e \simeq 0$, $\e' \simeq 0$). So this value contains a
larger cancellation error for smaller $\Re \delta$ and $\Im \delta$.
Note again that relative errors for the energy are always
$\CO(10^{-15})$.

\subsubsection{A complete list of solutions for $N=3$ and $M=1$}
We can present complete lists of eigenstates for all finite systems
tractable by our numerical technique. As a further example, we consider
all eigenstates for $N=3$ and $M=1$ in a 6-site system ($L=6$). The
list is shown in appendix~C. It confirms again the completeness of
the Bethe ansatz. 

Let us briefly discuss the numbers of eigenstates of different types.
First, we note that there are in total $6 {6 \choose 2} = 90$
eigenstates. Inspection of the tables in appendix~C shows that they
can be classified into the following four types. 
\begin{enumerate} 
\item 
40 eigenstates with three real charge momenta $k_1, k_2, k_3$ and 
one real spin rapidity $\Lambda$.
\item
24 eigenstates with one $k$-$\Lambda$ string with $k_1 = k_- = q - \i
\xi$, $k_2 = k_+ = q  + \i \xi$, and $\Lambda$ and $k_3$ real.
\item
20 eigenstates with $S = 3/2$ and $S_z = 1/2$ belonging to spin
quartets.
\item
6 eigenstates with one $\h$-pair and one real charge momentum.
\end{enumerate} 

Let us now confirm that these numbers agree with Takahashi's counting,
(\ref{r1})-(\ref{r3}): Case (i) was considered in section III.A below
Lemma III.2. There we showed that Takahashi's counting predicts a
number of $2 {L \choose 3}$ real solutions for three electrons and
one down spin. Inserting $L = 6$ we have $2 {6 \choose 3} = 40$
eigenstates, which is in accordance with our numerical calculation.
Case (ii) was considered below Lemma III.5. The number of eigenstates
obtained there by Takahashi's counting was $L(L - 2)$, which for
$L = 6$ gives as desired $6 \cdot 4 = 24$. Let us consider case (iii).
These states are the second highest states ($S_z = 1/2$) in spin
quartets. They are obtained from regular Bethe states with $N = 3$,
$M = 0$ by multiplication with the spin lowering operator $\z^\dagger$.
Since $N = 3$ and $M = 0$, we have ${6 \choose 3} = 20$ states of this
type. The $\h$-pair in case (iv) is obtained by acting with $\h^+$ on
regular Bethe states with $N = 1$ and $M = 0$. Hence, Takahashi's
counting gives 6 eigenstates of this type on a 6-site lattice.
\subsection{Summary}
In this section we have presented a thorough numerical study of the
Hubbard model. We calculated, in particular, all eigenstates and
eigenvectors for a six-site lattice with two and three electrons and
one down spin by direct numerical diagonalization of the Hamiltonian.
These data were compared with data obtained by numerical solution of
the Lieb-Wu equations. Both sets of data are in perfect numerical
agreement and confirm once again the results of our analytical
investigation in the previous sections. The structure of our numerical
data is fully consistent with Takahashi's string hypothesis. The
number and classification of the eigenstates is consistent with
Takahashi's counting (\ref{r1})-(\ref{r3}). Thus, our numerical data
confirm not only the existence of $k$-$\Lambda$ strings but also the
completeness of the Bethe ansatz. Without $k$-$\La$ strings the Bethe
ansatz would be incomplete.
@

%%%%%%%%%%%%%%%%%%%%%%%%%%%%%%%%%%%%%%%%%%%%%%%%%%%%%%%%%%%%%%%%%%%%%%%%
%   scd5.tex                                                           %
%%%%%%%%%%%%%%%%%%%%%%%%%%%%%%%%%%%%%%%%%%%%%%%%%%%%%%%%%%%%%%%%%%%%%%%%
\def\be{\begin{equation}}
\def\ee{\end{equation}}
\def\bdm{\begin{displaymath}}
\def\edm{\end{displaymath}}
\def\nn{\nonumber\\}
\def\up{\uparrow}
\def\da{\downarrow}
\def\sgn{{\rm sign}}
\def\eps{\epsilon}
\def\r#1{(\ref{#1})}

\section{Thermodynamics in the Yang-Yang approach and excitation 
spectrum in the infinite volume}

Let us now turn to the determination of thermodynamic quantities
and the zero-temperature excitation spectrum in the
infinite volume. A convenient way to construct the spectrum was
pioneered by C.N. Yang and C.P. Yang for the case of the
delta-function Bose gas \cite{YaYa69}. The starting point are
the Bethe Ansatz equations in the finite volume. They are used to
derive a set of coupled, nonlinear integral equations called
thermodynamic Bethe Ansatz (TBA) equations, which describe
the thermodynamics of the model at finite temperatures. 
The quantities entering these equations have a natural interpretation
in terms of dressed energies of elementary excitations. 
Yang and Yang's formalism is a natural generalization of the
thermodynamics of the free Fermi gas to interacting systems. 

In what follows we review Takahashi's derivation of the TBA equations
for the case of the repulsive Hubbard model \cite{Takahashi72}. The
analogous calculations for the attractive case can be found in
\cite{LeSc89}.

Our starting point are the discrete Takahashi equations \r{t1}-\r{t3}
and expressions for energy \r{en} and momentum \r{mom} for very large
but finite $L$. A very important property of \r{t1}-\r{t3} is that as we
approach the thermodynamic limit $L\to \infty$, $N/L$ and $M/L$ fixed
(finite densities of electrons and spin down electrons), the roots of
\r{t1}-\r{t3} become dense
\be
k_{j+1}-k_j={\cal O}(L^{-1}),\quad
\Lambda^n_{\alpha+1}-\Lambda^n_\alpha={\cal O}(L^{-1}),\quad
{\Lambda^\prime}^n_{\alpha+1}-{\Lambda^\prime}^n_\alpha=
{\cal O}(L^{-1}).
\label{dense}
\ee

We now define so-called {\sl counting functions}
$y,z_n,z^\prime_n$ as follows

\begin{eqnarray} 
     k L & = & L y(k) - \sum_{n=1}^\infty \sum_{\alpha = 1}^{M_n}
                 \theta \left(
		 \frac{\sin k - \Lambda_\alpha^n}{nU/4} \right)
                 - \sum_{n=1}^\infty \sum_{\alpha = 1}^{M_n'}
                 \theta \left(
		 \frac{\sin k - {\Lambda'}_\alpha^n}{nU/4} \right),
		 \\ 
     \sum_{j=1}^{N - 2M'} \theta \left(
		 \frac{\Lambda - \sin k_j}{nU/4} \right) & = &
		 L z_n(\Lambda) +
		 \sum_{m=1}^\infty \sum_{\beta = 1}^{M_m}
		 \Theta_{nm} \left(
		 \frac{\Lambda - \Lambda_\beta^m}{U/4} \right),
		 \\ 
     L [\arcsin({\Lambda'} + n \tst{\frac{\i U}{4}})
        + \arcsin({\Lambda'} - n \tst{\frac{\i U}{4}})] & = &
	         L z^\prime_n(\Lambda') +
		 \sum_{j=1}^{N - 2M'} \theta \left(
		 \frac{{\Lambda'} - \sin k_j}{nU/4} \right) +
		 \sum_{m=1}^\infty \sum_{\beta = 1}^{M_m'}
		 \Theta_{nm} \left(
		 \frac{{\Lambda'} - {\Lambda'}_\beta^m}{U/4}
		 \right).
\end{eqnarray}
By definition the counting functions satisfy the following equations
when evaluated for a given solution of the discrete Takahashi equations 
\be
y(k_j)=2\pi I_j/L\ ,\quad
z^\prime_n({\Lambda^\prime}^n_\alpha)=2\pi {J^\prime}^n_\alpha/L\ ,\quad 
z_n({\Lambda}^n_\alpha)=2\pi J^n_\alpha/L\ .
\ee
In the next step we define the so-called {\sl root densities},
which are related to the counting functions as follows. By definition
the counting functions ``enumerate'' the Bethe Ansatz roots e.g.
\be
L[y(k_j)-y(k_n)]=2\pi(I_j-I_n).
\ee
For a given solution of \r{t1}-\r{t3} certain of the (half-odd)
integers between $I_j$ and $I_n$ will be ``occupied'' i.e. there will
be a corresponding root $k$, whereas others will be omitted. We describe
the corresponding $k$-values in terms of a root density $\rho(k)$ for
``particles'' and a density $\rho^h(k)$ for ``holes''. In a very large
system we then have by definition (here the property \r{dense} is very
important)
\bea
L\rho(k)\ dk&=& {\rm number\ of\ k's\ in\ dk}\ ,\nn
L\rho^h(k)\ dk&=& {\rm number\ of\ holes\ in\ dk}\ .
\eea
It is then clear that in the thermodynamic limit we have
\be
2\pi[\rho(k)+\rho^h(k)]=\frac{dy(k)}{dk}\ .
\ee
The analogous equations for the other roots of \r{t1}-\r{t3} are
\bea
     2\pi[\sigma_n(\Lambda) + \sigma_n^h(\Lambda)] =
	\frac{dz_n(\Lambda)}{d\Lambda}\ , \qquad
     2\pi[\sigma^\prime_n(\Lambda)+{\sigma_n^\prime}^h(\Lambda)]=
        \frac{dz^\prime_n(\Lambda)}{d\Lambda}\ .
\eea

%\be
%\rho(k_j)=\lim_{L\to\infty}\frac{1}{L(k_{j+1}-k_j)}\ ,\quad
%\sigma_n(\Lambda^n_\alpha)=\lim_{L\to\infty}\frac{1}{L(\Lambda^n_{\alpha+1}
%-\Lambda^n_\alpha)}\ ,\quad
%\sigma^\prime_n(\Lambda^n_\alpha)=\lim_{L\to\infty}
%\frac{1}{L({\Lambda'}^n_{\alpha+1}-{\Lambda'}^n_\alpha)}.
%\ee

In the thermodynamic limit the discrete Takahashi equations can
now be turned into coupled integral equations involving both counting
functions and root densities

\begin{eqnarray} \label{z1}
     k & = &  y(k) - \sum_{n=1}^\infty \int_{-\infty}^\infty
		d\Lambda\ \theta \left(
		 \frac{\sin k - \Lambda}{nU/4} \right)
		\left[\sigma^\prime_n(\Lambda)+\sigma_n(\Lambda)\right],
		 \\ \label{z2}
     \int_{-\pi}^\pi dk\ \theta \left(
		 \frac{\Lambda - \sin k}{nU/4} \right)\ \rho(k) & = &
		  z_n(\Lambda) +
		 \sum_{m=1}^\infty \int_{-\infty}^\infty
		d\Lambda^\prime\  \Theta_{nm} \left(
		 \frac{\Lambda - \Lambda^\prime}{U/4} \right)\ 
			\sigma_m(\Lambda^\prime),
		 \\ \label{z3}
      \arcsin({\Lambda} + n \tst{\frac{\i U}{4}})
        + \arcsin({\Lambda} - n \tst{\frac{\i U}{4}}) & = &
	          {z^\prime_n}(\Lambda) +
		 \int_{-\pi}^\pi dk\ \theta \left(
		 \frac{{\Lambda} - \sin k}{nU/4} \right)\ \rho(k)\nn 
		&& +
		 \sum_{m=1}^\infty \int_{-\infty}^\infty d\Lambda^\prime\
		 \Theta_{nm} \left(
		 \frac{{\Lambda} - {\Lambda^\prime}}{U/4}
		 \right)\ \sigma^\prime_m(\Lambda^\prime).
\end{eqnarray}

As we are interested in the Hubbard model at finite temperatures we
need to express the entropy in terms of the root densities. This is
achieved by observing that e.g. the number of vacancies for $k$'s in
the interval $[k,k+dk]$ is simply $L(\rho(k)+\rho^h(k))dk$. Of these
``vacancies'' $L\rho(k)dk$ are occupied. The corresponding
contribution $dS$ to the entropy is thus formally
\be
e^{dS}=\frac{(L[\rho(k)+\rho^h(k)]dk)!}{(L\rho(k)dk)!(L\rho^h(k)dk)!}\ ,
\ee
where $!$ denotes the factorial. The differential $dS$ is
obtained via Stirling's formula. After integration we obtain the
following expression for the total entropy density of the Hubbard
model
\bea
S/L&=&\int_{-\pi}^\pi dk\left\lbrace
[\rho(k)+\rho^h(k)] \ln[\rho(k)+\rho^h(k)]
- \rho(k) \ln\rho(k) - \rho^h(k) \ln\rho^h(k)
\right\rbrace\nn
&&+\sum_{n=1}^\infty\int_{-\infty}^\infty d\Lambda\left\lbrace
[\sigma_n(\Lambda)+\sigma_n^h(\Lambda)]
\ln[\sigma_n(\Lambda)+\sigma_n^h(\Lambda)]-
\sigma_n(\Lambda)\ln\sigma_n(\Lambda) - 
\sigma_n^h(\Lambda)\ln\sigma_n^h(\Lambda)
\right\rbrace\nn
&&+\sum_{n=1}^\infty\int_{-\infty}^\infty d\Lambda\left\lbrace
[\sigma^\prime_n(\Lambda)+{\sigma_n^\prime}^h(\Lambda)]
\ln[\sigma_n^\prime (\Lambda)+{\sigma^\prime_n}^h(\Lambda)]-
\sigma^\prime_n(\Lambda)\ln{\sigma^\prime_n}(\Lambda) - 
{\sigma^\prime_n}^h(\Lambda)\ln{\sigma^\prime_n}^h(\Lambda)
\right\rbrace .
\eea
The Gibbs free energy per site is thus
\bea
f&=&\frac{E-\mu N-2BS^z-TS}{L}\nn
&=&\int_{-\pi}^\pi dk\left[-2\cos k -\mu-U/2-B\right]\rho(k)
+\sum_{n=1}^\infty\int_{-\infty}^\infty d\Lambda\ 2nB\ \sigma_n(\Lambda)\nn
&&+\sum_{n=1}^\infty\int_{-\infty}^\infty d\Lambda\left[ 4{\rm
Re}\sqrt{1-(\Lambda-inU/4)^2}-2n\mu-nU\right]\sigma^\prime_n(\Lambda)-T\ S/L\ .
\label{gibbs1}
\eea
Here $\mu$ is a chemical potential, $B$ is a magnetic field and $T$ is
the temperature.
The alert reader will have realized that we are still missing
a set of equations that allows us to completely determine the root
densities and counting functions (the thermodynamic limit
\r{z1}-\r{z3} of the discrete Takahashi equations are clearly
insufficient). This is the topic of the following subsection.

\subsection{Takahashi's Thermodynamic Equations}

Let us start by differentiating \r{z1}-\r{z3}, which yields
\bea
\rho(k)+\rho^h(k)&=&\frac{1}{2\pi}+\cos
k\sum_{n=1}^\infty\int_{-\infty}^\infty d\Lambda\frac{nU/4}{\pi}
\frac{\sigma^\prime_n(\Lambda)+\sigma_n(\Lambda)}{(nU/4)^2+(\sin
k-\Lambda)^2 }\ ,\nn
\sigma_n^h(\Lambda)&=&-\sum_{m=1}^\infty A_{nm}*\sigma_m\bigg|_\Lambda
+\int_{-\pi}^\pi dk \frac{nU/4}{\pi}\frac{\rho(k)}{(nU/4)^2+(\sin
k-\Lambda)^2 }\ ,\nn
{\sigma^\prime_n}^h(\Lambda)&=&\frac{1}{\pi}{\rm
Re}\frac{1}{\sqrt{1-(\Lambda -inU/4)^2}}
-\sum_{m=1}^\infty A_{nm}*\sigma^\prime_m\bigg|_\Lambda 
-\int_{-\pi}^\pi dk \frac{nU/4}{\pi}\frac{\rho(k)}{(nU/4)^2+(\sin
k-\Lambda)^2 }\ .
\label{densities}
\eea
Here $A_{nm}$ is an integral operator
acting on a function $f$ as
\bea
A_{nm}*f\bigg|_x&=&\delta_{nm}f(x)+\frac{d}{dx}\int_{-\infty}^\infty
\frac{dy}{2\pi} \Theta_{nm}\left(\frac{x-y}{U/4}\right)f(y).
%\nn
%\Theta_{n,m}(x) &=& (1-\delta_{m,n})\theta ({x\over{|n-m|}}) +
%2\ \theta ({x\over{|n-m|+2}})+\dots +2\ \theta ({x\over{n+m-2}}) +
%\theta ({x\over{n+m}})\ ,
\eea
%where $\theta(x)=2\arctan(x)$. 

Equations \r{densities} can be used to express the densities of holes
in terms of densities of particles.

A second set of equations is obtained by considering the Gibbs free
energy density \r{gibbs1} as a functional of the root densities. In
thermal equilibrium $f$ is stationary with respect to variations
of the root densities
\bea
0&=&\delta f \nn
&=&\frac{\delta f}{\delta\rho(k)}\delta\rho(k)
+\frac{\delta f}{\delta\rho^h(k)}\delta\rho^h(k)
+\sum_{n=1}^\infty\left[\frac{\delta
f}{\delta\sigma_n(\Lambda)}\delta\sigma_n(\Lambda)
+\frac{\delta f}{\delta\sigma^h_n(\Lambda)}\delta\sigma^h_n(\Lambda)
+\frac{\delta
f}{\delta{\sigma^\prime_n}(\Lambda')}\delta{\sigma^\prime_n}(\Lambda')
+\frac{\delta f}{\delta{\sigma^\prime_n}^h(\Lambda')}
\delta{\sigma^\prime_n}^h(\Lambda')\right],\nn
\eea
where we need to take into account \r{densities} as constraint
equations. In this way one obtains a set of equations for the ratios
$\zeta=\rho^h/\rho$, $\eta_n=\sigma_n^h/\sigma_n$ and
$\eta_n^\prime={\sigma^\prime_n}^h/\sigma_n^\prime$
\bea
&&\ln \zeta(k)=\frac{-2\cos k -\mu-U/2-B}{T}+\sum_{n=1}^\infty 
\int_{-\infty}^\infty\!\! d\Lambda \frac{nU}{4\pi}
\frac{1}{(nU/4)^2+(\sin k-\Lambda)^2}\left[
\ln\left(1+\frac{1}{\eta^\prime_n(\Lambda)}\right)
-\ln\left(1+\frac{1}{\eta_n(\Lambda)}\right)\right].\nn
\label{zeta}
\eea

\bea
&&\ln\left(1+\eta_n(\Lambda)\right) +\int_{-\pi}^\pi dk \frac{\cos k}{\pi}
\frac{nU/4}{(nU/4)^2+(\sin k-\Lambda)^2}\
\ln\left(1+\frac{1}{\zeta(k)}\right)=\frac{2nB}{T}+
\sum_{m=1}^\infty A_{nm}*\ln\left(1+\frac{1}{\eta_m}\right)\bigg|_\Lambda.
\label{lambda}
\eea
\bea
&&\ln\left(1+\eta^\prime_n(\Lambda)\right)
+\int_{-\pi}^\pi dk \frac{\cos k}{\pi}
\frac{nU/4}{(nU/4)^2+(\sin k-\Lambda)^2}\
\ln\left(1+\frac{1}{\zeta(k)}\right)=\nn
&&=\frac{4{\rm Re}\sqrt{1-(\Lambda -inU/4)^2}-2n\mu-nU}{T}
+\sum_{m=1}^\infty A_{nm}*\ln\left(1+\frac{1}{\eta^\prime_m}\right)
\bigg|_\Lambda.
\label{kl}
\eea
Note that \r{densities} together
with \r{zeta}-\r{kl} completely determine the densities of holes {\sl
and} particles in the state of thermal equilibrium. 

The Gibbs free energy per site is given in terms of solutions of 
\r{zeta}-\r{kl} as
\bea
f&=&-T\int_{-\pi}^\pi\frac{dk}{2\pi}\ \ln\left(1+\frac{1}{\zeta(k)}\right)
-T\sum_{n=1}^\infty\int_{-\infty}^\infty\frac{d\Lambda}{\pi}
\ \ln\left(1+\frac{1}{\eta^\prime_n(\Lambda)}\right) {\rm Re}\frac{1}{\sqrt{1-(\Lambda
-inU/4)^2}}\ .
\label{gibbs}
\eea

Following Takahashi we define 
\be
\kappa(k)=T\ln(\zeta(k))\ ,
\epsilon_n(\Lambda)=T\ln(\eta_n(\Lambda))\ ,
\epsilon^\prime_n(\Lambda)=T\ln(\eta^\prime_n(\Lambda))\ .
\ee
As was first shown by Yang and Yang for the delta-function Bose
gas \cite{YaYa69}, the quantities defined in this way describe the
dressed energies of elementary excitations in the zero temperature
limit. Before turning to this we will give a brief summary on how to 
calculate thermodynamic quantities in the framework of Takahashi's
approach.

\subsection{Thermodynamics}

The expression for the Gibbs free energy density \r{gibbs} can be simplified
\cite{Takahashi74}
\bea
f&=&E_0-\mu-U/2-T\left[\int_{-\pi}^\pi dk\ \rho_0(k)\ 
\ln\left(1+\zeta(k)\right)+\int_{-\infty}^\infty d\Lambda\ \sigma_0(\Lambda)
\ \ln\left(1+\eta_1(\Lambda)\right)
\right],
\label{gibbs2}
\eea
where 
\bea
\sigma_0(\Lambda)&=&\int_{-\pi}^\pi dk \frac{1}{U} 
\frac{1}{\cosh\frac{2\pi}{U}(\Lambda-\sin k)}\rho_0(k)\ ,\nn
\rho_0(k)&=&\frac{1}{2\pi}+\cos
k\int_{-\infty}^\infty \frac{d\Lambda}{\pi}
\frac{U/4}{(U/4)^2+(\sin k-\Lambda)^2 }\ \sigma_0(\Lambda)\ ,\nn
E_0&=&-2\int_{-\pi}^\pi dk\ \cos(k)\ \rho_0(k)
=-4\int_0^\infty d\omega\frac{J_0(\omega)J_1(\omega)}{1+\exp(\omega U/2)}\ .
\label{omega_simp}
\eea
We note that $\rho_0$, $\sigma_0$ and $E_0$ are the root density
for real $k$'s, the root density for real $\Lambda$'s and the ground
state energy for the half-filled repulsive Hubbard model, respectively.
Since the occurrence of quantities related to the half-filled Hubbard
model in \r{gibbs2} may be surprising, we would like to emphasize
that \r{gibbs2}, \r{omega_simp} holds for all negative values of the
chemical potential $\mu$ i.e.\ for all particle densities between zero
and one.

The representation \r{gibbs2} is convenient as it shows that the Gibbs
free energy is determined by the dressed energies for real $k$'s and
real $\Lambda$'s only. In order to derive \r{gibbs2} the following 
identities are useful
\bea
&&\int_{-\pi}^\pi dk \frac{nU}{4\pi}\frac{\rho_0(k)}{(nU/4)^2+(\sin k-\Lambda)^2}
=\frac{1}{\pi}{\rm Re}\frac{1}{\sqrt{1-(\Lambda-inU/4)^2 }}\ ,\nn
&&\int_{-\pi}^\pi dk \frac{nU}{4\pi}\frac{\rho_0(k)}{(nU/4)^2+(\sin k-\Lambda)^2}
=A_{n1}*\sigma_0\bigg|_\Lambda\ .
\eea

At very low temperatures $T\ll B$ it is possible to determine the
Gibbs free energy by using an expansion of the TBA equations 
\r{zeta}-\r{kl} for small $T$ \cite{Takahashi74}. The TBA equations
essentially reduce two only two coupled equations for $\zeta$ and 
$\eta_1$ in this limit. For generic values of $B$ and arbitrary 
temperatures one needs to resort to a numerical solution of 
\r{zeta}-\r{kl}. In order to do so, one needs to truncate the 
infinite towers of equations for $\Lambda$ and $k$-$\Lambda$ strings
at some finite value of their respective lengths.
In \cite{KUO89,UKO90} such a truncated set of equations was solved by
iteration. The integrals were discretized by using of the order of
$50$ ($100$) points for the $k$ ($\Lambda$) integrations.
The results of these computations are compared to the results of the
Quantum Transfer Matrix approach in the next section.

\subsection{Zero Temperature Limit}
For the remainder of this section we set the magnetic field to zero
$B=0$.

For $T\to 0$ the thermodynamic equations \r{zeta}-\r{kl} then reduce
to \cite{Takahashi72}
\bea
\kappa(k)&=&-2\cos k-\mu-U/2+\int_{-\infty}^\infty 
\frac{d\Lambda}{\pi} \frac{U/4}{(U/4)^2+(\sin
k-\Lambda)^2}\epsilon_1(\Lambda)\ ,
\label{kappa}
\eea
\bea
(1-\delta_{n,1})\epsilon_n(\Lambda)&=&\int_{-Q}^Qdk\ \frac{\cos
k}{\pi} \frac{nU/4}{(nU/4)^2+(\sin k-\Lambda)^2}\ \kappa(k)
-A_{n1}*\epsilon_1\bigg|_\Lambda
\ ,\label{eps}
\eea
\bea
\epsilon_n^\prime(\Lambda)&=&
{4{\rm Re}\sqrt{1-(\Lambda -inU/4)^2}-2n\mu-nU}+\int_{-Q}^Qdk\
\frac{\cos k}{\pi} \frac{nU/4}{(nU/4)^2+(\sin k-\Lambda)^2}\ \kappa(k)
\ . \label{epsprime}
\eea
Note that $\kappa(\pm Q)=0$, $\kappa(k)<0$ for $|k|<Q$ and
$\epsilon_1(\Lambda)<0$. Using Fourier transform, equations \r{kappa}
and \r{eps} can be simplified further with the result 
\bea
\kappa(k)&=&-2\cos k-\mu-U/2+\int_{-Q}^Qdk^\prime \cos k^\prime\ 
R(\sin k^\prime -\sin k) \ \kappa(k^\prime)\ ,\nn
\epsilon_1(\Lambda)&=&\int_{-Q}^Qdk \frac{\cos k}{U} 
\frac{1}{\cosh\frac{2\pi}{U}(\Lambda-\sin k)}\kappa(k)\ ,\nn
\epsilon_n(\Lambda)&=&0\quad n=2,3,\ldots
\label{dresseden0}
\eea
where 
\be
R(x)=\int_{-\infty}^\infty\frac{ d\omega}{2\pi} \frac{\exp(i\omega
x)}{1+\exp(U|\omega|/2)}\ .
\ee
The vanishing of the dressed energies of $\Lambda$-strings of lengths
greater than one i.e. $\epsilon_n(\Lambda)=0$ for $n\geq 2$ is due to
the absence of a magnetic field. For finite magnetic fields all
$\epsilon_n(\Lambda)$ will be nontrivial functions. 

In order to characterize the excitation spectrum we need to determine
the dressed momenta in addition to the dressed energies. This can be
done by considering the zero temperature limit or \r{densities}
\bea
\rho(k)&=&\frac{1}{2\pi}+\int_{-Q}^Qdk^\prime \cos k\ 
R(\sin k^\prime -\sin k) \ \rho(k^\prime)\ ,\quad |k|\leq Q\ ;
\qquad \rho(k) = 0 \ , \quad |k| > Q \ ,\nn
\rho^h(k)&=&\frac{1}{2\pi}+\int_{-Q}^Qdk^\prime \cos k\ 
R(\sin k^\prime -\sin k) \ \rho(k^\prime)\ ,\quad |k|> Q\ ;
\qquad \rho^h (k) = 0 \ , \quad |k| \le Q \ ,\nn
\sigma_1(\Lambda)&=&\int_{-Q}^Qdk \frac{1}{U} 
\frac{1}{\cosh\frac{2\pi}{U}(\Lambda-\sin k)}\rho(k)\ ,\nn
\sigma_n(\Lambda)&=&0\quad n=2,3,\ldots\qquad
\sigma_m^h(\Lambda)=0=\sigma_m^\prime(\Lambda)\quad m=1,2,\ldots\qquad\nn
{\sigma_n^\prime}^h(\Lambda)&=&\frac{1}{\pi}
{\rm Re}\frac{1}{\sqrt{1-(\Lambda -inU/4)^2}}-\int_{-Q}^Qdk\ \frac{nU/4}
{\pi} \frac{\rho(k)}{(nU/4)^2+(\sin k-\Lambda)^2}\ .
\label{densT=0}
\eea
Equations \r{densT=0} describe the ground state of the repulsive
Hubbard model at zero temperature and zero magnetic field. There is
one Fermi sea of $k$'s (charge degrees of freedom) with Fermi rapidity
$\pm Q$ and a second Fermi sea of $\Lambda^1$'s (spin degrees of
freedom), which are filled on the entire real axis. 

The total momentum \r{mom} can be rewritten
by using \r{t1}-\r{t3} in the following useful manner
\bea
P&=&\frac{2\pi}{L}\left(
\sum_{j=1}^{N-2M^\prime}I_j+\sum_{m=1}^\infty \sum_{\beta=1}^{M_m}
J^m_\beta
-\sum_{n=1}^\infty \sum_{\alpha=1}^{M_n'} {J^\prime}^n_\alpha
\right)+\pi\sum_{n=1}^\infty \sum_{\alpha=1}^{M'_n}(n+1)\nn
&=&
\sum_{j=1}^{N-2M^\prime}y(k_j)+\sum_{m=1}^\infty \sum_{\beta=1}^{M_m}
z_m(\Lambda^m_\beta)
-\sum_{n=1}^\infty \sum_{\alpha=1}^{M_n'}
z^\prime_n({\Lambda^\prime}^n_\alpha)
+\pi\sum_{n=1}^\infty \sum_{\alpha=1}^{M'_n}(n+1).
\label{mtmint}
\eea
Using this expression for the total momentum we now can identify the
dressed momenta of various types of excitations. We find that an
additional real $k$ with $|k|>Q$ (``particle'') or hole in the sea of
$k$'s ($|k|<Q$) carry momentum $\pm{\tt p}(k)$ respectively, where 
\be
{\tt p}(k)=y(k)=2\pi\int_0^k dk^\prime\ [\rho(k^\prime)+\rho^h(k^\prime)]
\ .
\label{pk}
\ee
Similarly, the dressed momentum of a hole in the sea of $\Lambda^1$'s
is
\bea
{\tt p}_1(\Lambda)&=&-z_1(\Lambda)=2\pi\int_\Lambda^\infty
d\Lambda^\prime \sigma_1(\Lambda^\prime)-z_1(\infty)
=2\int_{-Q}^Q {dk}
\arctan[\exp\left(-\frac{2\pi}{U}(\Lambda-\sin k)\right)]\ \rho(k)
-\pi\frac{N}{2L}\  .
\eea
This result was first obtained by Coll \cite{Coll74}.
Finally, a $k$-$\Lambda$ string of length $n$ has dressed momentum
\bea
{\tt p}^\prime_n(\Lambda)&=&-z^\prime_n(\Lambda)+\pi (n+1)\nn
&=&-2{\rm Re}\arcsin(\Lambda-inU/4)+\int_{-Q}^Qdk\ 
2\arctan\left(\frac{\Lambda-\sin k}{nU/4}\right)\rho(k)+\pi(n+1),
\label{klmom}
\eea
where the second line is obtained from the $T\to 0$ limit of \r{z3}.
We are now in a position to completely classify the excitation
spectrum at zero temperature. The dispersion relations of all elementary
excitations follow from \r{epsprime}-\r{dresseden0} and \r{pk}-%
\r{klmom}. These equations involve only the two unknown functions,
$\rho (k)$ and $\kappa (k)$, which are solutions of linear Fredholm
integral equations.

\subsection{Ground state for a less than half-filled band}
The integral equations describing the root densities of the ground
state are \r{densT=0}

\bea
\rho(k)&=&\frac{1}{2\pi}+\int_{-Q}^Qdk^\prime \cos k\ 
R(\sin k^\prime -\sin k) \ \rho(k^\prime)\ ,\quad |k|\leq Q\ ,\nn
\sigma_1(\Lambda)&=&\int_{-Q}^Qdk \frac{1}{U} 
\frac{1}{\cosh\frac{2\pi}{U}(\Lambda-\sin k)}\rho(k)\ ,
\label{densGS}
\eea
where
\be
\int_{-Q}^Q dk \rho(k)=N/L\ ,\qquad
\int_{-\infty}^\infty d\Lambda \sigma_1(\Lambda)=M_1/L=N/2L\ .
\ee
The ground state energy per site is given by

\be
(E-\mu N)/L=\int_{-Q}^Q dk\ (-2\cos k-\mu -U/2)\ \rho(k)\ .
\ee

\subsection{Excitations for a less than half-filled band}
There are three different kinds of {\sl elementary} excitations.
\begin{itemize}
\item{} The first type of elementary excitation is gapless and 
involves the charge degrees of freedom only. It
corresponds to adding a particle to or making a hole in the
distribution of $k$'s. 
%Such excitations are sometimes 
%called {\sl antiholons} and {\sl holons} respectively. 
Such excitations have dressed energy $\mp\kappa(k)$ and dressed
momentum ${\tt p}_{p,h}(k)$. 

\item{} The second type of elementary excitation is gapless and carries
spin but no charge. It corresponds to a hole in the distribution of
$\Lambda^1$'s. Such excitations are called {\sl spinons}. They have
dressed energy $-\epsilon_1(\Lambda)$ and dressed momentum ${\tt
p}_1(\Lambda)$. 
\item{} There is an infinite number of different types of gapped
excitations that carry charge but no spin. they correspond to adding a
$k$-$\Lambda$ string of length $n$ to the ground state. Their dressed
energies are  $\epsilon^\prime_n(\Lambda)$, their dressed momenta are
${\tt p}^\prime_n(\Lambda)$.
\end{itemize}
Let us emphasize that this is a classification of {\sl elementary}
excitations in the repulsive Hubbard model below half-filling.
It is important to distinguish these from ``physical'' excitations,
which are the {\sl permitted combinations} of elementary excitations.
In other words, not any combination of particle-hole excitations,
spinons and $k$-$\Lambda$ strings is allowed, but only those
consistent with the {\sl selection rules} \r{r1}. 
To illustrate how this works and relate our findings to known results
in the literature we consider several examples. We introduce the
following terminology: we call the set $\{
N,M_n,M^\prime_n|n=1\ldots\infty\}$ of the numbers of real $k$'s,
$\Lambda$-strings of length $n$ and $k$-$\Lambda$-strings of length
$n$ {\sl occupation numbers} of the corresponding excitation. This is
in contrast to our usage of the term {\sl quantum numbers}, which is
reserved for the eigenvalues of energy, momentum, $S^z, {\vec
S}\cdot{\vec S},\eta^z$ and ${\vec \eta}\cdot{\vec \eta}$.

\vskip .5cm
\underbar{Example 1}: Particle-hole excitation.

\begin{figure}

\begin{center}
\includegraphics[width = 8cm,
height = 8cm ,angle = 0]{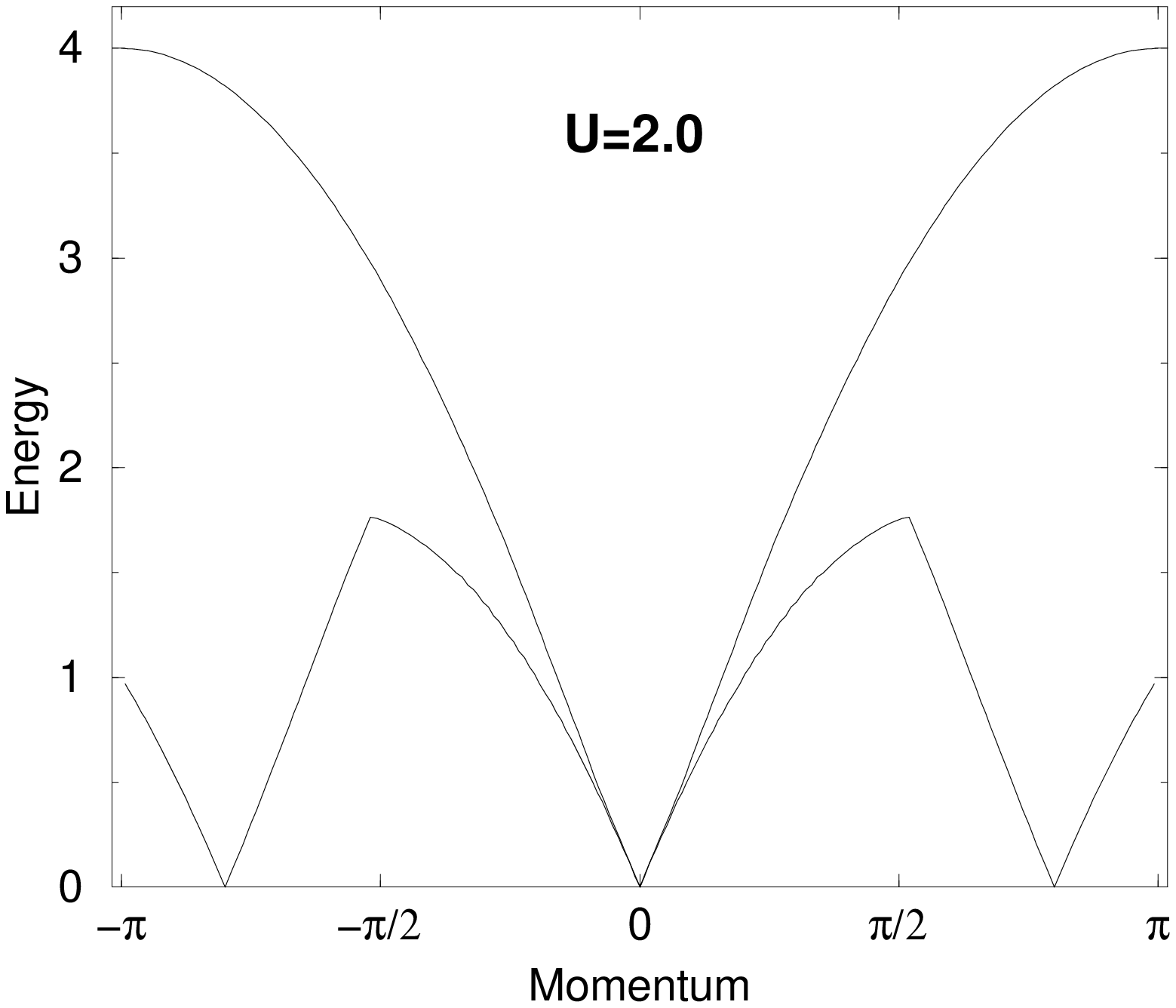}
\includegraphics[width = 8cm,
height = 8cm ,angle = 0]{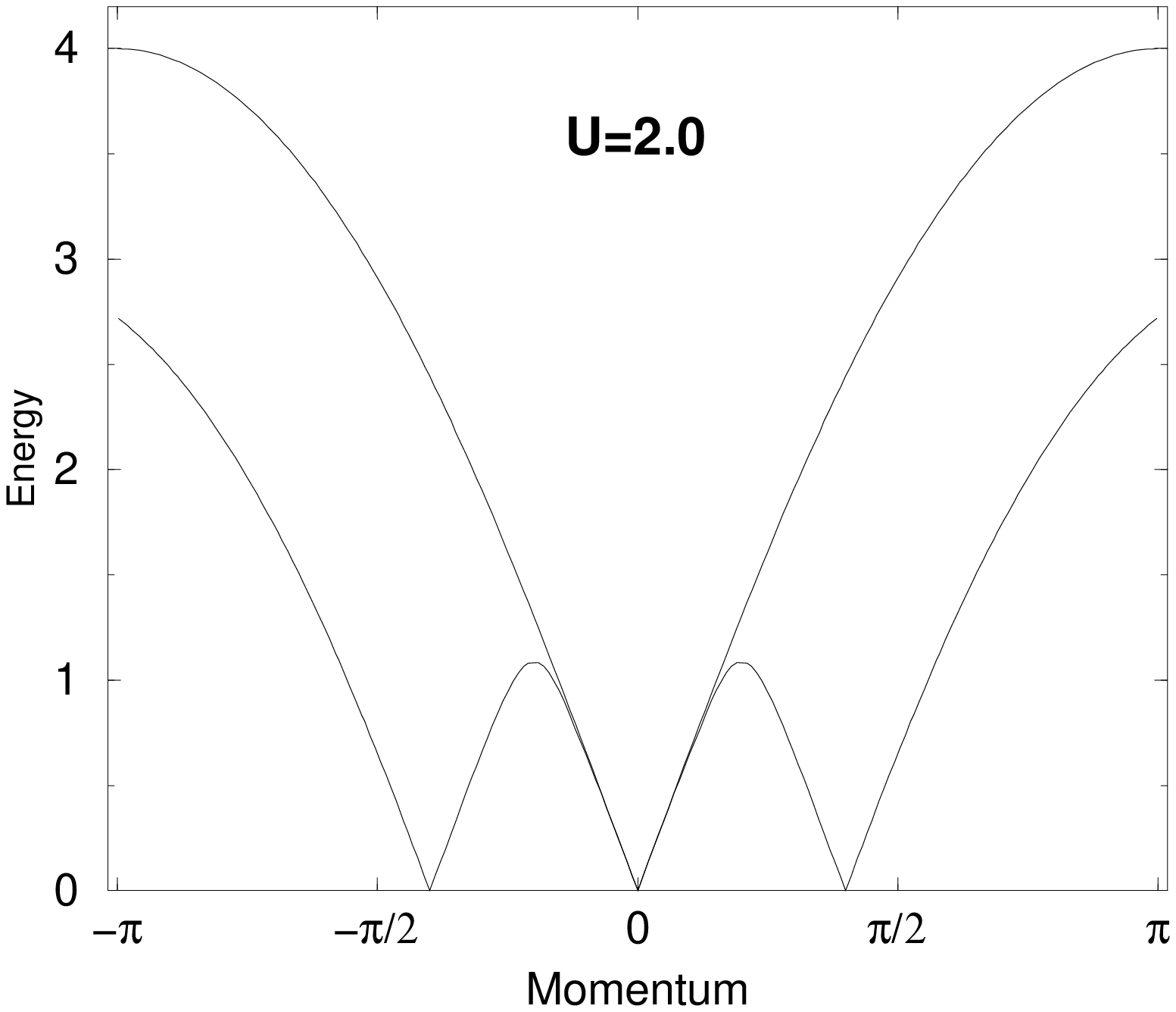}
\caption{
Particle-hole excitation for $U=2.0$ and densities $n=0.6$ and
$n=0.8$. Shown are the lower and upper boundaries of the continuum.
}
\label{fig:ph}
\end{center}

\end{figure}

This is a two-parametric gapless physical excitation with spin and
charge zero, i.e. its quantum numbers as well as its occupation
numbers are the same as the ones of the
ground state. It is obtained by removing a spectral parameter $k_h$ with
$|k_h|<Q$ from the ground-state distribution of $k$'s and adding a
spectral parameter $k_p$ with $|k_p|>Q$. Its energy and momentum are
\bea
E_{ph}&=&\kappa(k_p)-\kappa(k_h)\ ,\nn
P_{ph}&=&{\tt p}(k_p)-{\tt p}(k_h)\ .
\eea
This excitation is allowed by the selection rules \r{r1} as in the
ground state only the (half-odd) integers $|I_j|\leq (N-1)/2$ are
occupied and thus the possibility of removing a root corresponding to
$|I_h|\leq (N-1)/2$ and adding a root corresponding to $|I_p|>(N-1)/2$
exists. This excitation was first studied by a different approach in
\cite{Coll74}. In Fig.~\ref{fig:ph} we show the particle-hole spectrum for 
densities $n=0.6$ and $n=0.8$. As we approach half-filling the phase-space
for particles shrinks to zero.

\vskip .5cm
\underbar{Example 2a}: Spin triplet excitation.

Let us consider an excitation involving the spin degrees of freedom
next. If we change the number of down spins by one while keeping the
number of electrons fixed we obtain an excitation with spin
1. Recalling that in the ground state we have $N$ electrons out of
which $M_1=N/2$ have spin down, the excited state will have occupation
numbers $N$ and $M_1=N/2-1$. The selection rules \r{r1} then
read
\be
-\frac{L}{2}<I_j\leq \frac{L}{2}\ ,\qquad
|J^1_\alpha|\leq\frac{N}{4}\ .
\ee
The first condition is irrelevant as we are below half filling, but
the second one tells us that there are two more vacancies than there
are roots. In other words, flipping one spin leads to {\sl two} holes
in the distribution of $\Lambda^1$'s. There is one more subtlety we
have to take care of: changing the number of down spins by one, while
keeping the number of electrons fixed leads to a shift of all $I_j$ in
\r{t1} by either $\frac{1}{2}$ or $-\frac{1}{2}$. The consequence of
this shift is a constant contribution of $\pm \pi \frac{N}{L}$ to the
momentum of the excited state. This leads to two ``branches'' of the
same excitation.

\begin{figure}

\begin{center}
\includegraphics[width = 8cm,
height = 8cm ,angle = 0]{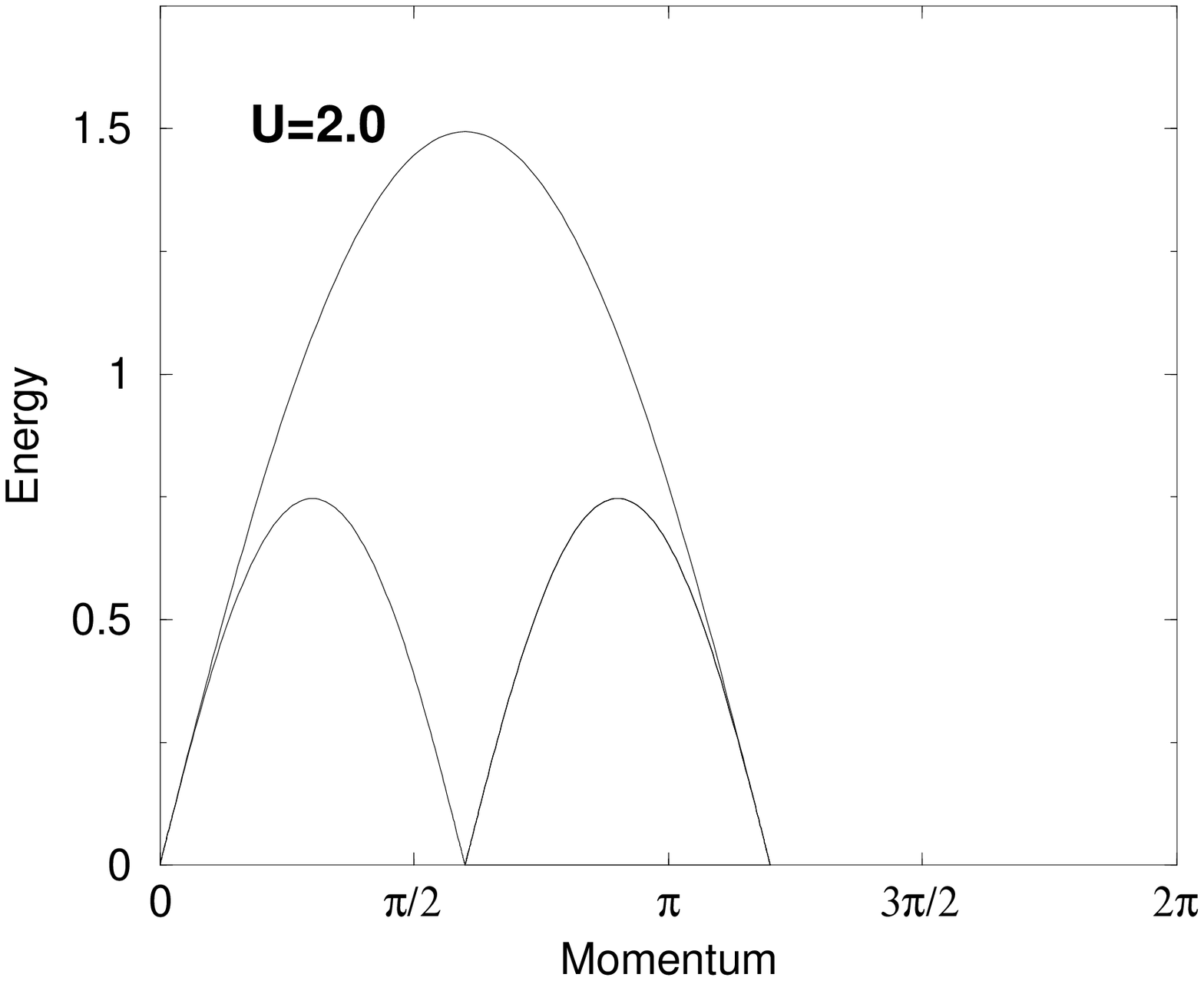}
\includegraphics[width = 8cm,
height = 8cm ,angle = 0]{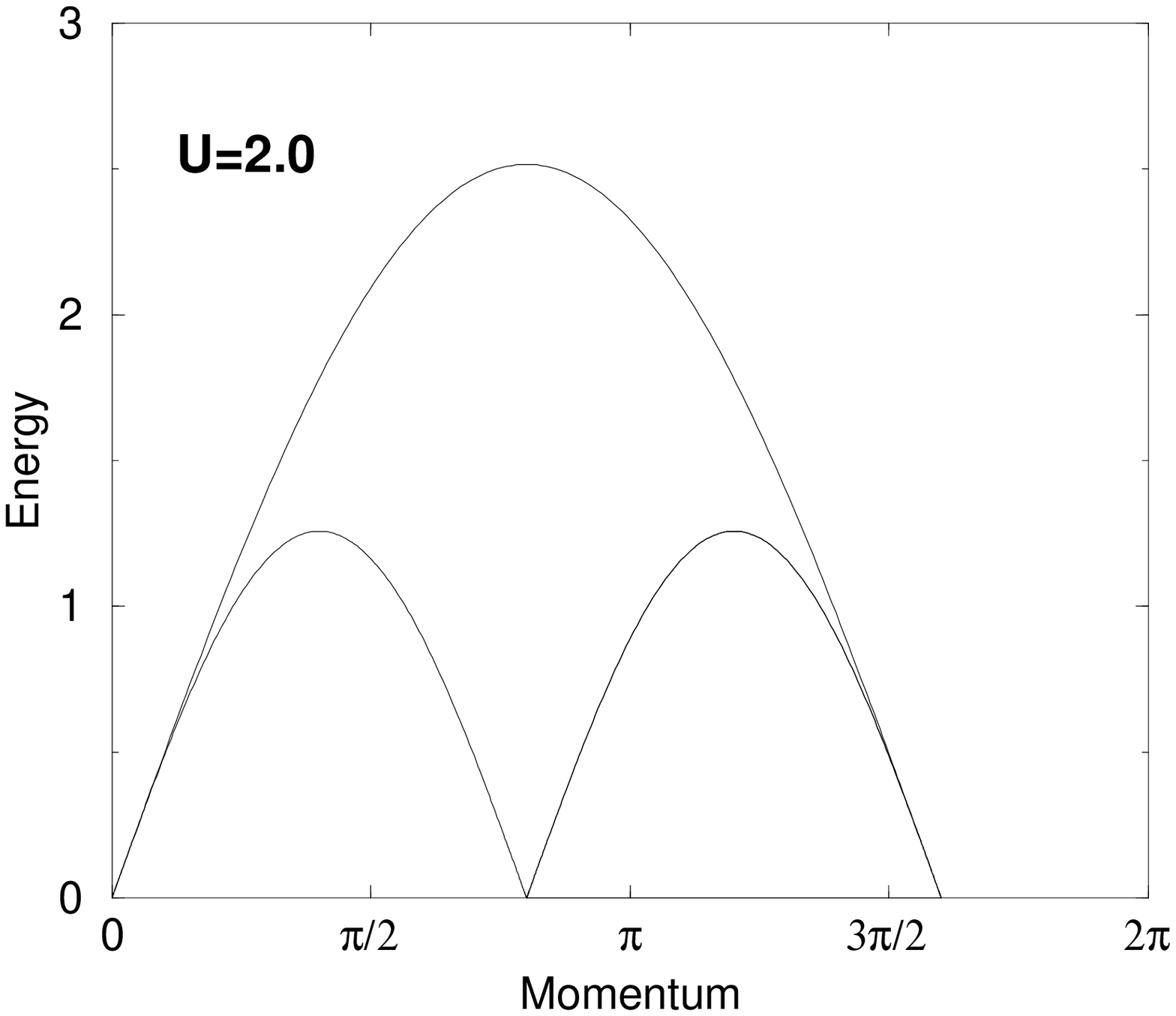}
\caption{
Spin-triplet excitation for $U=2.0$ and densities $n=0.6$ and $n=0.8$.
Shown are the lower and upper boundaries of the continuum for the positive
branch. The negative branch is obtained by reversing the sign of the
momentum. Note that we have not folded back to the first Brillouin zone.
}
\label{fig:st}
\end{center}

\end{figure}

The physical excitation obtained in this way is a gapless two-spinon
scattering state with energy and momentum 
\bea
E_{\rm trip}&=&-\epsilon_1(\Lambda_1)-\epsilon_1(\Lambda_2)\ ,\nn
P_{\rm trip}&=&{\tt p}_1(\Lambda_1)+{\tt
p}_1(\Lambda_2)\pm\pi\frac{N}{L}
\ .
\eea
In Fig.~\ref{fig:st} we show the spin-triplet spectrum for 
densities $n=0.6$ and $n=0.8$. 

In the Hubbard model the spin-triplet excitations were first studied
by Ovchinnikov \cite{Ovchinnikov70} and by Coll \cite{Coll74}. 
The situation encountered here is similar to the spin-1/2 Heisenberg
chain \cite{FaTa84} in the sense that the spin-triplet excitation is a
scattering continuum of two spin-1/2 objects. Furthermore there is a
spin-singlet excitation, which is precisely degenerate with the
triplet (see Example 2b below). This fits nicely into a
picture based on spin-1/2 objects: scattering states of two spinons
give precisely one spin $1$ and one spin $0$ multiplet
$\frac{1}{2}\otimes\frac{1}{2}=1\oplus 0$. Finally, when we
approach half-filling, the spin-triplet continuum constructed above
goes over into the $S=1$ two-spinon scattering continuum of the
half-filled Hubbard model \cite{Woynarovich83a,EsKo94b}.

On the other hand there are differences as well: in the less than
half-filled Hubbard model the Fermi momentum is generally
incommensurate, which leads to incommensurabilities in the spin
excitations (see Fig.~\ref{fig:st}).

More importantly, it is always possible to combine {\sl
any} type of excitation with a particle-hole excitation. It is
therefore not possible to distinguish the two-parametric spin-triplet
excitation constructed above from the special case of a
four-parametric excitation, where a particle-hole excitation sits
``on top'' of the spin-triplet excitation and where the momenta of the
particle and the hole are fixed at the Fermi rapidity.
In other words, due to the presence of gapless particle-hole
excitations there is an inherent ambiguity in the interpretation of the
excitation spectrum on the basis of an ${\cal O}(1)$ calculation of
energy eigenvalues of the Hamiltonian.

\vskip .5cm
\underbar{Example 2b}: Spin singlet excitation.

Let us now choose the occupation numbers as $N$, $M_1=\frac{N}{2}-2$, 
$M_2=1$. The corresponding state has the same quantum numbers 
as the ground state. From \r{r1} we find that there are $N/2$ vacancies
for real $\Lambda$'s and thus two holes corresponding to rapidities
$\Lambda_1$ and $\Lambda_2$. In other words the excitation considered
involves two spinons. As far as the 2-string is 
concerned we find that the associated integer must be zero 
$J^2_1=0$. The same shift as in example 2a occurs for the $I_j$'s.
Using \r{dresseden0} and \r{mtmint} we obtain the 
energy and momentum of the associated excitation
\bea
E_{\rm sing}&=&-\epsilon_1(\Lambda_1)-\epsilon_1(\Lambda_2)\ ,\nn
P_{\rm sing}&=&{\tt p}_1(\Lambda_1)+{\tt p}_1(\Lambda_2)\pm\pi\frac{N}{L}\ .
\eea
We see that the spin singlet is precisely degenerate with the 
spin triplet considered above. This is a consequence of the
spin $SU(2)$ symmetry of the Hamiltonian in zero magnetic field.

\vskip .5cm
\underbar{Example 3}: $k$-$\Lambda$ string of length $2$.

Let us consider the simplest excitation involving a
$k$-$\Lambda$-string. One possibility is to choose the occupation numbers
as $N$, $M_1=N/2-1$, $M^\prime_1=1$. In addition we keep the distribution
of $I_j$ fixed in such a way that $I_{j+1}-I_j=1$. It is easily
checked that this excitation is allowed by \r{r1}. 
Its energy and momentum are
\be
E_{k-\Lambda}=\eps_1^\prime(\Lambda)\ ,\quad
P_{k-\Lambda}={\tt p}^\prime_1(\Lambda)\ ,
\label{klenmom}
\ee
where $\Lambda\in(-\infty,\infty)$.
\begin{figure}
\begin{center}
\includegraphics[width = 8cm,
height = 8cm ,angle = 0]{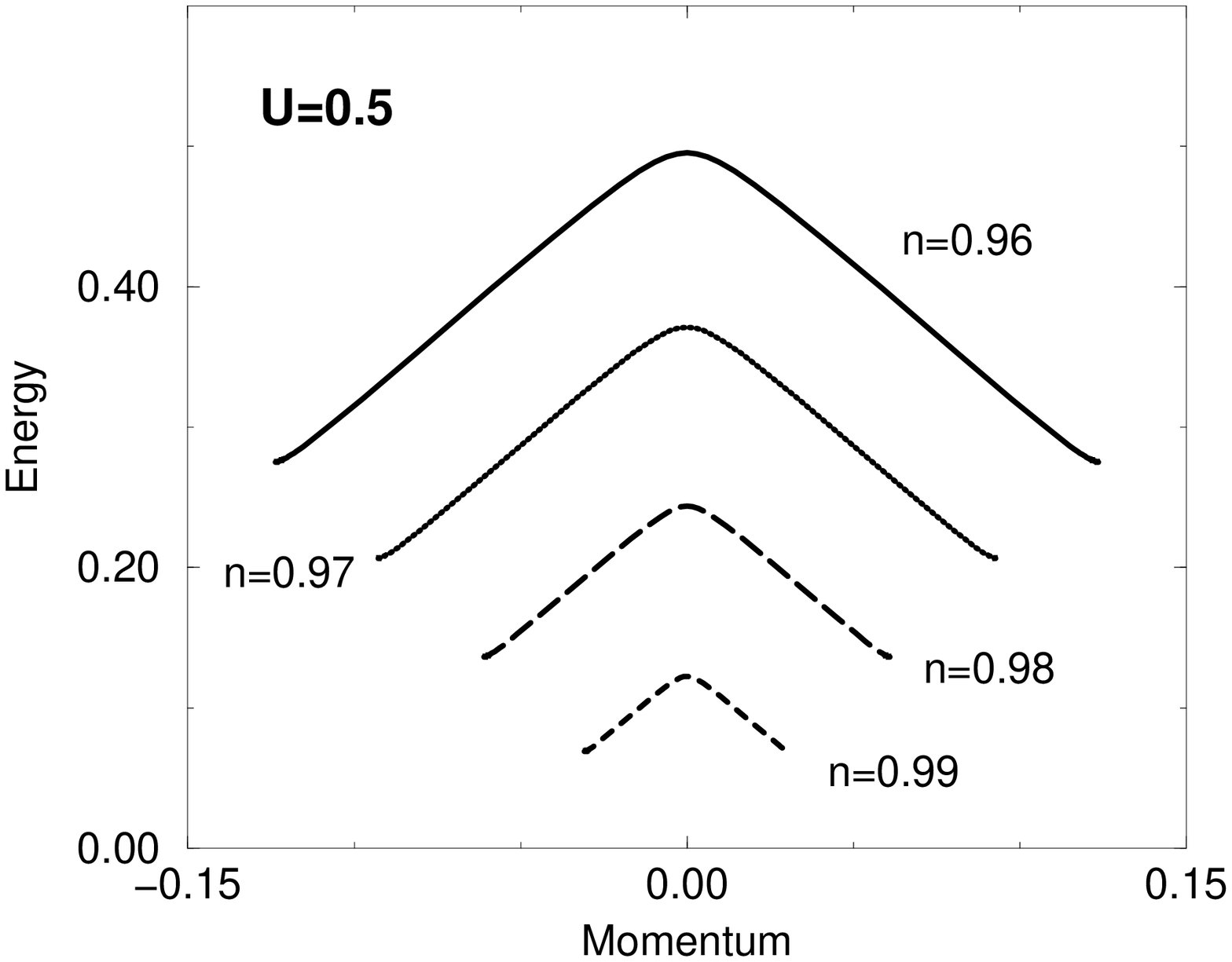}
\includegraphics[width = 8cm,
height = 8cm ,angle = 0]{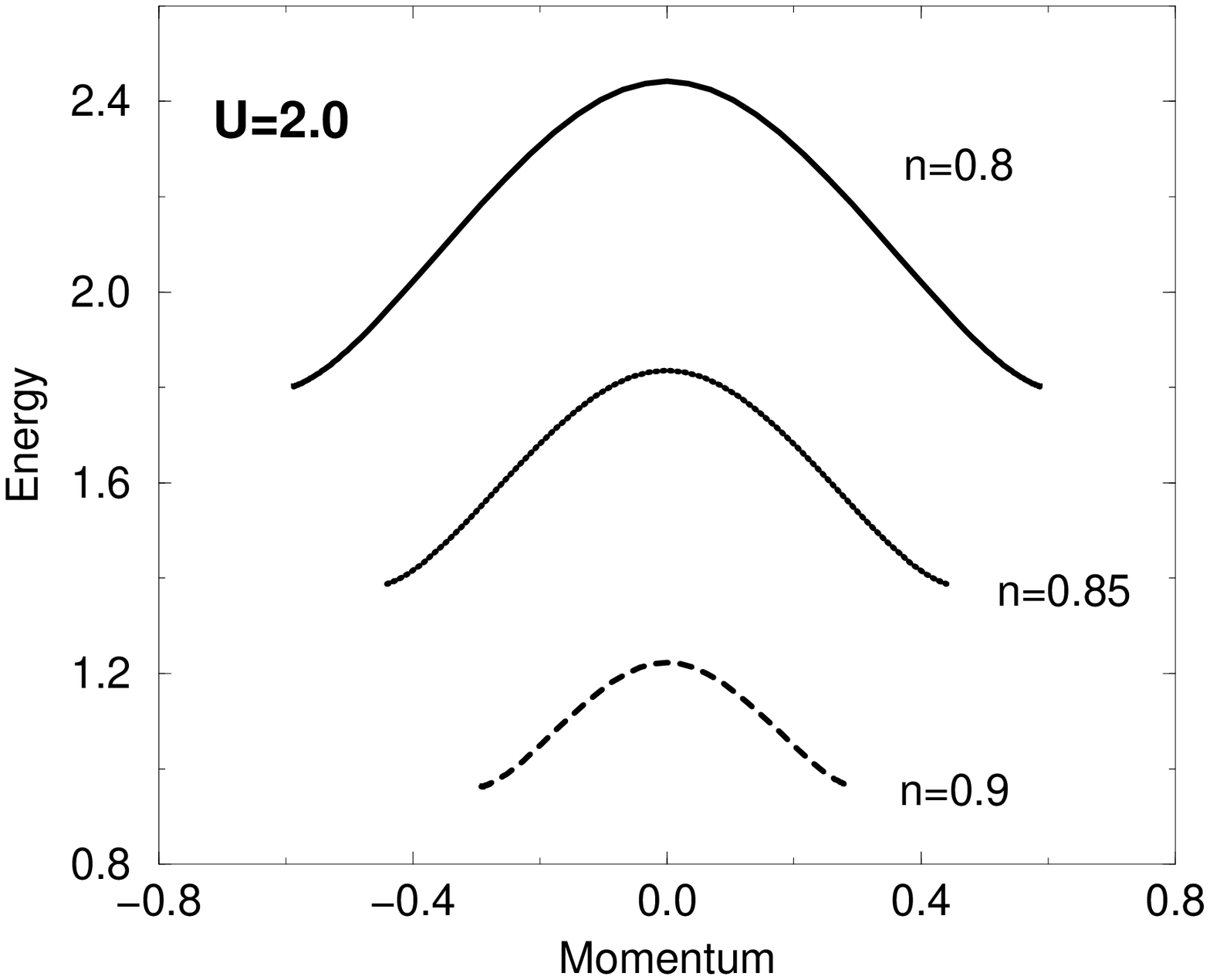}
\caption{
Dispersion of a $k$-$\Lambda$ string excitation of length $2$ for
several values of $U$ and density $n$. 
}
\label{fig:kl1}
\end{center}

\end{figure}
In Fig.~\ref{fig:kl1} the dispersion of a $k$-$\Lambda$ string of
length 2 is shown for $U=0.5$ and $U=2.0$ and several values of
density $n$. We see that the range of momenta collapses to zero as we
approach half-filling. At the same time the dressed energy approaches
zero. This is in agreement with the results for a half-filled band
\cite{EsKo94b}, where both the dressed energy and the range of
momentum are identically zero. 

%The same type of excitation was studied recently in \cite{BrAn98c} by
%a different approach (we show below that the resulting dressed energies
%and momenta coincide with \r{klenmom}, \r{epsprime} and \r{klmom} up
%to a constant).
%There only excitations at fixed density were considered and therefore
%the dispersions were plotted without the chemical potential 
%contribution. In order to facilitate a comparison we subtract the
%chemical potential piece of the dispersion as well as the constant
%$U$ and display the resulting curves in Fig.~\ref{fig:kl3}.

\begin{figure}
\begin{center}
\includegraphics[width = 8cm,
height = 8cm ,angle = 0]{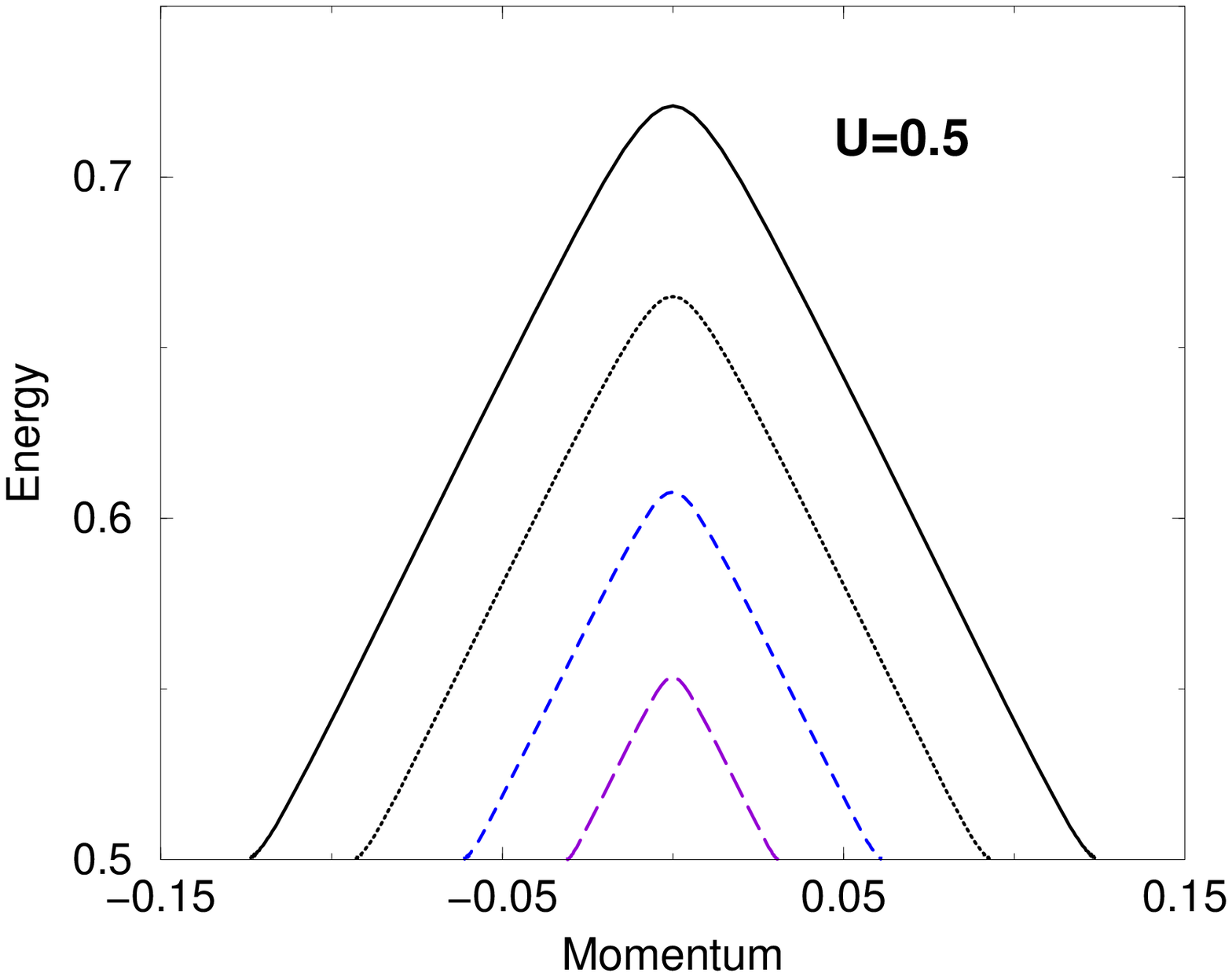}
\includegraphics[width = 8cm,
height = 8cm ,angle = 0]{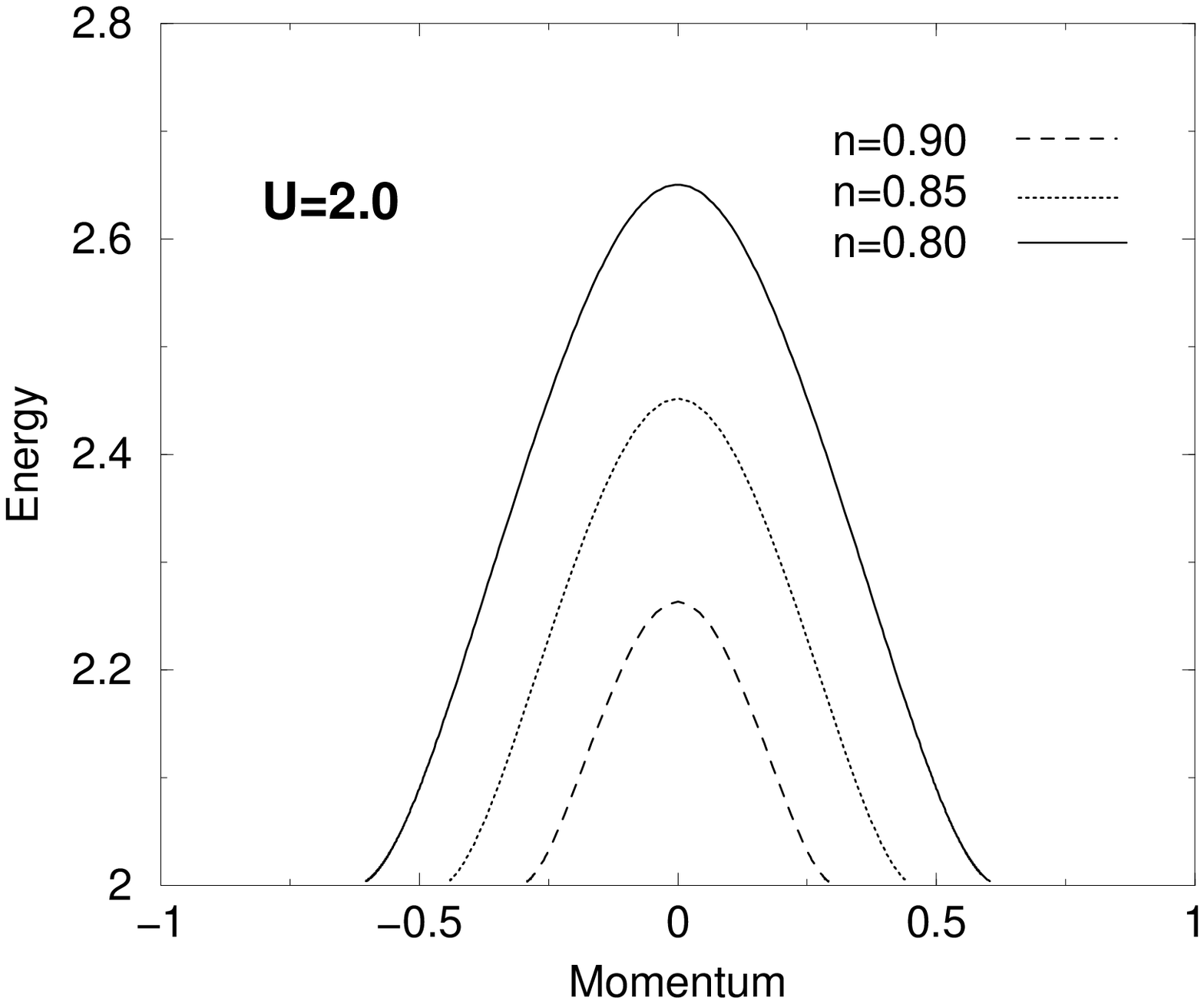}
\caption{
Dispersion of a $k$-$\Lambda$ string excitation of length $2$ for
several values of $U$ and density $n$, where the contribution
$-2\mu-U$ has been subtracted.
}
\label{fig:kl3}
\end{center}

\end{figure}

In order to further exhibit this collapse we subtract the offset
$-2\mu-U$. The resulting curves are displayed in Fig.~\ref{fig:kl3}.

%Unlike \cite{BrAn98c} we observe no ``oscillatory'' features in the
%dispersions. We believe that the findings of \cite{BrAn98c} are due to
%inaccurate numerical solutions of the integral equations.

\subsection{Excitations in the half-filled band}
In the limit of a half-filled band $\mu\to 0$ the Fermi-rapidity $Q$
tends to $\pi$ and the excitation spectrum simplifies drastically
\cite{EsKo94b}. We find that $\eps^\prime_n(\Lambda)=0\ \forall n$ and
the only non-vanishing dressed energies are
\bea
\eps_1(\Lambda) &=&-
2\int_0^\infty\frac{d\omega}{\omega}\frac{J_1(\omega)\
\cos(\omega\Lambda)}{\cosh(\omega U/4)}\ ,\nn
\kappa(k) &=& -2\ {\rm cos}k -U/2 - 2 \int_{0}^\infty {d\omega\over \omega}
{J_1(\omega) {\rm cos}(\omega\ {\rm sin}k)e^{-\omega U/4}\over {\rm
cosh}(\omega U/4)}\nn
\eea
where $J_{0,1}$ are Bessel functions. The charge excitations are now
gapped as a result of the Mott-Hubbard transition. 
The {\sl complete} spectrum of physical excitations is derived in
detail in \cite{EsKo94b} (see in particular the appendix). 
It is given in terms of spinon and holon scattering states forming
representations of SO(4). 

\subsection{Relation to the root-density formalism}

There exists a second standard approach to the zero temperature excitation spectrum
in Bethe Ansatz solvable models, which in the case of the Hubbard model has been 
employed in for example in \cite{Ovchinnikov70,Coll74}. We call this the {\sl
root-density formalism} (RDF).
In this approach one specifies a distribution of (half-odd) integers $I_j$, $J^n_\alpha$,
$J'^n_\alpha$ in the discrete Takahashi equations \r{t1}-\r{t3}. One then
takes the thermodynamic limit using that the distribution of roots 
becomes dense, so that \r{t1}-\r{t3} turn into a set of coupled linear integral equations
for root densities. From these one calculates energy and momentum via the thermodynamic
limit of \r{mom} and \r{en}. 
We will now show how to relate the results of the Yang-Yang approach we implemented above
to the RDF. For definiteness we consider the example of a $k$-$\Lambda$ string excitation
of length $n$.

We start by rewriting the integral equation for $\kappa(k)$ in the following way
\be
(1-K)*\kappa\bigg|_k=
\int_{-Q}^Q dk^\prime\left[ \delta(k-k^\prime)-\cos k^\prime\ R(\sin
k^\prime - \sin k)\right] \kappa(k^\prime) = -2 \cos k - \mu -U/2\ .
\label{kappa2}
\ee
Here the kernel of $K$ is given by $K(x,y)=\cos y R(\sin y-\sin x)$.
Equation \r{kappa2} is solved by the Neumann series
\be
\kappa(k)=-\sum_{l=0}^\infty \int_{-Q}^Qdk^\prime (2\cos
k^\prime+\mu+U/2) K^l(k,k^\prime)\ ,
\label{neumann}
\ee
where $K^l(x,y)$ denotes the $l$-fold convolution of the kernel
$K(x,y)$. Using \r{neumann} in \r{epsprime} we obtain the following representation
for the dressed energy of a $k$-$\Lambda$ string of length $n$

\bea
&&\eps^\prime_n(\Lambda)=
{4{\rm Re}\sqrt{1-(\Lambda -inU/4)^2}-2n\mu-nU}\nn
&&-\sum_{l=0}^\infty\int_{-Q}^Qdk\int_{-Q}^Qdk^\prime\ 
\frac{\cos k}{\pi} \frac{nU/4}{(nU/4)^2+(\sin k-\Lambda)^2}
K^l(k,k^\prime)\ (2\cos k^\prime+\mu+U/2)\ .
\label{epsprime2}
\eea
Let us now define a function $\rho^{bs}_n(k|\Lambda)$ by
\be
\int_{-Q}^Q dk^\prime\left[ \delta(k-k^\prime)-K(k^\prime,k)\right]
\rho^{bs}_n(k^\prime|\Lambda) = \frac{\cos k}{\pi}
\frac{nU/4}{(nU/4)^2+(\sin k-\Lambda)^2}\ .
\label{rbs}
\ee
%We note that for the case $n=1$ this precisely reduces to the root
%density defined in \cite{BrAn98c} in order to describe an excitation
%involving a $k$-$\Lambda$ string of length $2$. 
The solution of \r{rbs} is given by the Neumann series
\be
\rho^{bs}_n(k|\Lambda)=\sum_{l=0}^\infty\int_{-Q}^Q\frac{dk^\prime}{\pi}
 K^l(k^\prime,k)\ \cos k^\prime 
\frac{nU/4}{(nU/4)^2+(\Lambda-\sin k^\prime)^2}\ .
\label{rhobs}
\ee
Using \r{rhobs} in \r{epsprime2} we finally obtain
\be
\eps^\prime_n(\Lambda)=
{4{\rm Re}\sqrt{1-(\Lambda -inU/4)^2}-2n\mu-nU}
-\int_{-Q}^Qdk (2\cos k+\mu+U/2)\ \rho^{bs}_n(k|\Lambda)\ .
\ee

%Apart from the constant contribution $-2n\mu-U$ the
%special case $n=1$ agrees with the result obtained in
%\cite{BrAn98c}. In \cite{BrAn98c} only excitations at fixed density
%were considered i.e. the chemical potential term of the $k$-$\Lambda$
%string was cancelled by the chemical potential terms of two holes in the
%distribution of real $k$'s. The difference of $U$ is due to
%differences in the definition of Hamiltonian.

\subsection{Summary}
In this section we have reviewed the Yang-Yang approach \cite{YaYa69}
to the thermodynamics of the Hubbard model \cite{Takahashi72}. We have
shown how to express the Gibbs free energy per site \r{gibbs} in terms
of the solutions of the infinite set of coupled nonlinear integral
equations \r{zeta}-\r{kl}.

By taking the zero-temperature limit of the thermodynamic equations
we have obtained the complete classification of the spectrum of
elementary excitations of the Hubbard model below half-filling for
vanishing magnetic field. 

We emphasize that this approach is based on Takahashi's string
hypothesis \cite{Takahashi72} (see section II.C). It does not only
provide the dispersion curves of all elementary excitations in the
zero temperature limit, but also a set of `selection rules'
\r{t1}-\r{t3}. These rules determine the set of {\sl physical}
excitations, i.e.\ the allowed combinations of elementary excitations
(see section VI.E and, for the half-filled case, \cite{EsKo94a,%
EsKo94b}).

The numerical calculation of the Gibbs free energy from \r{gibbs} or
\r{gibbs2} is difficult, since it involves the solution of an infinite
number of coupled non-linear integral equations \r{zeta}-\r{kl}.
A truncation scheme has to be introduced \cite{KUO89,UKO90}, which
restricts the numerical accuracy of the calculation. 

In the next section we present a different approach to the 
thermodynamics of the Hubbard model, which circumvents such
difficulties. 

%Dear Friends,
%
%here is the amended version of my file. The new text is marked by
%a preceding line starting with %New
%Please let me know if there is more need for cross-references and
%citations and more salesman's rhetoric.
%
%With best regards,
%
%Andreas
%
%P.S.: I have some suggestions regarding the rest of the manuscript.
%This has to wait till tomorrow.
%
%
%\bibitem{Klu92}
%A.~Kl\"umper, Ann. Physik 1 (1992) 540, see also appendix.
%
%\bibitem{Klu93}
%A.~Kl\"umper, Z. Physik B 91 (1993) 507.
%
%\bibitem{Tsune91}
%K.~Tsunetsugu, J. Phys. Soc. Japan 60 (1991) 1460.
%
%\bibitem{KluBar96}
%A.~Kl\"umper and R.Z. Bariev, Nucl. Phys. B 458 (1996) 623.

%%%%%%%%%%%%%%%%%%%%%%%%%%%%%%%%%%%%%%%%%%%%%%%%%%%%%%%%%%%%%%%%%%%%%%%%
%   scd6.tex                                                           %
%%%%%%%%%%%%%%%%%%%%%%%%%%%%%%%%%%%%%%%%%%%%%%%%%%%%%%%%%%%%%%%%%%%%%%%%
\def\ba {\begin{array}}
\def\ea {\end{array}}
\def\e{{\rm e}}
\def\bl#1{{\overline {l_{#1}}}}
\def\b{{\frak b}}
\def\bb{\overline{{\frak b}}}
\def\c{{\frak c}}
\def\bc{\overline{{\frak c}}}
\def\t{\tau}
\def\bt{\overline{\tau}}
\def\B{{\frak B}}
\def\bB{\overline{{\frak B}}}
\def\C{{\frak C}}
\def\bC{\overline{{\frak C}}}
\def\T{T}
\def\bT{\overline{T}}
\def\squareforqed{\hbox{\rlap{$\sqcap$}$\sqcup$}}
\def\qed{\ifmmode\squareforqed\else{\unskip\nobreak\hfil
\penalty50\hskip1em\null\nobreak\hfil\squareforqed
\parfillskip=0pt\finalhyphendemerits=0\endgraf}\fi}
\def\wcirc{\squareforqed}
\def\L#1{${\cal L}_#1$}
\def\sing{\equiv_s}

%\documentstyle[12pt,amsmath,amsfonts,amssymb,graphicx]{article}
%\begin{document}

\section{Thermodynamics in the quantum transfer approach}

In this section we present a treatment of the thermodynamic
properties of the Hubbard chain in a lattice path integral
formulation with a subsequent eigenvalue analysis of the matrix
describing transfer along the chain direction
(quantum transfer matrix, QTM). This approach has several advantages.
First, the analysis is very simple as only the largest 
eigenvalue of the QTM is necessary in order to calculate the free
energy. This has to be compared with the traditional TBA
\cite{Takahashi72} 
%New
(see Sect.\ VI) where all eigenvalues of the Hamiltonian have to be
taken into account. Second, the number of `density functions' and
integral equations obtained below is {\it finite} in contrast to
\cite{Takahashi72}
%New
(see Sect.\ VI A). Finally, the finite temperature correlation lengths
can be derived through a calculation of next-largest eigenvalues of the
QTM
%New
\cite{Klu92,Klu93}. This however, will not be explored in this report.

%New
Within the QTM approach we obtain several results of which we want
to point out the conceptual achievements. First, the data obtained 
for various physical quantities agree with those obtained in
\cite{KUO89,UKO90} based on Takahashi's string hypothesis. Judging from
this we do not see any evidence for a failure of Takahashis's
formulation based on strings. Second, in the low temperature limit our
integral equations quite naturally yield the Tomonaga-Luttinger liquid
picture of separate spin and charge contributions to the free energy as
suggested in \cite{FrKo90,FrKo91}. Mathematically, the dressed
energy formalism known from the ground state analysis is recovered.

\subsection{The classical counterpart}

There are many direct path integral formulations of the Hubbard model,
see for instance \cite{Shastry86a,Bariev82}. For our purposes we want
to keep the integrability structure as far as possible. To this end it
proved useful to employ the exactly solvable classical model
corresponding to the Hubbard chain, namely Shastry's model which is a
two-sublattice six-vertex model with decoration \cite{Shastry88b}.

More precisely, the vertex weights of the classical model for the case
$U=0$ are given by the product of the vertex weights of two six-vertex 
models (with components $\sigma$ and $\tau$): $\ell_{1,2} (u) =
\ell^{\sigma}_{1,2}(u) \otimes  \ell^{\tau}_{1,2}(u)$, where
\begin{equation}
     \ell^{\sigma}_{1,2}(u) = \tst{\2} (\cos(u) + \sin(u)) + 
        \tst{\2} (\cos(u) - \sin(u)) \sigma^z_1 \sigma^z_2 +
        (\sigma_1^+ \sigma_2^- + \sigma_1^- \sigma_2^+) \qd.
\end{equation}
Taking account of the $U\not=0$ interactions, the following local
vertex weight operator (denoted by $S$) was found \cite{Shastry88b} 
\begin{eqnarray}
S_{1,2}(v,u) 
&=& \cos(u+v)\, \ch(h(v,U)-h(u,U))\, \ell_{1,2} (v-u)
+ \cos(v-u)\, \sh(h(v,U)-h(u,U))\, \ell_{1,2} (u+v)\,
    \sigma_2^z \tau_2^z ,
\end{eqnarray}
where $\sh(2 h(u,U)) := \frac{U}{4} \sin(2u)$. 
The Yang-Baxter equation for triple $S$ matrices
was conjectured \cite{Shastry88b}, but only recently proved in
\cite{ShWa95}.

The so-called $L$-operator is related to $S$ by
\begin{equation} \label{shastryl}
     L_{i,g}(u)=S_{i,g}(u,0) \qd,
\end{equation}
where $i$ and $g$ are indices referring to the $i$th lattice site and
the auxiliary space, respectively. For the $L$-operator the proof of
the Yang-Baxter equation with $S$ as intertwiner
was already given in \cite{Shastry88b}.
The commutativity of the row-to-row transfer matrix
\begin{equation} \label{shastryt}
     {\cal T}(u) := \tr\prod^{\leftarrow}_i L_{i, g} (u)
\end{equation}
is a direct consequence. 

Next we define $R_{1,2}(u,v) = \Pi_{1,2}  S_{1,2}(v,u)|_{U \rightarrow
- U}$, where $\Pi_{1,2}$ is the permutation matrix. The matrix elements
$R_{\alpha,\beta}^{\mu,\nu}(u,v)$ will be considered as the local
Boltzmann weights associated with vertex configurations $\alpha$,
$\beta$, $\mu$, $\nu$ on the lower, upper, left, and right bond, where
the spectral parameters $u$ and $v$ ``live'' on the vertical and
horizontal bonds, respectively.
For later use we introduce $\overline{R}(u,v)$ and $\widetilde{R}(u,v)$
($u$ and $v$ associated with the vertical and horizontal bond)
by clockwise and anticlockwise 90$^\circ$ rotations of $R$, 
or in matrix notation
\begin{equation}
\overline{R}_{\alpha,\beta}^{\mu,\nu}(u,v)=
R_{\mu,\nu}^{\beta,\alpha}(v,u) \qd, \qd
\widetilde{R}_{\alpha,\beta}^{\mu,\nu}(u,v)= 
R_{\nu,\mu}^{\alpha,\beta}(v,u) \qd.
\end{equation}
Similar to (\ref{shastryl}) and (\ref{shastryt}) we can associate a
row-to-row transfer matrix with $\overline{R}$. We note the Hamiltonian
limits ${\cal T}(u)=\exp(iP+uH+O(u^2))$ and $\overline{\cal T}(u) =
\exp(-iP+uH+O(u^2))$. Consequentially, the partition function of the
Hubbard chain at finite temperature $T = 1/ \beta$ is given by
\begin{equation}
Z =  \lim_{L\rightarrow \infty} \tr \, e^{-\beta H} = 
 \lim_{L\rightarrow \infty}\lim_{N\rightarrow \infty}
  \hbox{ tr } [{\cal T}(u) \overline{\cal T}(u)]^{N/2} |_{u=\beta/N}
  \qd.
\end{equation}
We regard the resulting system as a fictitious two-dimensional model 
on a $L\times N$ square lattice. Here $N$ is the extension in the
fictitious (imaginary time) direction, sometimes referred to as the
Trotter number. The lattice consists of alternating rows each being a
product of only $R$ weights or of only $\overline{R}$ weights,
respectively. 
Now by looking at the system in a 90$^\circ$ rotated frame which turns
$\overline{R}$ and $R$ weights into $R$ and $\widetilde{R}$ weights, 
it is natural to define
the `quantum transfer matrix' (QTM) by 
\begin{equation}
\{  {\cal T}_{{\rm QTM}} (u,v)   \}^{\{\beta\}}_{\{\alpha\}}
:= \sum_{\{ \mu \}} \prod_{i=1}^{N/2} 
   R^{\mu_{2i-1},\mu_{2i}}_{\alpha_{2i-1},\beta_{2i-1}} (-u,v) \,
   \widetilde{R}^{\mu_{2i},\mu_{2i+1}}_{\alpha_{2i},\beta_{2i}}(u,v)
   \qd,
\label{QTM}
\end{equation}
which is identical to the column-to-column transfer matrix of the
square lattice for $v=0$. The interchangeability of the two limits
($ L, N \rightarrow \infty$) \cite{SuIn87,SAW90} leads to the
following expression for the partition function,
\begin{equation}
Z= \lim_{N\rightarrow \infty} \lim_{L\rightarrow \infty}
      \; \tr \left[{\cal T}_{{\rm QTM}}\left(u=\frac{\beta}{N},0\right)
\right]^{L} \qd.
\end{equation}
There is a gap between the largest and the second largest eigenvalues
of ${\cal T}_{{\rm QTM}}(u,0)$ for finite $\beta$. Therefore the free
energy per site is expressed just by the largest eigenvalue
$\Lambda_{{\rm max}}(u,0)$ of ${\cal T}_{{\rm QTM}}(u,0)$,
\begin{equation} \label{freeen}
f = - \, T \lim_{N\rightarrow \infty}\ln\Lambda_{{\rm max}}
                   \left(u=\frac{\beta}{N},0\right) \qd.
\end{equation}
It is relatively simple to see that (\ref{QTM}) is integrable, i.e.\ a
family of commuting operators for variable $v$ and fixed $u$.
A non-vanishing chemical potential $\mu$ and magnetic field $B$
can be incorporated\footnote{We note that our conventions for the
magnetic field in this article are different from \cite{JKS98}. In
\cite{JKS98} the magnetic field was denoted by $H = 2B$.}
as they merely lead to trivial modifications due to twisted boundary
conditions for the QTM (cf.\ \cite{Klu93}).

\subsection{Diagonalization of the Quantum Transfer Matrix}
Here we summarize the main results of \cite{JKS98} where the 
diagonalization of (\ref{QTM}) on the basis of an algebraic Bethe ansatz
was performed. Note that the general expression for the eigenvalue
of the quantum transfer matrix is quite complicated \cite{JKS98}, but
simplifies considerably at $v=0$ and $u \rightarrow 0$,
\begin{equation}
\Lambda(v=0) = e^{\beta U/4}
          (1+e^{\beta (\mu+B)} )(1+e^{\beta (\mu-B)} )
          u^N \prod_{j=1}^{m} z_j \qd.
\label{eig_prac}
\end{equation}
The numbers $z_j$ are charge rapidities satisfying Bethe ansatz
equations which are most transparently written in terms of the related
quantities
\begin{equation}
s_j =\frac{1}{2i} \left(z_j-\frac{1}{z_j}\right) \qd.
\end{equation}
For these rapidities $s_j$ and additional rapidities $w_{\alpha}$ the
coupled eigenvalue equations read
\begin{equation}
     e^{-\beta(\mu - B)} \phi(s_j) =
         -\frac{q_2(s_j - \i U/4)}{q_2(s_j + \i U/4)} \qd, \qd
         e^{- 2\beta\mu} \,
         \frac{q_2(w_\alpha + \i U/2)}{q_2(w_\alpha - \i U/2)} =
         - \frac{q_1(w_\alpha + \i U/4)}{q_1(w_\alpha - \i U/4)} \qd,
         \label{QTMBA}
\end{equation}
where we have employed the abbreviations for products over rapidities
\beq
     q_1(s)=\prod_j(s-s_j) \qd, \qd q_2(s)=\prod_\alpha(s-w_\alpha) \qd.
\eeq
The function $\phi$ is defined by
%New
\begin{equation}
  \phi(s) =
  \left( \frac{(1-z_-/z(s))(1-z_+/z(s))}
              {(1+z_-/z(s))(1+z_+/z(s))} \right)^{N/2} \qd, \qd
   z(s) = \i s \left( 1+\sqrt{(1-1/s^2)} \right) \qd,
\end{equation}
where $z(s)$ possesses two branches. The standard (``first'') branch 
is chosen by the requirement $z(s)\simeq 2\i s$ for large values of
$s$, and the branch cut line $[-1,1]$. 
%New
The numbers $z_\pm$ are defined by $z_\pm=\exp(\alpha)(\tan u)^{\pm 1}$,
where $\sinh(\alpha)= -{U\over 4}\sin 2u$.

As usual, equations like (\ref{QTMBA}) which are identical in structure 
to (\ref{bak}) and (\ref{bas}) have many solutions. In our approach to
the thermodynamics just the largest eigenvalue of the QTM matters. The
corresponding distribution of rapidities is relatively simple. For
$\mu=B=0$ the rapidities are all situated on the real axis. Naturally,
for finite $\mu$ and $B$ we have modifications which, however, do not
affect qualitative aspects of the distribution.

%New
In a similar way a path integral formulation of the Hubbard chain
and a diagonalization of the corresponding QTM was employed in 
\cite{Tsune91,KlBa96}. In \cite{Tsune91} the eigenvalue equations 
were studied numerically for finite Trotter number and the case of 
half-filling. In \cite{KlBa96} an analytic attempt was undertaken to
study the limit of infinite Trotter number and to derive a set of
non-linear integral equations. Unfortunately, these equations were
rather ill posed with respect to numerical evaluations. In the next
section we present a formulation of non-linear integral equations on
the basis of recent work \cite{JKS98}.

\subsection{Non-linear integral equations}
\label{sec:non-linear-integral-equations}
In this section we are concerned with the derivation of well posed
integral equations equivalent to (\ref{QTMBA}) for the largest
eigenvalue of the QTM and thus for the free energy per site (see
(\ref{freeen})). We introduce a set of auxiliary functions ($\b$, $\c$,
and $\bc$) described in more detail in (\ref{auxFunct}) below. These
auxiliary functions are complex functions, however mostly evaluated
close to the real axis.

In terms of the auxiliary functions the Gibbs free energy per site is
expressed in several ways
\bea
   f & = & - \mu - \frac{U}{4} - \frac{T}{2 \p \i}
      \int_{\cal L}[\ln z(s)]' \ln\left(1+\c+\bc\right)ds \nonumber \\
      & & \qqqd + \, \frac{T}{4 \p \i} \int_{\cal L}
          \left[\ln\frac{z(s-\i U/2)}{z(s)}\right]'\ln(1+\b(s))ds
        + \frac{T}{4 \p \i} \int_{\cal L}
         \left[\ln\frac{z(s+\i U/2)}{z(s)}\right]'\ln(1+1/\b(s))ds
	 \qd, \label{int-eig-p} \\
   & = & \frac{U}{4} - \frac{T}{2 \p \i} \int_{\cal L}[\ln z(s)]'
         \ln\frac{1+\c+\bc}{\bc}ds
         - \frac{T}{2 \p \i}
	 \int_{\cal L} \left[\ln{z(s-\i U/2)}\right]'\ln(1+\c(s))ds
	 \qd.
\label{int-eig}
\eea
For yet another expression see \cite{JKS98}.

Equations (\ref{int-eig-p}) and (\ref{int-eig}) have to be compared
with equations (\ref{gibbs}) and (\ref{gibbs2}) which give the free
energy in the string based approach of Takahashi. In contrast to the
string based approach, the auxiliary functions $\b$, $\c$ and $\bc$
entering (\ref{int-eig-p}) and (\ref{int-eig}) satisfy a closed set of
finitely many (non-linear) integral equations,
\bea
\ln\b&=&-2 \beta B+K_2\wcirc\ln(1+\b)-{\overline{K_1}}\circ
\ln(1+1/\bc) \qd, \nonumber \\
\ln\c&=&-\beta{U}/{2}+\beta(\mu+B)+\varphi
-{\overline{K_1}}\wcirc\ln(1+1/\b)-{\overline{K_1}}\circ\ln(1+\bc) \qd,
   \nonumber \\
\ln\bc&=&-\beta{U}/{2}-\beta(\mu+B)-\varphi
+K_1\wcirc\ln(1+\b)+K_1\circ\ln(1+\c) \qd.
\label{all}
\eea
Here we have used the definition
\begin{equation}
     \varphi(x)=-2\beta \i x\sqrt{1-1/x^2}
\end{equation}
and have introduced the notation 
\begin{equation}
     (g\circ f)(s)=\int_{\cal L}g(s-t)f(t)dt
\end{equation}
for the convolution of two functions $g$ and $f$
with contour ${\cal L}$ surrounding the real axis 
at infinitesimal distance above and below in anticlockwise
manner. The definition of $\wcirc$ is similar, with integration
contour surrounding the real axis at imaginary parts $\pm U/4$.

The kernel functions are rational functions,
\beq
     K_1(s)=\frac{U/4 \pi}{s(s + \i U/2)} \qd, \qd
     {\overline{K_1}}(s)=\frac{U/4 \pi}{s(s - \i U/2)} \qd, \qd
     K_2(s)=\frac{U/2 \pi}{s^2 + U^2/4} \qd.
\label{kernels}
\eeq

%We want to note that (\ref{all}) is complete though it does not 
%provide any explicit equation for $\bb$. In fact, $\bb$ is nothing but
%the reciprocal of $\b$
%\be
%\bb(s)={1\over\b(s)}
%\ee
%Note that $\c$ and $\bc$ are not related in any similar manner. This 
%point will become clear in (\ref{auxFunct}).

Next, we want to point out that the function $\b$ will be evaluated 
on the lines $\Im s=\pm U/4$. The functions $\c$ and $\bc$ need only
be evaluated on the real axis infinitesimally above and below the
interval $[-1,1]$. Also the convolutions involving the ``$\c$
functions'' in (\ref{all}) can be restricted to a contour surrounding
$[-1,1]$ as these functions are analytic outside.

Lastly, we want to comment on the derivation of (\ref{int-eig}),
(\ref{all}). 
The explicit expressions of the functions $\b$, $\c$, $\bc$
are
\bea
    \b&=&
    \frac{\bl{1}+\bl{2}+\bl{3}+\bl{4}}{l_1+l_2+l_3+l_4} \qd, \nonumber
       \\
    \c&=&\frac{l_1+l_2}{l_3+l_4}\cdot
    \frac{\bl{1}+\bl{2}+\bl{3}+\bl{4}}
         {l_1+l_2+l_3+l_4+\bl{1}+\bl{2}+\bl{3}+\bl{4}} \qd,
         \label{auxFunct} \\
    \bc&=&\frac{\bl{3}+\bl{4}}{\bl{1}+\bl{2}}\cdot
    \frac{l_{1}+l_{2}+l_{3}+l_{4}}
         {l_1+l_2+l_3+l_4+\bl{1}+\bl{2}+\bl{3}+\bl{4}} \qd, \nonumber
\eea
where the functions $l_j$ and $\bl{j}$ are 
\beq
l_j(s)=\lambda_j(s- \i U/4)\cdot\e^{2 \beta B}{\phi^+(s)\phi^-(s)} \qd,
\qd \bl{j}(s)=\lambda_j(s+ \i U/4) \qd,
\eeq
and the $\lambda_j$ are defined in terms of the $q_1$ and $q_2$
functions, i.e. in terms of the Bethe ansatz rapidities
\bea
    \lambda_1(s) &=&{\rm e}^{\beta(\mu+B)}
      \frac{\phi(s- \i U/4)}{q_1(s- \i U/4)} \qd, \qd
    \lambda_2(s) ={\rm e}^{2\beta\mu}
      \frac{q_2(s- \i U/2)}{q_2(s)q_1(s- \i U/4)} \qd, \nonumber \\
    \lambda_3(s) &=&\frac{q_2(s+ \i U/2)}{q_2(s)q_1(s+ \i U/4)} \qd,
                    \qquad\quad
    \lambda_4(s) ={\rm e}^{\beta(\mu-B)}
      \frac{1}{\phi(s+ \i U/4)\,q_1(s+ \i U/4)} \qd.
\label{eig-aux}
\eea
The functions defined in (\ref{auxFunct}) are proven to 
satisfy a set of closed functional
equations which can be transformed into integral form (\ref{all}),
cf.\ \cite{JKS98}.
Also (\ref{int-eig}) follows from (\ref{auxFunct}) 
after a lengthy yet direct
calculation. The merit of (\ref{int-eig}),
(\ref{all}) is that this formulation does no longer make any reference
to the Bethe ansatz equations! Hence the calculation of an infinite
set of discrete rapidities is replaced by the computation of analytic
functions for which much more powerful tools are available.

\subsection{Analytical solutions of the integral equations}
\label{sec:analytical-solutions}

Before numerically studying 
various thermodynamic properties for general temperatures
and particle concentrations we want to give some analytic treatments
of limiting cases of the Hubbard model.

\subsubsection{Strong-coupling limit}
\label{sec:strong-coupling-limit}

In the strong-coupling limit \mbox{$U\to\infty$} at half-filling
($\mu=0$) the Hubbard model is expected to reduce to the
Heisenberg chain. Indeed, in the strong-coupling limit 
we find that $\c$, $\bc\to 0$. Hence, the only non-trivial 
function determining the eigenvalue of the QTM is $\b$,
see the first expression of (\ref{int-eig}). The integral equation
for $\b$ as obtained from (\ref{all}) is identical to
that obtained directly for the thermodynamics of the Heisenberg model
\cite{JKS98}.

\subsubsection{Free-Fermion limit}
\label{sec:free-fermion-limit}

Next, let us consider the limit $U\to{}0$ leading to a free fermion
model, however representing a non-trivial consistency check of the
equations. Indeed, the auxiliary functions can be calculated
explicitly. Finally, the free energy per site reads
\beq
     f = - \frac{T}{2\pi} \int_{-\pi}^{\pi}
         \ln \Big\{ \big[1 + \exp((\mu + B + 2\cos{k})/T) \big]
                    \big[1 + \exp((\mu - B + 2\cos{k})/T) \big] \Big\}
		    \, {\rm d}k \qd,
\eeq
which, as desired, is the result for free tight binding electrons.

\subsubsection{Low-temperature asymptotics}
\label{sec:low-temperature-limit}

The  low-temperature regime is the most interesting limit as the system
shows Tomonaga-Luttinger liquid behavior. We want to describe the
relation of the non-linear integral equations to the known dressed
energy formalism \cite{FrKo90,FrKo91} of the Hubbard model. This
represents a further and in fact the most interesting consistency check.

%New
For $T=1/\beta\to 0$ we  can simplify the non-linear integral equations
as they turn into {\it linear} integral equations, however in the
generic case ($B$ and $\mu\not=0$) with {\it finite} integration
contours.  This can be seen as follows. To be specific let us adopt
fields $B>0$, $\mu\le 0$ (particle density $n\le 1$). In the
low-temperature limit several auxiliary functions tend to zero, namely
${\b(s)}\to{}0$ on the line $\Im\, s=-U/4$, $\c(s)\to 0$ above and
below the real axis, and $1/\bc(s)\to{}0$ just below the real axis, 
i.e.\ on the line $\Im\, s=-\epsilon$. The remaining non-trivial
functions are ${\b}$ on the line $\Im\, s=+U/4$ and $1/\bc$ just above
the real axis for which we introduce the notation
\beq
     b(\lambda)=\b(\lambda+ \i U/4) \qd, \qd
     c(\lambda)=1/\bc(\lambda+ \i \epsilon) \qd.
\eeq
We note that
\beq
  |b|,\, |c|\gg{}1
                 \quad{\rm for}\quad |x|<\lambda_0,\, {\sigma}_0    
  \quad{\rm and}\quad
  |b|,\, |c|\ll{}1\quad{\rm for}\quad |x|>\lambda_0,\, {\sigma}_0 \qd,
\eeq
for certain crossover values $\lambda_0,\, {\sigma}_0$. The slopes for
the crossover are steep, so that the following approximations to the
integral equations (\ref{all}) are valid,
\bea \label{linearbc-p}
 \ln b & = & \phi_b - \int_{-\lambda_0}^{\lambda_0} 
             k_2 (\lambda - \lambda')\, \ln b (\lambda')\,
             {\rm d}\lambda'
        + \int_{-k_0}^{k_0} k_1 (\lambda - \sin k')\, \cos k' \,
			    \ln c(k')\, {\rm d}k' \qd, \\
 \ln c & = & \phi_c + \int_{-\lambda_0}^{\lambda_0} 
             k_1 (\sin k - \lambda')\, \ln b (\lambda')\,
             {\rm d}\lambda' \qd.
\label{linearbc}
\eea
where $k_1(\lambda)=K_1(\lambda- \i U/4)=\overline{K_1}(\lambda+
\i U/4)$ and $k_2(\lambda)=K_2(\lambda)$. In order to facilitate
comparison with the dressed energy formalism we also introduced
a new integration variable $k$ leading to a change of the boundaries
of integration, $\s_0 \rightarrow k_0 = \arcsin \s_0$. The driving
terms in (\ref{linearbc-p}) and (\ref{linearbc}) are related to the
bare energies
\beq
     \varepsilon_s^0 = B \qd, \qd
     \varepsilon_c^0 = - 2 \cos k - \m - U/2 - B
\eeq
by
\begin{equation}
 \phi_b =- \beta\,\varepsilon_s^0+ \CO(1/\beta) \quad{\rm and}\quad
 \phi_c =- \beta\,\varepsilon_c^0+ \CO(1/\beta) \qd.
\end{equation}
Therefore, we find the following connections between auxiliary
functions and the dressed energy functions,
\begin{equation}
 \ln b = - \beta\,\varepsilon_s + \CO(1/\beta) \qd, \qd
 \ln c = - \beta\,\varepsilon_c + \CO(1/\beta) \qd.
\label{BC-relations-in-low-T-limit}
\end{equation}
For a comparison with \cite{FrKo90,FrKo91} note the different
normalization of the chemical potential.

For a comparison with the results in section VI we set the magnetic
field equal to zero. Then $\varepsilon_s^0 = B = 0$, and it can be seen
that $\lambda_0 = \infty$. Setting $Q = k_0$ we can identify
(\ref{linearbc-p}), (\ref{linearbc}) in the zero temperature limit
with (\ref{dresseden0}). We find
\bea
     && - \lim_{T \rightarrow 0} T \ln b (\la) =
        \lim_{T \rightarrow 0} \eps_1 (\la) =
        \lim_{T \rightarrow 0} T \ln \h_1 (\la) \qd, \\
     && - \lim_{T \rightarrow 0} T \ln c (\la) =
        \lim_{T \rightarrow 0} \k (k) =
        \lim_{T \rightarrow 0} T \ln \z (k) \qd.
\eea

The free energy also admits the above approximation scheme, yielding
up to $\CO(T^2)$-terms the low-temperature expansion
\begin{equation}
     f=\varepsilon_0 -\frac{\pi}{6}\,
       \left(\frac{1}{v_c}+\frac{1}{v_s}\right)T^2 \qd.
\label{lowTsep}
\end{equation}
Here the definitions of the sound velocities and the ground state 
energy are standard. The additive occurrence of $1/v_c$ and $1/v_s$
on the right hand side of (\ref{lowTsep}) is a manifestation of
spin-charge separation in the one-dimensional Hubbard model, due to
which each elementary excitation contributes independently to
(\ref{lowTsep}). The velocities $v_c$ and $v_s$ typically take
different values.

\subsubsection{High-temperature limit}

Finally, we consider the high-temperature limit $T\to\infty$ with $B$,
$U$ as well as $\beta\mu$ fixed ensuring a fixed particle density $n$.
The integral equations turn into algebraic equations which are easily
solved, resulting in the high-temperature limit
\begin{equation}
     S = 2 \ln \left( \frac{2}{2-n} \right) -
	 n \ln \left( \frac{n}{2-n} \right)
\end{equation}
for the entropy, as expected by counting the degrees of freedom per
lattice site. Especially at half-filling, $n=1$, this equals to
$S=\ln(4)$.

\subsection{Numerical Results}

Here we show numerical results for 
specific heat $C$, magnetic susceptibility
$\chi_m$, and charge susceptibility $\chi_c$ 
for half-filling and small doping, 
see Figs.\ \ref{fig:fig1},\ \ref{fig:fig2},\ \ref{fig:fig3}.
In addition, we aim at a comparison of results 
\cite{KUO89,UKO90} obtained within the thermodynamic Bethe ansatz
\cite{Takahashi72} based on the string hypothesis, and results
\cite{JKS98} obtained within the quantum transfer matrix approach.

In essence, we observe convincing agreement of the data obtained
within the two absolutely different approaches. We conclude that 
there is no indication 
%New of -> for
for any failure of Takahashi's
formulation of thermodynamics based on the string hypothesis.
Of course rather small differences exist in the sets of data. A more
thorough comparison shows deviations of the data presented in
Fig.\ \ref{fig:fig1} and\ \ref{fig:fig2}.
This is nothing but expected due to the truncation procedure adopted
in \cite{KUO89,UKO90}. Instead of dealing with an infinite
%New
set of integral equations for 
density functions (\ref{densities},\ref{zeta}-\ref{kl})
for string excitations of spin rapidities
%New
(\ref{ideal1})
and charge rapidities
%New
(\ref{idealk}), a finite subset was taken into account. In the case
of complex charge rapidities, strings describe excitations with gap.
These degrees of freedom are less sensitive to errors introduced by the
truncation than strings involving only 
spin rapidities which describe gapless excitations. In fact, the 
agreement of the data for the charge susceptibility $\chi_c$
is best, small deviations are observed in $C$ and
$\chi_m$.

The advantage of the QTM approach over the traditional TBA is threefold.
First, in \cite{JKS98} the Hubbard chain with very strong doping 
was analyzed (apparently not possible in the traditional TBA) and quite
unexpected structures in the susceptibility data were found. In the 
magnetic susceptibility only traces of spinon excitations were visible.
The charge susceptibility, however, exposed holon signatures and
additional maxima due to spinon excitations indicating deviations from
the concept of spin-charge separation. Second, the accuracy of
numerical data obtained within the QTM approach is much higher as it
is more efficient to deal with a set of integral equations which is
strictly finite from beginning. Finally, even correlation lengths can
be calculated within the QTM approach although not presented in detail
yet.

\newlength{\Breite}
\setlength{\Breite}{5cm}

\begin{figure}
\begin{picture}(0,200)
\put(13.0,16.0){\includegraphics[width=1.461\Breite,
height=1.265\Breite,angle=0]{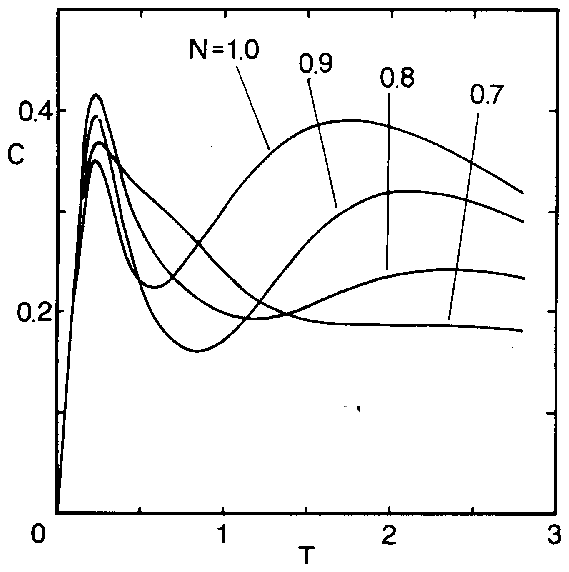}}
\put(220,0){
\includegraphics[width=1.5\Breite,
height=1.5\Breite,angle=0]{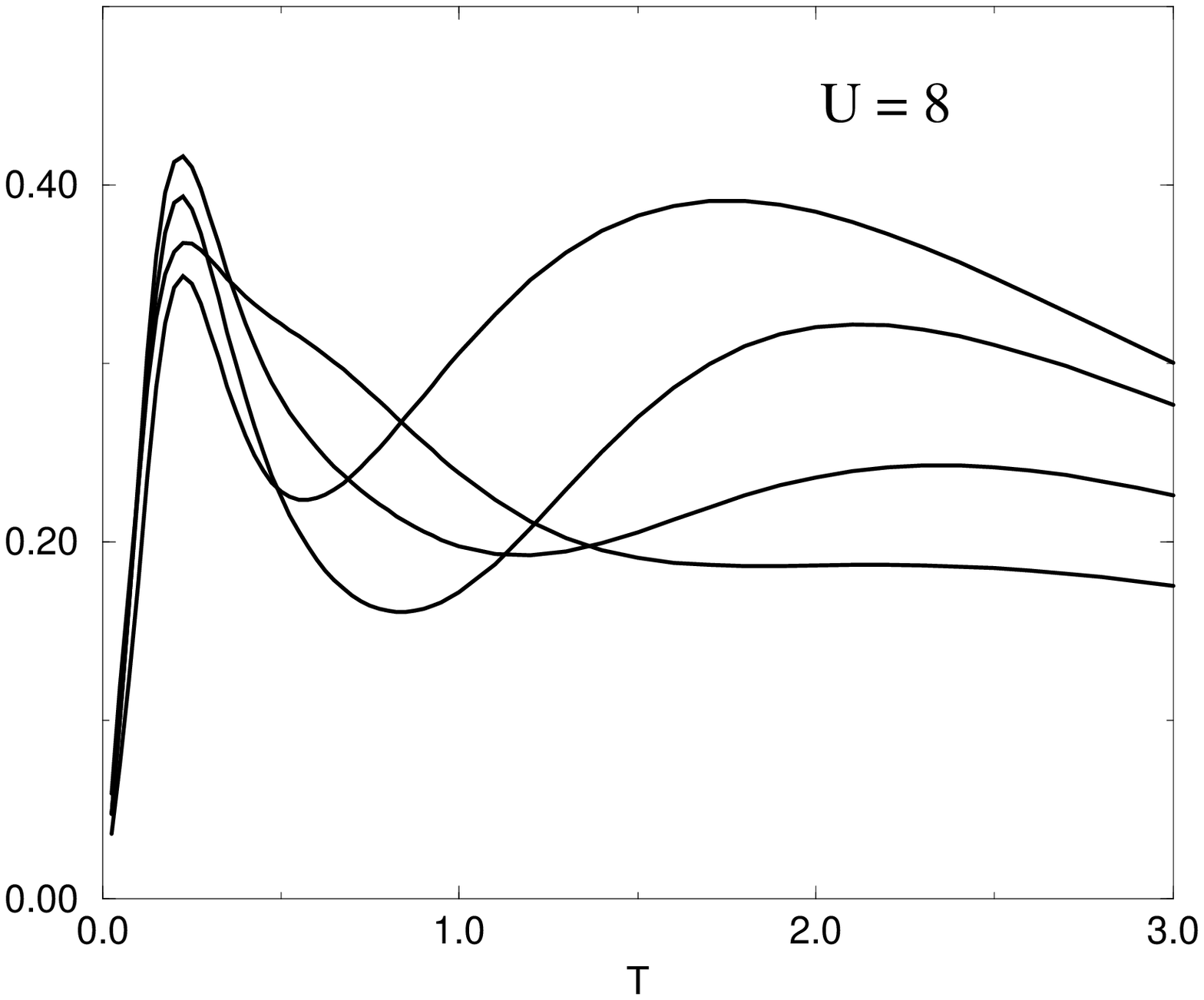}}
\end{picture}
\caption{Specific heat $C$ versus temperature for $U = 8$ and particle
densities $n=1$, $n=0.9$, $n=0.8$ and $n=0.7$. The left graph shows
results obtained within the string based approach,
%{\cite{KUO89,UKO90}}
the right graph shows results obtained by the quantum transfer
matrix method.
%{\cite{JKS98}}
}
  \label{fig:fig1}
\end{figure}

\begin{figure}
\begin{picture}(0,200)
\put(1,11){
\includegraphics[width=1.529\Breite,
height=1.32\Breite,angle=0.5]{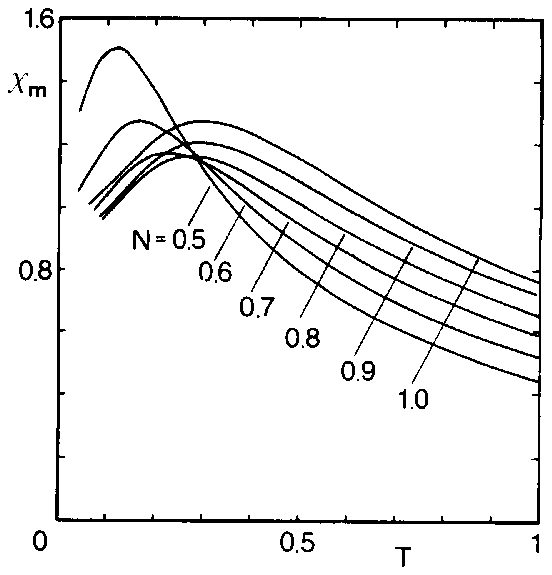}}
\put(220,0){
\includegraphics[width=1.5\Breite,
height=1.5\Breite,angle=0]{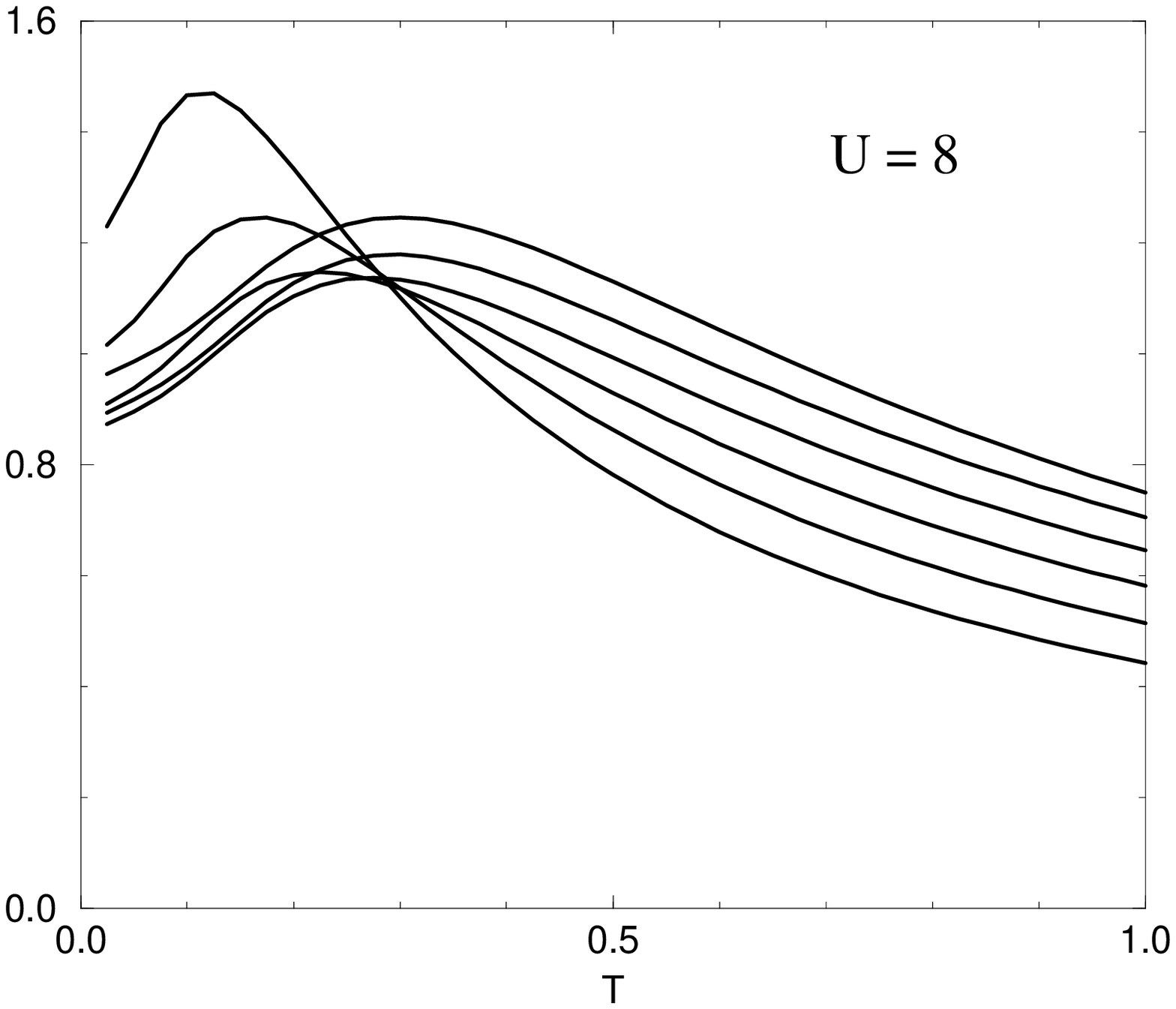}}
\end{picture}
\caption{Magnetic susceptibility $\chi_m$ versus temperature for $U = 8$
and particle densities $n=1$, $n=0.9$, \dots, $n=0.5$. The left graph
shows results obtained within the string based approach,
%{\cite{KUO89,UKO90}}
the right graph shows results obtained by the quantum transfer
matrix method.
%{\cite{JKS98}}
}
  \label{fig:fig2}
\end{figure}

\begin{figure}
\begin{picture}(0,200)
\put(-2.5,7){
\includegraphics[width=1.532\Breite,
height=1.336\Breite,angle=0]{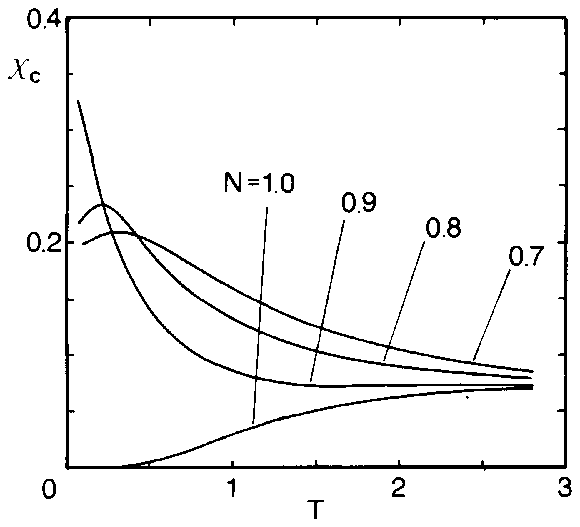}}
\put(220,0){
\includegraphics[width=1.5\Breite,
height=1.5\Breite,angle=0]{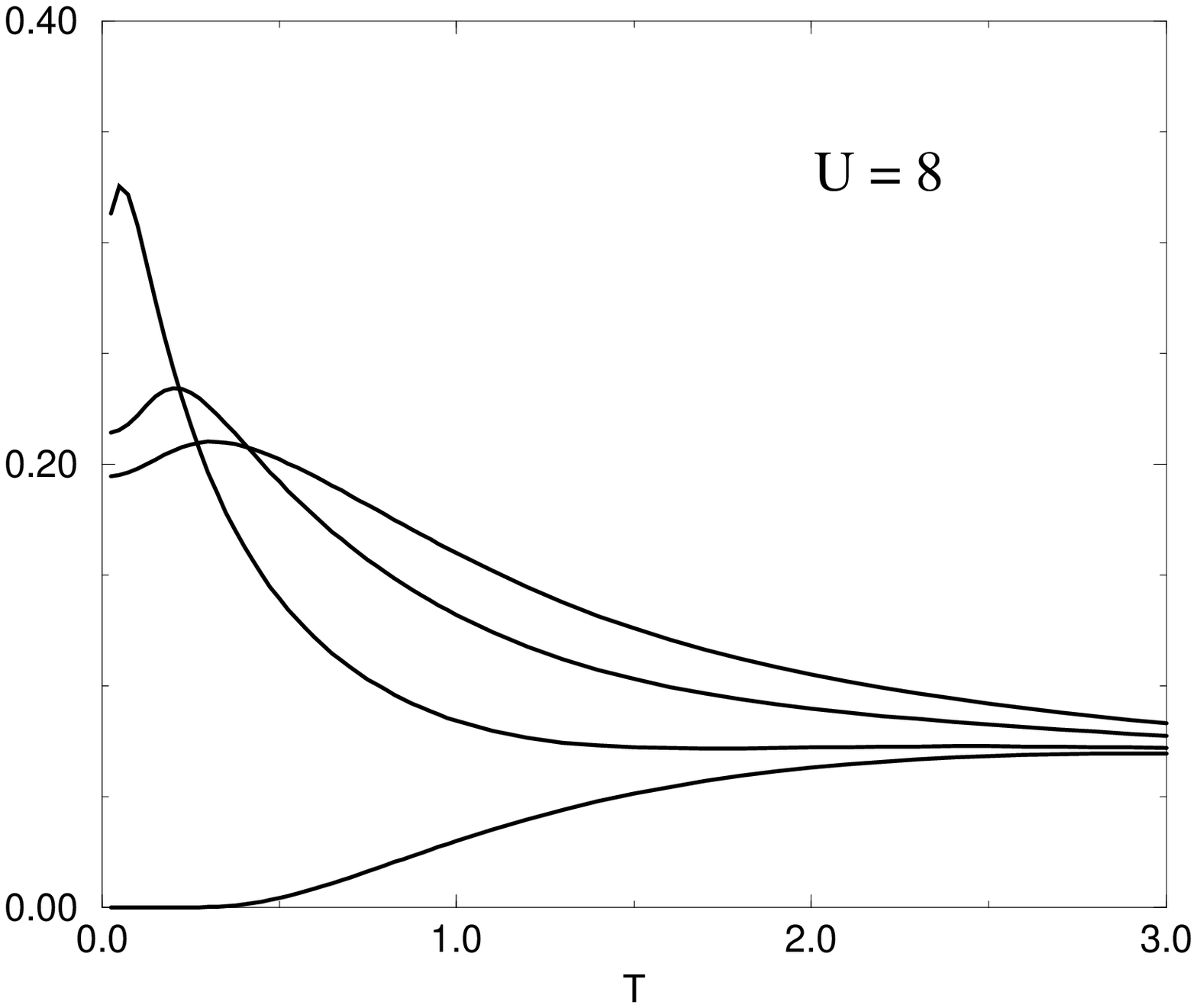}}
\end{picture}
\caption{Charge susceptibility $\chi_c$ versus temperature for
$U = 8$ and particle densities $n=1$, $n=0.9$, $n=0.8$ and $n=0.7$.
The left graph shows results obtained within the string based approach,
%{\cite{KUO89,UKO90}}
the right graph shows results obtained by the quantum transfer matrix
method.
%{\cite{JKS98}}
}
  \label{fig:fig3}
\end{figure}
\subsection{Equivalence of string based thermodynamics with QTM
approach}

Here we want to address the fundamental and perhaps puzzling question
how two exact, however completely different formulations of the 
thermodynamics of the Hubbard model as presented in sections~VI and VII 
can exist.
It is very important to understand this problem as the 
means of derivation and the mathematical properties of both approaches
are seemingly lacking any similarities. 

In the remainder of this section we restrict ourselves to a sketch
of the general mathematical structures underlying
the relation of the traditional thermodynamical formulation based on
typically infinitely many density functions and the new formulation 
with strictly finitely many auxiliary functions. This relation
has been worked out for a number of models including spin chains
\cite{Klu92} and correlated fermion models like $t$-$J$ systems
\cite{JKS98b}, not yet however for the Hubbard chain. In the following
we want to introduce the general technique for spin-1/2 Heisenberg
models and related systems.

The starting point is an integrable Hamiltonian corresponding to an
exactly solvable classical model as discussed in section~VII.A. For
this model a path integral representation is formulated leading to an
integrable QTM {\boldmath $T$}$(v)$ with eigenvalue equation
\be
T(v)={f(v-\lambda)q(v+\lambda)+f(v)q(v-\lambda)\over q(v)} \qd,
\label{Tq}
\ee
where $f(v)$ is a known function and $q(v)$ is a polynomial or a
product over trigonometric functions with zeros exactly at the
Bethe ansatz rapidities. By a line of reasoning similar to that
presented in VII.C {\it one} non-linear integral equation can be
derived which yields full information on the free energy of the model.

The key to understanding the relation of the standard TBA to the above 
approach is the notion of fusion algebras and inversion identities.
By repeated use of the fusion procedure a set of transfer matrices 
{\boldmath $T$}$(v)=$ {\boldmath $T$}$^1(v)$,
{\boldmath $T$}$^2(v)$, {\boldmath $T$}$^3(v)$,...,
is generated satisfying
\be
\mbox{\boldmath $T$}^{q}(v)\mbox{\boldmath $T$}^{q}(v+\lambda)=
f(v-\lambda) f(v+\lambda)\mbox{\boldmath $I$}
+\mbox{\boldmath $T$}^{q+1}(v)\mbox{\boldmath
$T$}^{q-1}(v+\lambda) \qd, \label{func}
\ee
where $q$ takes integer values, $q = 1, 2, \dots$, and
{\boldmath $T$}$^0(v)$ is proportional to the identity operator.
Introducing operators
\be
\mbox{\boldmath $Y$}^{q}(v)={\mbox{\boldmath $T$}^{q-1}(v+\lambda)
\mbox{\boldmath $T$}^{q+1}(v)\over
f(v-\lambda) f(v+q\lambda)}\label{littlet}
\ee
we find the extremely useful inversion identity hierarchy
\be
\mbox{\boldmath $Y$}^{q}(v)\mbox{\boldmath $Y$}^{q}(v+\lambda)
=[\mbox{\boldmath $I$}+\mbox{\boldmath $Y$}^{q-1}(v+\lambda)]
   [\mbox{\boldmath $I$}+\mbox{\boldmath $Y$}^{q+1}(v)] \qd.
   \label{invIdHier}
\ee
This set of functional equations can be transformed into a set of
non-linear integral equations for the largest eigenvalue of
{\boldmath $T(v)$}. These integral equations appear to be identical to
the equations obtained along the traditional TBA approach!

Several remarks are in order. Relations (\ref{invIdHier}) were derived
in \cite{Zamolodchikov91a,Zamolodchikov91b} starting with the TBA
equations, however the connection to the fusion hierarchies was not
made. In contrast to the TBA integral equations the functional
equations (\ref{invIdHier}) (as of course equation (\ref{Tq})) admit
more than one solution. The physical meaning of these solutions
becomes clear only through the microscopic construction of transfer
matrices. The next-leading eigenvalues describe the correlation
lengths of static correlation functions at finite temperature. Such
applications are investigated in current research. Some results have
been reported in \cite{JKS98b,FKM98,SSSU99,Sakai99}.

In summary, the traditional TBA equations are not in contradiction with
the new thermodynamics based on the QTM. Rather on the contrary, the
QTM approach represents a unified approach to both sets of integral
equations. The central object is the study of the largest eigenvalue
of the QTM.  This can be achieved by two different methods, either by
a Bethe ansatz (\ref{Tq}) or by use of the fusion hierarchy
(\ref{func}). 

In physical terms, the appearance of the fusion hierarchy is related 
to the pole structure of the intertwiner, i.e. the $R$-matrix satisfying
the YBE. In turn the $R$-matrix is related to the $S$-matrix of
particles, the poles are related to bound states (strings) which in
turn are the central objects of the traditional analysis of the
thermodynamics of integrable systems. In this way we observe
mathematical and physical connections between the initially so
differently looking approaches. 

%\bibitem{Zam91} Al. B. Zamolodchikov, Phys. Lett. B 253, 39 (1991);
%Nucl. Phys. B 358, 497 (1991)

%\bibitem{JKS98b} G.~J\"uttner, A. Kl\"umper, J. Suzuki, 
%Nucl. Phys. B 512, 581, 1998

%\bibitem{FKM98} K. Fabricius, A. Kl\"umper, B. M. McCoy,
%cond-mat/9810278

%\bibitem{SSSU99} K. Sakai, M. Shiroishi, J. Suzuki, Y. Umeno,
%cond-mat/9902296

%\bibitem{S99} K. Sakai, cond-mat/9903112

%%%%%%%%%%%%%%%%%%%%%%%%%%%%%%%%%%%%%%%%%%%%%%%%%%%%%%%%%%%%%%%%%%%%%%%%
%     scdcon.tex                                                       %
%%%%%%%%%%%%%%%%%%%%%%%%%%%%%%%%%%%%%%%%%%%%%%%%%%%%%%%%%%%%%%%%%%%%%%%%
\section{Conclusions}
In this article we considered fundamental questions concerning
the exact solution of the one-dimensional Hubbard model. In appendices
A and B  we presented a detailed derivation of the wave function. We
studied solutions of the Lieb-Wu equations. We concentrated our
attention on $k$-$\La$ strings. Unlike $\La$ strings they were never
carefully studied in the literature before. We arrived to the
conclusion that, for fixed coupling and large lattice size, $k$-$\La$
strings deviate from their ideal positions in a controllable way. In
the infinite chain limit $k$-$\La$ strings approach their ideal
positions. We also reviewed the thermodynamics of the model. There are
two different approaches to thermodynamics: one is based on strings,
whereas the other one is not. Both approaches show convincing agreement
for the calculation of the bulk thermodynamic properties of the Hubbard
model. Passing to the zero temperature limit we obtained the dispersion
curves of all elementary excitations at zero magnetic field, below
half-filling.

We would like to conclude with pointing out some interesting open
problems:
\begin{enumerate}
\item
Perhaps the most fundamental open problem is the calculation of the
norm of the wave function (\ref{sectorq})-(\ref{wwfsa}). Formulae for
norms of wave functions are known for Bethe ansatz solvable models with
only one level of Bethe ansatz equations, like the Bose gas with delta
interaction \cite{Gaudin83,Korepin82} or the XXX and XXZ spin-$\2$
chains \cite{Korepin82} (see also \cite{KBIBo}). For these models the
norms are expressed as determinants of the Jacobians of the Bethe ansatz
equations. The calculation of the norm of the Bethe ansatz wave
functions may be considered as a first step towards the calculation of
determinant representations of correlation functions \cite{KBIBo}.
\item
Another interesting open problem which we already mentioned above is
the calculation of the finite temperature correlation length within
the quantum transfer matrix approach. This requires to calculate the
second largest eigenvalue of the quantum transfer matrix. One has to
overcome certain technical subtleties coming from the fact that the
Hubbard model is a model of Fermions. A formalism capable of
calculating the correlation length of integrable fermionic systems has
recently been developed \cite{USW98,SSSU99,Sakai99}.
\item
At zero temperature some of the correlation functions of the Hubbard
model show a power law decay. Conformal field theory\footnote{In the
context of condensed matter physics conformal field theory is
equivalent to Luttinger liquid theory \cite{Haldane81}.} naturally
describes these powers (the set of conformal dimensions) \cite{FrKo90}.
Still, there are open problems within the conformal approach. For
example, since the expansion of the lattice operators in terms of
conformal fields is not known explicitly, the resulting expressions
contain unknown amplitudes. Some of these amplitudes may actually
vanish. Recently \cite{EsFr99} the vanishing of the amplitudes
corresponding to density correlations for the half-filled model was
shown by use of the SO(4) symmetry. For a more thorough understanding
of correlation functions an identification of the operators which
are of interest for the Hubbard model with the standard operators of
conformal field theory would be important. Recent work on a scaling
limit of the Hubbard model \cite{Melzer95,WoFo97,WoFo99} may prove to
be useful in this context.
\item
The predictions of the conformal approach to correlation functions
are limited to large distances, corresponding to very low energies.
Theoretically as well as from the point of view of recent experiments
on quasi one-dimensional structures in solids (e.g.\ \cite{KimEtal96,%
KimEtal97}) it would be highly appreciable to have a method for the
calculation of correlation functions at all energy scales. This problem
was recently tackled in \cite{EsKo99} within the form factor approach
\cite{KaWe78,BKW79,Smirnov92}, which was originally designed for
integrable 1+1 dimensional quantum field theories. In \cite{EsKo99} it
was argued that the form factor approach might as well apply to the
half-filled Hubbard model, and a formula for the two-spinon form
factor of the spin-operator $S_j^+$ was presented. It would be
interesting to extend the result to form factors of electronic creation
and annihilation operators, as such kind of extension could be directly
applied to the interpretation of the angle-resolved photo emission
spectroscopy data of \cite{KimEtal96,KimEtal97}.
\item
Despite the progress in the understanding of the mathematical structure
of the Hubbard model, which was achieved over the past few years and
which we briefly discussed in the introduction, we still feel
uncomfortable with the present stage of our knowledge. Shastry's
$R$-matrix \cite{Shastry86b,Shastry88b,OWA87}, which is the key for our
present understanding of the algebraic structure behind the Hubbard
model, is unusual as compared to $R$-matrices of other integrable
models. It does not possess the so-called difference property, i.e. it
is not a function of the difference of the spectral parameters alone.
The $S$-matrix at half-filling \cite{EsKo94a}, on the other hand,
possesses the difference property and can therefore be associated with a
Y(su(2))$\oplus$Y(su(2)) Yangian \cite{UgKo94}. The precise relation
between $R$-matrix and $S$-matrix is only understood in the rather
simple situation of an empty band \cite{MuGo97a,MuGo98a}. Because of
the lack of the difference property we can neither find a boost
operator for the Hubbard model by the reasoning of \cite{Tetelman82}
nor can we associate a spectral curve with it.

Another problem is the dimension of the elementary $L$-operator related
to Shastry's $R$-matrix, which is $4 \times 4$ (rather than $3 \times
3$ as one could guess naively from the fact that the Bethe ansatz for
the Hubbard model has two levels). For this reason there are too many
candidates for creation and annihilation operators in the algebraic
Bethe ansatz \cite{Shastry88b,RaMa97,MaRa98}. Again this redundancy
has only been partially understood in the empty band case
\cite{MuGo98a}. The known algebraic Bethe ansatz \cite{RaMa97,MaRa98}
is of involved structure and hopefully will be simplified in the future.
\end{enumerate}
\noindent
{\bf Acknowledgment.} We would like to thank C. M. Hung for drawing
figures 1-3 for us. We are indebted to H. Frahm, N. Kawakami,
B. M. McCoy, A. Schadschneider, J. Suzuki, A. M. Tsvelik and
M. Takahashi for helpful and stimulating discussions. This work was
supported by the EPSRC (F.H.L.E), by the Deutsche Forschungsgemeinschaft
under grant numbers Go 825/2-1 (F.G.) and Kl 645/3 (A.K.) and by the
National Science Foundation under grant number PHY-9605226 (V.E.K).
A.K. acknowledges support by the Sonderforschungsbereich 341,
K\"oln-Aachen-J\"ulich.

\clearpage

%%%%%%%%%%%%%%%%%%%%%%%%%%%%%%%%%%%%%%%%%%%%%%%%%%%%%%%%%%%%%%%%%%%%%%%%
%     scdapp.tex                                                       %
%%%%%%%%%%%%%%%%%%%%%%%%%%%%%%%%%%%%%%%%%%%%%%%%%%%%%%%%%%%%%%%%%%%%%%%%
\def\be{\begin{equation}}
\def\ee{\end{equation}}
\def\bdm{\begin{displaymath}}
\def\edm{\end{displaymath}}
\def\nn{\nonumber\\}
\def\up{\uparrow}
\def\da{\downarrow}
\def\sgn{{\rm sign}}
\def\eps{\epsilon}
\def\r#1{(\ref{#1})}
\def\vac{|0\rangle}
\appendix
\section{Derivation of the wave function}

\subsection{General setting}
The expression (\ref{sectorq})-(\ref{wwfsa}) for the Bethe ansatz wave
function of the Hubbard model was first presented by Woynarovich
\cite{Woynarovich82a}. The purpose of this appendix is to give a
detailed derivation of Woynarovich's wave function. We will make use
of results obtained in references \cite{LiWu68}, \cite{Yang67} and of
the quantum inverse scattering method \cite{KBIBo,EKS92a}. Other
derivations of the Bethe Ansatz wave functions for the Hubbard model
can be found for example in \cite{Sutherland85} and in \cite{IzSk:88}.

The outline of this appendix is as follows.
In section 1 we define the wave function $\psi(x_1,\ldots, x_N)$,
for $N$ electrons and derive the first quantized Schr\"odinger
equation for the Hubbard model.

In sections $2$ and $3$ we present the explicit solution of the
Schr\"odinger equation with periodic boundary conditions for the cases
of $2$ and $3$ electrons, respectively. We treat these cases in
considerable detail for pedagogical reasons. Finally, in section $4$
we discuss the general case of $N$ electrons. 

An important ingredient in the construction of the wave function is
the exact solution of an inhomogeneous spin-1/2 Heisenberg model. The
essence of the nested Bethe Ansatz procedure employed in constructing
wave-functions for the Hubbard Hamiltonian is the reduction of this
problem to a simpler one, which involves only the spin degrees of
freedom. The dynamics of the spin degrees of freedom are described by
an inhomogeneous Heisenberg model, and its exact solution constitutes
the ``nesting'' of the Bethe Ansatz procedure. We summarize the
algebraic Bethe Ansatz solution of the inhomogeneous Heisenberg model
in Appendix \ref{section:XXX}.

Let us recall the explicit form of the Hamiltonian
\beq \label{hamilapp}
     H = - \sum_{j=1}^L \sum_{\s = \auf, \ab}
	   (c_{j, \s}^+ c_{j+1, \s} + c_{j+1, \s}^+ c_{j, \s})
	   + U \sum_{j=1}^L
	   (n_{j \auf} - \tst{\2})(n_{j \ab} - \tst{\2})\ .
\eeq
As the total number of electrons $N$ and the number of electrons with
spin down $M$ are good quantum numbers, we can use them to label
eigenstates of \r{hamilapp} 
\bea
|N,M   \rangle = \sum_{\{\sigma_j\}} \sum_{\{x_k\}}
\psi(x_1\ldots x_N; \sigma_1,\ldots,\sigma_N)
 c_{x_1, \sigma_1}^{\dagger} \cdots c_{x_N, \sigma_N}^{\dagger} |0
\rangle\ .
\label{eigenstate}
\eea
Here $\sum_{\{\sigma_j\}}$ denotes summation over all $N!/((N-M)!M!)$
possible spin-configurations with $M$ down-spins.
Due to the anticommutation relations between the Fermion operators,
we may assume without loss of generality that the amplitudes $\psi$
are totally antisymmetric
\beq
\psi(x_{P_1},\ldots,x_{P_N} ; \sigma_{P_1},\ldots,\sigma_{P_N})
= \sign(P)  \psi(x_{1},\ldots,x_{N} ; \sigma_{1},\ldots,\sigma_{N})
\ ,
\label{antisymmetry}
\eeq
where $P=(P_1,P_2,\ldots, P_N)$ is a permutation of the labels
$\{1,2,\ldots, N\}$, i.e. an element of the symmetric group $S_N$. 
The antisymmetry property \r{antisymmetry} implies that the summation
over spin configurations in \r{eigenstate} is redundant. Indeed one
finds that 
\beq
|N,M  \rangle = {\frac {N!} {(N-M)!M!}}  \sum_{\{x_j\}}
\psi(x_{1},\ldots,x_{N} ; \sigma_{1},\ldots,\sigma_{N})
 c_{x_1, \sigma_1}^{\dagger} \cdots
c_{x_N, \sigma_N}^{\dagger} |0  \rangle\ ,
\eeq
where $(\sigma_1,\ldots,\sigma_N)\in S_N$ is arbitrary. In order to
derive the Schr\"odinger equations it is therefore convenient to work
with the following simplified expression for general eigenstates of $H$
\beq
|N,M;{\vec \sigma}  \rangle = \sum_{\{ x_j\}}
\psi(x_{1},\ldots,x_{N} ; \sigma_{1},\ldots,\sigma_{N})
 c_{x_1, \sigma_1}^{\dagger} \cdots c_{x_N, \sigma_N}^{\dagger} |0
\rangle \ .
\eeq
It is now straightforward to show, that the eigenvalue problem
\beq
H |N,M;{\vec \sigma}\rangle = E |N,M;{\vec \sigma}\rangle\ ,
\eeq
implies the following Schr\"odinger equation for the wave function
$\psi$ \cite{LiWu68}
\bea
&& - \sum_{j=1}^{N} \sum_{s=\pm 1}
\psi(x_1, \ldots, x_j + s, \ldots, x_N; {\vec \sigma} ) +
U \sum_{j<k} \delta(x_j, x_k) 
\psi(x_{1},\ldots,x_{N} ; \sigma_{1},\ldots,\sigma_{N})
\nn
&&\qquad = (E+\frac{UN}{2}-\frac{UL}{4}) \psi(x_{1},\ldots,x_{N};
\sigma_{1},\ldots,\sigma_{N}) \ .
\label{SG}
\eea
Here $\delta(a,b)$ denotes the Kronecker delta.

\subsection{\boldmath Two electrons}

Let us now explicitly construct the wave function for the case of
two electrons, $N=2$. The Schr\"odinger equation is of the form

\bea
&&-\psi(x_1-1, x_2; \sigma_1,\sigma_2 ) 
-\psi(x_1+1, x_2; \sigma_1,\sigma_2 ) 
-\psi(x_1, x_2-1; \sigma_1,\sigma_2 )
-\psi(x_1, x_2+1; \sigma_1,\sigma_2 )\nn
&&\qquad +U \delta(x_1, x_2) \psi(x_1,x_2;\sigma_1,\sigma_2)
 = (E+U-\frac{UL}{4}) \psi(x_1,x_2;\sigma_1,\sigma_2)\quad .
\label{2SG}
\eea
As long as $x_1<x_2$ or $x_1>x_2$ \r{2SG} reduces to the Schr\"odinger
equation for free electrons on a lattice and its solutions are
therefore just superpositions of plane waves. When the electrons
occupy the same site, they interact. This can be thought of in terms
of a scattering process. Due to integrability this scattering is
purely elastic, which means that the momenta of the two electrons are
individually conserved. Thus, the most that can happen is that the
electrons exchange their momenta. These considerations lead to the
famous ``nested'' Bethe ansatz form for the wave functions, which we
will discuss next.

Let $Q$ be a permutation of the labels of coordinates i.e. 
$Q=(Q_1,Q_2)\in\{ (1,2), (2,1)\}$. In the ``sector'' $Q$ defined by
the condition $x_{Q_1}\leq x_{Q_2}$ the nested Bethe Ansatz for the
wave function is
\beq
\psi(x_1, x_2; \sigma_1,\sigma_2) =
\sum_{P \in S_2}\sign(PQ) A_{\sigma_{Q_1}\sigma_{Q_2}}(k_{P_1},k_{P_2})
\exp(i \sum_{j=1}^{2} k_{Pj}x_{Qj}) \quad .
\label{BAWF2}
\eeq

Substituting \r{BAWF2} into \r{2SG} for the case $x_1\neq x_2$ we obtain
\beq
E = - ( 2 \cos k_1 + 2 \cos k_2) -U+\frac{UL}{4}\qd.
\eeq

When $x_1=x_2$ we have to ``match'' the wave function defined in the
two sectors $Q=(12)$ and $Q=(21)$. This requires single valuedness
\bea
\psi(x,x;\sigma_1,\sigma_2)&=& 
\left[A_{\sigma_1\sigma_2}(k_1,k_2)-A_{\sigma_1\sigma_2}(k_2,k_1)\right]
\exp(i[k_1+k_2]x)\nn
&=& \left[A_{\sigma_2\sigma_1}(k_2,k_1)-A_{\sigma_2\sigma_1}(k_1,k_2)\right]
\exp(i[k_1+k_2]x)\ .
\label{cont}
\eea
In addition, the Schr\"odinger equation \r{2SG} for $x=x_1=x_2$ needs
to be fulfilled, which yields the condition
\bea
&&-e^{-ik_1}A_{\sigma_1\sigma_2}(k_1,k_2)
+e^{-ik_2}A_{\sigma_1\sigma_2}(k_2,k_1)
+e^{ik_2}A_{\sigma_2\sigma_1}(k_1,k_2)
-e^{ik_1}A_{\sigma_2\sigma_1}(k_2,k_1)\nn
&&-e^{ik_2}A_{\sigma_1\sigma_2}(k_1,k_2)
+e^{ik_1}A_{\sigma_1\sigma_2}(k_2,k_1)
+e^{-ik_1}A_{\sigma_2\sigma_1}(k_1,k_2)
-e^{-ik_2}A_{\sigma_2\sigma_1}(k_2,k_1)\nn
&&+U\left[A_{\sigma_1\sigma_2}(k_1,k_2) - 
A_{\sigma_1\sigma_2}(k_2,k_1)\right]\nn
&&= -2(\cos k_1+\cos k_2)\left[A_{\sigma_1\sigma_2}(k_1,k_2) - 
A_{\sigma_1\sigma_2}(k_2,k_1)\right].
\label{2SG2}
\eea
By means of \r{2SG2} and \r{cont} we can express two of the four
amplitudes $A_{\sigma_{Q_1}\sigma_{Q_2}}(k_{P_1},k_{P_2})$ in terms of
the other two. A short calculation gives
\bea
A_{\sigma_1\sigma_2}(k_2,k_1)&=&
\frac{-U/2i}{\sin k_1 - \sin k_2 -U/2i} A_{\sigma_1\sigma_2}(k_1,k_2)
+\frac{\sin k_1 - \sin k_2}{\sin k_1 - \sin k_2 -U/2i} 
A_{\sigma_2\sigma_1}(k_1,k_2)\ .
\label{ybe0}
\eea
Equation \r{ybe0} has a natural interpretation in terms of a
scattering process of two particles. In order to see this we rewrite
it as

\bea
A_{\sigma_2\sigma_1}(k_2,k_1)&=&
\sum_{\tau_1,\tau_2} S^{\sigma_1\tau_1}_{\sigma_2\tau_2}(k_1,k_2)\ 
A_{\tau_1\tau_2}(k_1,k_2)\ ,
\label{ybe1}
\eea
where $S(k_1,k_2)$ is the two-particle S-matrix with elements
\be
S^{\sigma_1\tau_1}_{\sigma_2\tau_2}(k_1,k_2)=
\frac{-U/2i}{\sin k_1 - \sin k_2 -U/2i}
\Pi^{\sigma_1\tau_1}_{\sigma_2\tau_2}
+\frac{\sin k_1 - \sin k_2}{\sin k_1 - \sin k_2 -U/2i} 
I^{\sigma_1\tau_1}_{\sigma_2\tau_2}\ .
\label{smat}
\ee
Here $I$ is the identity operator
$I^{\sigma_1\tau_1}_{\sigma_2\tau_2}=\delta_{\sigma_1,\tau_1}
\delta_{\sigma_2,\tau_2}$ and $\Pi$ is a permutation operator
$\Pi^{\sigma_1\tau_1}_{\sigma_2\tau_2}=\delta_{\sigma_1,\tau_2}
\delta_{\sigma_2,\tau_1}$.

In the next step we want to impose periodic boundary conditions on the
wave function
\bea
\psi(L+1,x_2;\sigma_1,\sigma_2)&=&\psi(1,x_2;\sigma_1,\sigma_2)\ ,\nn
\psi(0,x_2;\sigma_1,\sigma_2)&=&\psi(L,x_2;\sigma_1,\sigma_2)\ ,\nn
\psi(x_1,L+1;\sigma_1,\sigma_2)&=&\psi(x_1,1;\sigma_1,\sigma_2)\ ,\nn
\psi(x_1,0;\sigma_1,\sigma_2)&=&\psi(x_1,L;\sigma_1,\sigma_2)\ .
\eea
A short calculation shows that this imposes the following conditions on
the amplitudes
\be
A_{\sigma_{Q_1}\sigma_{Q_2}}(k_{P_1},k_{P_2})=\exp\left(ik_{P_1}L\right)
A_{\sigma_{Q_2}\sigma_{Q_1}}(k_{P_2},k_{P_1})\ ,
\label{pbc2A}
\ee
where $P,Q\in S_2$ are arbitrary.
In principle we now could simply solve \r{pbc2A} and thus determine
the quantization conditions for the momenta $k_{1,2}$. Rather than
proceeding in this way, we will introduce some seemingly unnecessary
formalism, which however will be very useful for treating the case of
more than two electrons. 

We define an auxiliary spin model on a lattice with $N$ sites i.e. a
lattice formed by the electrons. On every site $j$ there are two
allowed configurations $|\uparrow\rangle_j$ and $|\downarrow\rangle_j$,
corresponding to spin up and down respectively. 
Next we define spin operators $S_j^{\pm,z}$ acting on the resulting
Hilbert space as follows
\bea
S_j^{-}|\downarrow\rangle_j =0=S_j^{+}|\uparrow\rangle_j\ ,\quad
S_j^{-}|\uparrow\rangle_j =|\downarrow\rangle_j\ ,\quad
S_j^{+}|\downarrow\rangle_j =|\uparrow\rangle_j\ ,\quad
S_j^{z}|\downarrow\rangle_j =-\frac{1}{2}|\downarrow\rangle_j\ ,\quad
S_j^{z}|\uparrow\rangle_j =\frac{1}{2}|\uparrow\rangle_j\ .
\label{spinops}
\eea
We now define a particular set of states in the spin model in the
following way \cite{EKS92a}      
\beq 
|k_{P_1},\ldots,k_{P_N} \rangle =
\sum_{\sigma_1\ldots\sigma_N=\pm 1}A_{\sigma_1\ldots \sigma_N}
(k_{P_1},\ldots,k_{P_N})
\prod_{\ell =1 } ^{N} \left(S_{\ell}^{-} 
\right)^{(1- \sigma_{\ell})/2}  |0 \rangle \qd.
\label{Astates}
\eeq
Here we have used conventions where $\sigma_j=\uparrow$ corresponds to $1$
and $\sigma_j=\downarrow$ corresponds to $-1$. The set of equations
\r{ybe1}, due to the Schr\"odinger equation, now induces the following
between states in the spin model 
\beq
|k_{P_1},k_{P_2} \rangle = 
Y^{1,2}(\sin k_{P_2}, \sin k_{P_1})
|k_{P_2},k_{P_1} \rangle \quad ,
\label{ybey}
\eeq
where the operator $Y^{j,k}$ is given by
\be
Y^{j,k}(v_1, v_2)
 =  {\frac {-U/2i} {v_1 -v_2 - U/2i}} I
+ {\frac {v_1-v_2} {v_1 -v_2 - U/2i}}
\Pi^{(j,k)} \quad.
\label{yjk}
\ee
Here $I$ is the identity operator over the Hilbert space of the spin
model and $\Pi^{(j,k)}$ is a permutation operator
\beq
\Pi^{(j,k)} = {\frac 1 2}
\left( I + 4 {\vec S}_j \cdot {\vec S}_k \right) \qd.
\label{pi12}
\eeq
Here $S_k^{x,y,z}$ are the spin operators defined in \r{spinops},
where $S^\pm_k=S^x_k\pm i S^y_k$.
In order to derive \r{ybey} one simply uses the explicit form \r{smat}
of the S-matrix and the following properties of permutation operators
\be
\Pi^{(1,2)}|0\rangle=|0\rangle\ ,\quad
\Pi^{(1,2)}S^-_1\Pi^{(1,2)}=S^-_2\ ,\quad
\Pi^{(1,2)}S^-_2\Pi^{(1,2)}=S^-_1\ .
\ee
The $Y$-operators (which are related to the S-matrix \r{smat} via
multiplication by a permutation operator) were first introduced by
C.N. Yang in his seminal 1967 paper \cite{Yang67}.

On the level of the auxiliary spin model the periodic boundary
conditions \r{pbc2A} translate into 
\be
|k_{P_1},k_{P_2}\rangle=\exp\left(ik_{P_1}L\right) \Pi^{(12)}
|k_{P_2},k_{P_1}\rangle\ .
\ee
With the use of \r{ybey} this is then transformed into
\be
|k_{P_1},k_{P_2}\rangle=\exp\left(ik_{P_1}L\right) 
X^{1,2}(\sin k_{P_1},\sin k_{P_2}) |k_{P_1},k_{P_2}\rangle\ ,
\label{pbc2}
\ee
where the operator $X^{j,k}$ is given by
\bea
X^{j,k}(v_1, v_2)& =& \Pi^{(j,k)}Y^{j,k}(v_1, v_2)
= {\frac {-U/2i} {v_1 -v_2 - U/2i}}
\Pi^{(j,k)}+
{\frac {v_1-v_2} {v_1 -v_2 - U/2i}} I \qd.
\label{xjk}
\eea
In other words, $X^{1,2}$ is the S-matrix \r{smat} viewed as an
operator in the auxiliary spin model.
We now make the crucial observation that the operator $X^{1,2}$ is 
precisely the transfer matrix of an inhomogeneous spin-1/2
Heisenberg model on a 2-site lattice
\be
X^{1,2}(\sin k_{P_1},\sin k_{P_2})=
\tau(\sin k_{P_1}|\{\sin k_{P_1},\sin k_{P_2}\})\ ,
\ee
where $\tau$ is given by \r{transferm} with $N=2$. This is shown in
full generality in \r{xtau}. 

The periodic boundary conditions \r{pbc2A} can thus be
rewritten as an eigenvalue problem for the transfer matrix
$\tau(\Lambda=\sin k_{P_1}|\{\sin k_{P_1},\sin k_{P_2} \})$ of an
inhomogeneous Heisenberg model on 2-site lattice 
\beq
|k_{P_1},k_{P_2}\rangle =e^{i Lk_{P_1}}\ 
\tau(\sin k_{P_1}|\{\sin k_{P_1},\sin k_{P_2} \})\
|k_{P_1},k_{P_2} \rangle \qd.
\label{pbc22}
\eeq
The diagonalization of the transfer matrix $\tau(\Lambda|\{\sin
k_{P_1},\sin k_{P_2} \})$ is carried out in Appendix
\ref{section:XXX}. {From} the point of view of constructing 
eigenstates of the Hubbard Hamiltonian with periodic boundary
conditions, we have succeeded in reducing the problem to a much
simpler one, namely diagonalizing the transfer matrix of an
inhomogeneous Heisenberg model. This is the essence of C.N. Yang's
nested Bethe Ansatz procedure.

For the problem at hand we need to distinguish two cases, depending on
the spins of the two electrons. 
\begin{itemize}
\item{} \underline{two electrons with spin up}

Here the appropriate eigenstate of $\tau(\Lambda|\{\sin k_{P_1},\sin
k_{P_2} \})$ is found in the sector with no overturned spins, i.e. it
is the ferromagnetic state with all spins up. From \r{eigen} we see
that its eigenvalue is equal to $1$. The periodic boundary conditions
for the case of two electrons with spin up thus take the simple form 
\be
e^{ik_j L}=1\ ,\qquad j=1,2.
\ee
These are precisely the Lieb-Wu equations \r{bak},\r{bas} for the case
$N=2$, $M=0$.
Using that the appropriate eigenstate of $\tau(\Lambda|\{\sin
k_{P_1},\sin k_{P_2} \})$ is equal to $|0\rangle$,
we infer by comparison with \r{Astates} that all amplitudes are equal
to $1$. This then implies the following explicit form for the wave
function in sector $Q$
\beq
\psi(x_1, x_2; \sigma_1,\sigma_2) =
\sum_{P \in S_2}\sign(PQ) \exp(i \sum_{j=1}^{2} k_{Pj}x_{Qj}) \quad .
\eeq
This agrees with Woynarovich's result \r{wswf}.

\item{} \underline{one electron with spin up, one with spin down}

Now the appropriate eigenstate of $\tau$ is found in the sector with
one overturned spin. Using \r{eigen} we obtain the corresponding
eigenvalue 
\be
1-iU/[2(\sin k_{P_1}-\lambda_1+iU/4)] = \frac{\sin k_{P_1}-\lambda_1-iU/4}
{\sin k_{P_1}-\lambda_1+iU/4}\ ,
\label{a31}
\ee
where $\lambda_1$ fulfills the Bethe Ansatz equations of the
inhomogeneous Heisenberg model on a 2-site lattice
\bea
1 & = & \prod_{j=1}^{2} {\frac
{\lambda_1 - \sin k_j + iU/4}
{\lambda_1 - \sin k_j - iU/4}} \quad .
\label{lw21}
\eea
Inserting \r{a31} into \r{pbc22} we obtain the following
quantization conditions for the momenta $k_j$ due to periodic boundary
conditions
\bea
\exp(i L k_{P_1}) & = & {\frac {\lambda_1 - \sin k_{P_1} - iU/4}
{\lambda_1 - \sin k_{P_1} + iU/4}  } \quad , \qd {\rm for} \, P \in S_2
\quad .
\label{lw22}
\eea
Equations \r{lw21} and \r{lw22} coincide with the Lieb-Wu equations
\r{bak},\r{bas} for the case $N=2$ and $M=1$. 

Let us also determine an explicit expression for the amplitudes
$A_{\sigma_1\sigma_2}(k_1,k_2)$. Comparing the result \r{1spin} for
the eigenstate of $\tau(\sin k_{P_1}|\{\sin k_{P_1},\sin k_{P_2} \})$
with \r{Astates} we see that
\be
A_{\sigma_1\sigma_2}(k_{P_1},k_{P_2})=
\prod_{j=1}^{y-1} \left({\lambda_1 - \sin k_{P_j} - {iU\over
4}\over {\lambda_1 - \sin k_{P_j} + {iU\over 4}}}\right){1\over\lambda_1 -
\sin k_{P_y} + {iU\over 4}}\ ,
\label{aexpl}
\ee
where $y$ is the position of the down spin in the sequence
$\sigma_1\sigma_2$. 
The wave functions \r{BAWF2} with amplitudes \r{aexpl} coincide with
Woynarovich's result \r{wswf} for the case $N=2$, $M=1$.

\end{itemize}

\subsection{\boldmath Three electrons}

Let us now explicitly construct the wave function for the case of
three electrons $N=3$. The Schr\"odinger equation is of the form

\bea
&&- \sum_{s=\pm 1}\left[\psi(x_1+s, x_2, x_3; \sigma_1,\sigma_2,\sigma_3 ) 
+\psi(x_1, x_2+s, x_3; \sigma_1,\sigma_2,\sigma_3 ) 
+\psi(x_1, x_2, x_3+s; \sigma_1,\sigma_2,\sigma_3 )\right] \nn
&&+U \sum_{j<k} \delta(x_j, x_k) \psi(x_1,x_2,x_3; \sigma_1,\sigma_2,\sigma_3)
 = (E+\frac{3U}{2}-\frac{UL}{4}) \psi(x_1,x_2,x_3;
\sigma_1,\sigma_2,\sigma_3)\quad . 
\label{3SG}
\eea
In the sector defined by $x_{Q_1}\leq x_{Q_2}\leq x_{Q_3}$, where $Q$
is a permutation of the labels $\{1,2,3\}$, the Bethe Ansatz
wavefunction reads
\beq
\psi(x_1, x_2,x_3; \sigma_1,\sigma_2,\sigma_3) =
\sum_{P \in S_3}\sign(PQ)
A_{\sigma_{Q_1}\sigma_{Q_2}\sigma_{Q_3}}(k_{P_1},k_{P_2},k_{P_3})
\exp(i \sum_{j=1}^{3} k_{Pj}x_{Qj}) \quad .
\label{BAWF3}
\eeq
Substituting \r{BAWF3} into \r{3SG} for the case $x_1\neq x_2\neq
x_3\neq x_1$ we obtain the following expression for the energies
\beq
E = - ( 2 \cos k_1 + 2 \cos k_2 +2\cos k_3) -\frac{3U}{2}+\frac{UL}{4}\qd.
\eeq

Single-valuedness of the wave-function now leads to a larger number of
relations between the amplitudes. We have to consider the three cases
$\psi(x,x,x_3;\sigma_1,\sigma_2,\sigma_3)$,
$\psi(x,x_2,x;\sigma_1,\sigma_2,\sigma_3)$, 
$\psi(x_1,x,x;\sigma_1,\sigma_2,\sigma_3)$.
A simple calculation yields the following conditions on the amplitudes
\be
A_{\sigma_{Q_1}\sigma_{Q_2}\sigma_{Q_3}}(k_{P_1},k_{P_2},k_{P_3})
-A_{\sigma_{Q_1}\sigma_{Q_2}\sigma_{Q_3}}(k_{P'_1},k_{P'_2},k_{P'_3})
=
A_{\sigma_{Q'_1}\sigma_{Q'_2}\sigma_{Q'_3}}(k_{P'_1},k_{P'_2},k_{P'_3})
-A_{\sigma_{Q'_1}\sigma_{Q'_2}\sigma_{Q'_3}}(k_{P_1},k_{P_2},k_{P_3}),
\label{cont3}
\ee
where $Q$ and $P$ are arbitrary permutations of $\{1,2,3\}$ and
$Q'=Q(j,j+1)$, $P'=P(j,j+1)$. Our notations are such that for any
permutation of $N$ elements $S=(S_1,\ldots ,S_N)$
\be
S(j,j+1)=(S_1,\ldots, S_{j-1},S_{j+1},S_j,S_{j+2},\ldots,S_N).
\ee
Let us consider a specific example of \r{cont3} in more detail. The
wave-function $\psi(x,x,x_3;\sigma_1,\sigma_2,\sigma_3)$ with $x_3> x$
can be expressed alternatively in sector $Q=(1,2,3)$ or in sector
$Q'=(2,1,3)$. Equating the respective expressions \r{BAWF3} and using
the orthogonality property of plane waves we obtain
\be
A_{\sigma_{1}\sigma_{2}\sigma_{3}}(k_{P_1},k_{P_2},k_{P_3})
-A_{\sigma_{1}\sigma_{2}\sigma_{3}}(k_{P_2},k_{P_1},k_{P_3})
=
A_{\sigma_{Q_2}\sigma_{Q_1}\sigma_{Q_3}}(k_{P_2},k_{P_1},k_{P_3})
-A_{\sigma_{Q_2}\sigma_{Q_1}\sigma_{Q_3}}(k_{P_1},k_{P_2},k_{P_3}).
\ee
This indeed coincides with the general result \r{cont3}.

We note that the Bethe ansatz wave function \r{BAWF3} by construction
is antisymmetric under simultaneous exchange of spin and space
variables. It is easy to see that this fact assures the Schr\"odinger
equation \r{3SG} to be satisfied, when the three electron are occupying
the same site. Moreover, this reasoning readily generalizes to the case
of an arbitrary number of electrons in the following subsection. The
only non-trivial case left to consider is the case of two electrons at
the same site.

Let us start with $x_1=x_2=x<x_3$. Using \r{BAWF3} in \r{3SG} we obtain
the following condition on the amplitudes 

\bea
&&-[e^{-ik_{P_1}}+e^{ik_{P_2}}]A_{\sigma_1\sigma_2\sigma_3}(k_{P_1},k_{P_2},k_{P_3})
+[e^{-ik_{P_2}}+e^{ik_{P_1}}]A_{\sigma_1\sigma_2\sigma_3}(k_{P_2},k_{P_1},k_{P_3})\nn
&&+[e^{ik_{P_2}}+e^{-ik_{P_1}}]A_{\sigma_2\sigma_1\sigma_3}(k_{P_1},k_{P_2},k_{P_3})
-[e^{ik_{P_1}}+e^{-ik_{P_2}}]A_{\sigma_2\sigma_1\sigma_3}(k_{P_2},k_{P_1},k_{P_3})\nn
&&+U\left[A_{\sigma_1\sigma_2\sigma_3}(k_{P_1},k_{P_2},k_{P_3}) - 
A_{\sigma_1\sigma_2\sigma_3}(k_{P_2},k_{P_1},k_{P_3})\right]\nn
&&= -2(\cos k_{P_1}+\cos k_{P_2})\left[A_{\sigma_1\sigma_2\sigma_3}(k_{P_1},k_{P_2},k_{P_3}) - 
A_{\sigma_1\sigma_2\sigma_3}(k_{P_2},k_{P_1},k_{P_3})\right].
\label{3sg}
\eea
Now making use of the fact that \r{3sg} and \r{cont3} are of the same
structure as \r{2SG2} and \r{cont}, we conclude that the following
relation between amplitudes holds 
\bea
A_{\sigma_2\sigma_1\sigma_3}(k_{P_2},k_{P_1},k_{P_3})&=&
\sum_{\tau_1,\tau_2} S^{\sigma_1\tau_1}_{\sigma_2\tau_2}(k_{P_1},k_{P_2})\ 
A_{\tau_1\tau_2\sigma_3}(k_{P_1},k_{P_2},k_{P_3})\ ,
\eea
where $S$ is the two-particle S-matrix \r{smat}. All other cases of
coinciding coordinates can be analysed in exactly the same way.
The final result is
\bea
A_{\sigma_{Q'_1}\sigma_{Q'_2}\sigma_{Q'_3}}(k_{P'_1},k_{P'_2},k_{P'_3})&=&
\sum_{\tau_1,\tau_2} S^{\sigma_{Q_j}\tau_1}_{\sigma_{Q_{j+1}}\tau_2}(k_{P_j},k_{P_{j+1}})\ 
A_{\sigma_{Q_1}\ldots \sigma_{Q_{j-1}}\tau_1\tau_2
\sigma_{Q_{j+2}}\ldots\sigma_{Q_3}}(k_{P_1},k_{P_2},k_{P_3})\ ,
\label{3asa}
\eea
where $Q$ and $P$ are arbitrary permutations and
$Q'=Q(j,j+1)$,$P'=P(j,j+1)$. Equation \r{3asa} has important
consequences. In order to exhibit these clearly, it is convenient to
express \r{3asa} in the framework of the auxiliary spin model
introduced above. Inserting \r{3asa} into \r{Astates} we obtain
\beq
|k_{P'_1},k_{P'_2},k_{P'_3} \rangle = 
Y^{j,j+1}(\sin k_{P_{j}}, \sin k_{P_{j+1}})
|k_{P_1},k_{P_2},k_{P_3} \rangle \quad ,
\label{3ybey}
\eeq
where $Y^{j,k}$ is given by \r{yjk}. There are altogether
$(N-1)N!=12$ equations \r{3ybey} and all of them have to be consistent
with one another! This puts severe constraints on the operators
$Y^{j,k}$.

Let us explain this in more detail. We recall that the symmetric
group $S_N$ is generated by the identity $\id$ and the transpositions
of nearest neigbours $(j, j + 1)$, $j = 1, \dots, N - 1$, modulo the
relations
\bea \label{braid}
     (j, j + 1) (j + 1, j + 2) (j, j + 1) & = &
     (j + 1, j + 2) (j, j + 1) (j + 1, j + 2) \qd, \\ \label{nnncom}
     (j, j + 1) (k, k + 1) & = & (k, k + 1) (j, j + 1) \qqd \mbox{for}
				 \ |j - k| > 1 \qd, \\
     (j, j + 1) (j, j + 1) & = & \id \qd. \label{idem}
\eea
Equation (\ref{braid}) is called the braid relation. Note that
(\ref{nnncom}) is non-trivial only for $N > 3$.

By inspection of (\ref{3ybey}) we see that the $Y$-operators act on
the states $|k_1, k_2, k_3\>$ by exchanging neighbouring components of
the vector $\vec k = (k_1, k_2, k_3)$, 
that determines the state $|k_1, k_2, k_3\>$ of our auxiliary spin
system. Hence all states $|k_{P_1}, k_{P_2}, k_{P_3}\>$, $P \in S_3$,
can be obtained from $|k_1, k_2, k_3\>$ by repeated use of
(\ref{3ybey}). Equivalently, (\ref{3ybey}) allows us to obtain the
state corresponding to any permutation $\bar P \in S_3$ from a state
corresponding to any other permutation $P \in S_3$.

It follows from (\ref{braid})-(\ref{idem}) that a representation of a
permutation as a product of transpositions of nearest neighbours is
not unique. 
For the case at hand, $N = 3$, we have for instance, $(1,3) =
(1,2) (2,3) (1,2) = (2,3) (1,2) (2,3)$. We are thus facing a consistency
problem for equations (\ref{3ybey}): The relations (\ref{braid})-%
(\ref{idem}) impose consistency conditions on the $Y$-operators.

Let us study these consistency conditions. For $N = 3$ we only have to
consider (\ref{braid}) for the case $j=1$ and (\ref{idem}). Thus
(\ref{braid}) implies that
\bea
|k_{\bar{P}_1},k_{\bar{P}_2},k_{\bar{P}_3} \rangle &=& 
Y^{1,2}(\sin k_{P_{2}}, \sin k_{P_{3}})
Y^{2,3}(\sin k_{P_{1}}, \sin k_{P_{3}})
Y^{1,2}(\sin k_{P_{1}}, \sin k_{P_{2}})
|k_{P_1},k_{P_2},k_{P_3} \rangle \ ,\nn
&=&
Y^{2,3}(\sin k_{P_{1}}, \sin k_{P_{2}})
Y^{1,2}(\sin k_{P_{1}}, \sin k_{P_{3}})
Y^{2,3}(\sin k_{P_{2}}, \sin k_{P_{3}})
|k_{P_1},k_{P_2},k_{P_3} \rangle ,
\label{yyy}
\eea
where $\bar{P}= P (1,3) = P (1,2) (2,3) (1,2) = P (2,3) (1,2) (2,3) =
(P_3,P_2,P_1)$. Assuming $|k_1, k_2, k_3\>$ to be arbitrary, we conclude
that
\be
Y^{1,2}(s_2,s_3)
Y^{2,3}(s_1,s_3)
Y^{1,2}(s_1,s_2)
=
Y^{2,3}(s_1,s_2)
Y^{1,2}(s_1,s_3)
Y^{2,3}(s_2,s_3)\ ,
\label{yyy2}
\ee
where $s_{1,2,3}$ are arbitrary complex numbers. This is the famous
Yang-Baxter equation. It is now crucial that (\ref{yyy2}) can be
verified by direct calculation. Therefore (\ref{3ybey}) is consistent
with (\ref{braid}).

Similarly, the use of equation (\ref{idem}) in (\ref{3ybey}) yields
the requirement
\be
\left(Y^{j,j+1}(u,v)\right)^{- 1}=Y^{j,j+1}(v,u)\ .
\label{yinv}
\ee
It is easy to verify by direct calculation that the operators $Y^{j,k}$
defined in\r{yjk} indeed fulfil \r{yinv}. Hence (\ref{3ybey}) is
consistent with (\ref{idem}), and we conclude that the entire set of
equations (\ref{3ybey}) is consistent.

Note that the above considerations easily generalize to the case of
arbitrary $N$, which will be treated in the next subsection. In addition
to (\ref{braid}) and (\ref{idem}) we then will have to consider
(\ref{nnncom}) which leads to the condition
\be
     Y_{j, j+1} (s_1, s_2) Y_{k, k+1} (s_3, s_4) =
     Y_{k, k+1} (s_3, s_4) Y_{j, j+1} (s_1, s_2) \qqd \mbox{for}\
	|j - k| > 1 \qd,
\ee
which is trivially satisfied.

%%%%%%%%%%%%%%%%%%%%%%%%%%%%%%%%%%%%%%%%%%%%%%%%%%%%%%%%%%%%%%%%%%%%%%%%

Let us now impose periodic boundary conditions on the wave function
\bea
\psi(0,x_2,x_3;\sigma_1,\sigma_2,\sigma_3)&=&\psi(L,x_2,x_3;\sigma_1,\sigma_2,\sigma_3)\ ,\nn
\psi(1,x_2,x_3;\sigma_1,\sigma_2,\sigma_3)&=&\psi(L+1,x_2,x_3;\sigma_1,\sigma_2,\sigma_3)\ ,\nn
\psi(x_1,0,x_3;\sigma_1,\sigma_2,\sigma_3)&=&\psi(x_1,L,x_3;\sigma_1,\sigma_2,\sigma_3)\ ,\nn
\psi(x_1,1,x_3;\sigma_1,\sigma_2,\sigma_3)&=&\psi(x_1,L+1,x_3;\sigma_1,\sigma_2,\sigma_3)\ ,\nn
\psi(x_1,x_2,0;\sigma_1,\sigma_2,\sigma_3)&=&\psi(x_1,x_2,L;\sigma_1,\sigma_2,\sigma_3)\ ,\nn
\psi(x_1,x_2,1;\sigma_1,\sigma_2,\sigma_3)&=&\psi(x_1,x_2,L+1;\sigma_1,\sigma_2,\sigma_3)\ .
\label{3pbc}
\eea
Inserting \r{BAWF3} into \r{3pbc} yields
\be
A_{\sigma_{Q_1}\sigma_{Q_2}\sigma_{Q_3}}(k_{P_1},k_{P_2},k_{P_3})=
\exp(i k_{P_1}L)
A_{\sigma_{Q_2}\sigma_{Q_3}\sigma_{Q_1}}(k_{P_2},k_{P_3},k_{P_1})\ ,
\label{3pbcA}
\ee
where $Q$ and $P$ are arbitrary permutations of $\{1,2,3\}$.
In terms of the auxiliary spin model \r{3pbcA} is expressed as
\bea
|k_{P_1},k_{P_2},k_{P_3}\rangle&=&
\exp(i k_{P_1}L)
\Pi^{(1,2)}\Pi^{(2,3)}|k_{P_2},k_{P_3},k_{P_1}\rangle\nn
&=&\exp(i k_{P_1}L)
\Pi^{(1,2)}\Pi^{(2,3)}Y^{2,3}(\sin k_{P_1},\sin k_{P_3})
Y^{1,2}(\sin k_{P_1},\sin k_{P_2})|k_{P_1},k_{P_2},k_{P_3}\rangle\nn
&=&\exp(i k_{P_1}L)
\Pi^{(1,2)}X^{2,3}(\sin k_{P_1},\sin k_{P_3})\Pi^{(1,2)}
X^{1,2}(\sin k_{P_1},\sin k_{P_2})|k_{P_1},k_{P_2},k_{P_3}\rangle\nn
&=&\exp(i k_{P_1}L)
X^{1,3}(\sin k_{P_1},\sin k_{P_3})
X^{1,2}(\sin k_{P_1},\sin k_{P_2})|k_{P_1},k_{P_2},k_{P_3}\rangle\ ,
\eea
where we have used the identities \r{ppp}. Using \r{xtau} we now
obtain in complete analogy with the 2-electron case
\beq
|k_{P_1},k_{P_2},k_{P_3}\rangle =e^{i Lk_{P_1}}\ 
\tau(\sin k_{P_1}|\{\sin k_{P_j}; j=1, \ldots,3\})\
|k_{P_1},k_{P_2},k_{P_3} \rangle \qd.
\label{pbc33}
\eeq
We again can use the results for the diagonalization of the
inhomogeneous transfer matrix $\tau(\sin k_{P_1}|\{\sin k_{P_j}
;j=1, \ldots,3 \})$
derived in Appendix \ref{section:XXX}. We need to distinguish two
cases, depending on the spins of the two electrons
(recall that we consider only states for which the number of
down spins not larger than the number of up spins, all other states are
obtained by using the spin-reversal symmetry).
\begin{itemize}
\item{} \underline{Three electrons with spin up}

Here the appropriate eigenvector of $\tau(\Lambda|\{\sin
k_{P_j};j=1,\ldots ,3 \})$ is the ferromagnetic state, with eigenvalue $1$.
The periodic boundary conditions for the case of three electrons with
spin up thus take the form
\be
e^{ik_j L}=1\ ,\qquad j=1,2,3.
\ee
These are precisely the Lieb-Wu equations \r{bak},\r{bas} 
for the case $N=3$, $M=0$.

Using \r{Astates} we see that all amplitudes are trivial
\be
A_{\sigma_1\sigma_2\sigma_3}(k_{P_1},k_{P_2},k_{P_3})= 1\ .
\ee
The corresponding wave-function \r{BAWF3} coincides with Woynarovich's
result \r{wswf}.

\item{} \underline{two electrons with spin up, one with spin down}

Now the appropriate eigenvector of $\tau$ is found in the sector with
one overturned spin. From \r{eigen} its eigenvalue is given by
\be
1-iU/[2(\sin k_{P_1}-\lambda_1+iU/4)] = \frac{\sin k_{P_1}-\lambda_1-iU/4}
{\sin k_{P_1}-\lambda_1+iU/4}\ ,
\label{eval}
\ee
where $\lambda_1$ fulfills
\bea
1 & = & \prod_{j=1}^{3} {\frac
{\lambda_1 - \sin k_j + iU/4}
{\lambda_1 - \sin k_j - iU/4}} \quad .
\label{lw31}
\eea
Inserting \
r{eval} into \r{pbc33} we obtain the following
quantization conditions due to periodic boundary conditions 
\bea
\exp(i L k_{P_1}) & = & {\frac {\lambda_1 - \sin k_{P_1} - iU/4}
{\lambda_1 - \sin k_{P_1} + iU/4}  } \quad , \qd {\rm for} \, P \in
S_3\ .
\label{lw32}
\eea
Equations \r{lw31} and \r{lw32} are precisely the Lieb-Wu equations
\r{bak},\r{bas} for the case $N=3$ and $M=1$. 

An explicit expression for the amplitudes again is obtained from
\r{Astates} and \r{1spin}, with the result

\be
A_{\sigma_1\sigma_2\sigma_3}(k_{P_1},k_{P_2},k_{P_3})=
\prod_{j=1}^{y-1} \left({\lambda_1 - \sin k_{P_j} - {iU\over
4}\over {\lambda_1 - \sin k_{P_j} + {iU\over 4}}}\right){1\over\lambda_1 -
\sin k_{P_y} + {iU\over 4}}\ ,
\label{aexpl3}
\ee
where $y$ is the position of the down spin in the sequence
$\sigma_1\sigma_2\sigma_3$. 
Inserting \r{aexpl3} into \r{BAWF3} we obtain \r{wswf} for the case
$N=3$, $M=1$.

\end{itemize}

\subsection{\boldmath $N$ electrons}

It is now clear how to generalize the above results to the case of $N$
electrons. The Bethe Ansatz for the solution $\psi$ of the 
Schr\"odiger equation \r{SG} in the sector $Q$ with
$x_{Q_1}\leq x_{Q_2}\leq \ldots\leq x_{Q_N}$ is

\beq
\psi(x_1, \ldots,x_N; \sigma_1,\ldots,\sigma_N) =
\sum_{P \in S_N}\sign(PQ)
A_{\sigma_{Q_1}\ldots\sigma_{Q_N}}(k_{P_1},\ldots,k_{P_N})
\exp(i \sum_{j=1}^{N} k_{Pj}x_{Qj}) \quad .
\label{BAWFN}
\eeq
Substituting \r{BAWFN} into \r{SG} for the case $x_j\neq x_k, (j,
k=1,\ldots,N; j\neq k)$ we obtain the following expression for the energies
\beq
E = -2\sum_{j=1}^N \cos k_j -\frac{NU}{2}+\frac{UL}{4}\qd.
\eeq
Using the single valuedness of the wave function and solving the matching
conditions at the sector boundaries i.e. the Schr\"odinger equation
for the cases where two of the coordinates coincide, we obtain
the following set of equations
\bea
A_{\sigma_{Q'_1}\ldots\sigma_{Q'_N}}(k_{P'_1},\ldots,k_{P'_N})&=&
\sum_{\tau_1,\tau_2} 
S^{\sigma_{Q_j}\tau_1}_{\sigma_{Q_{j+1}}\tau_2}(k_{P_j},k_{P_{j+1}})\ 
A_{\sigma_{Q_1}\ldots
\sigma_{Q_{j-1}}\tau_1\tau_2\sigma_{Q_{j+2}}\ldots\sigma_{Q_N}}
(k_{P_1},\ldots,k_{P_N})\ ,
\label{Nasa}
\eea
where $Q$ and $P$ are arbitrary permutations and
$Q'=Q(j,j+1)$,$P'=P(j,j+1)$. 
In terms of the auxiliary spin model \r{Nasa} reads
\beq
|k_{P'_1},\ldots,k_{P'_N} \rangle = 
Y^{j,j+1}(\sin k_{P_{j}}, \sin k_{P_{j+1}})
|k_{P_1},\ldots,k_{P_N} \rangle \quad .
\label{Nybey}
\eeq
The mutual consistency of equations \r{Nybey} follows from \r{yyy2}
and \r{yinv}. 
We now impose periodic boundary conditions on the wave function 
\bea
\psi(x_1,\ldots,x_{j-1},0,x_{j+1},\ldots,x_N;\sigma_1,\ldots,\sigma_N)&=&
\psi(x_1,\ldots,x_{j-1},L,x_{j+1},\ldots,x_N;\sigma_1,\ldots,\sigma_N)\
,\nn
\psi(x_1,\ldots,x_{j-1},1,x_{j+1},\ldots,x_N;\sigma_1,\ldots,\sigma_N)&=&
\psi(x_1,\ldots,x_{j-1},L+1,x_{j+1},\ldots,x_N;\sigma_1,\ldots,\sigma_N)\
,
\label{Npbc}
\eea
where $j=1,\ldots, N$. Inserting \r{BAWFN} into \r{Npbc} yields
\be
A_{\sigma_{Q_1}\ldots\sigma_{Q_N}}(k_{P_1},\ldots,k_{P_N})=
\exp(i k_{P_1}L)
A_{\sigma_{Q_2}\ldots\sigma_{Q_N}\sigma_{Q_1}}(k_{P_2},\ldots,k_{P_N},k_{P_1})
\ , 
\label{NpbcA}
\ee
where $Q,P\in S_N$ are arbitrary.
In terms of the auxiliary spin model \r{NpbcA} is expressed as
\bea
|k_{P_1},\ldots,k_{P_N}\rangle&=&
\exp(i k_{P_1}L)
\Pi^{(1,2)}\Pi^{(2,3)}\ldots\Pi^{(N-1,N)}
|k_{P_2},\ldots,k_{P_N},k_{P_1}\rangle\nn
&=&\exp(i k_{P_1}L)
\Pi^{(1,2)}\Pi^{(2,3)}\ldots\Pi^{(N-1,N)}
\left[\prod_{m=0}^{N-2} Y^{N-m-1,N-m}(\sin k_{P_1},\sin k_{P_{N-m}})\right]
|k_{P_1},\ldots,k_{P_N}\rangle\nn
&=&\exp(i k_{P_1}L)
X^{1,N}(\sin k_{P_1},\sin k_{P_{N}})
X^{1,N-1}(\sin k_{P_1},\sin k_{P_{N-1}})\ldots\nn
&&\times \ldots X^{1,3}(\sin k_{P_1},\sin k_{P_{3}})
X^{1,2}(\sin k_{P_1},\sin k_{P_{2}})
|k_{P_1},\ldots,k_{P_N}\rangle\ .
\eea
where we have used the identities \r{ppp}. Using \r{xtau} we now
obtain in complete analogy with the 2 and 3 electron cases
\beq
|k_{P_1},\ldots,k_{P_N}\rangle =e^{i Lk_{P_1}}\ 
\tau(\sin k_{P_1}|\{\sin k_{P_j}; j=1, \ldots,N \})\
|k_{P_1},\ldots,k_{P_N} \rangle \qd.
\label{pbc3N}
\eeq
Next we again use the results for the diagonalization of the
inhomogeneous transfer matrix $\tau(\sin k_{P_1}|\{\sin
k_{P_j};j=1,\ldots , N \})$ derived in Appendix \ref{section:XXX}. 
We now need to distinguish $[N/2]+1$ cases ($[x]$ is the integer part
of $x$), corresponding to the possible values of $M$.
In the sector with $M$ down spins the eigenvalue of 
$\tau(\sin k_{P_1}|\{\sin k_{P_j};j=1,\ldots , N \})$ is given by
\r{eigen}
\be
\prod_{j=1}^M \frac{\sin k_{P_1}-\lambda_j-iU/4}
{\sin k_{P_1}-\lambda_j+iU/4}\ ,
\ee
where the rapidities $\lambda_j$ fulfill
\bea
\prod_{l=1}^{N} {\frac
{\lambda_j - \sin k_l + iU/4}
{\lambda_j - \sin k_l - iU/4}}&=&
\prod_{\scriptstyle k=1\atop \scriptstyle k\neq j}^M
{\frac{\lambda_j - \lambda_k + iU/2}
{\lambda_j - \lambda_k - iU/2}} \quad .
\label{lwN1}
\eea
Inserting this result into \r{pbc3N} we finally obtain 
\bea
\exp(i L k_{P_1}) & = & \prod_{j=1}^M{\frac {\lambda_j - \sin k_{P_1} - iU/4}
{\lambda_j - \sin k_{P_1} + iU/4}  } \quad , \qd {\rm for} \, P \in
S_N\ .
\label{lwN2}
\eea
Equations \r{lwN1} and \r{lwN2} are precisely the Lieb-Wu equations
\r{bak},\r{bas} for the case of $N$ electrons, $M$ of which have spin down.
In order to obtain an explicit expression for the amplitudes we now
need to make use of the general result \r{stateswoy} for eigenstates
of the transfer matrix $\tau(\sin k_{P_1}|\{\sin k_{P_j};j=1,\ldots ,
N \})$ of the inhomogeneous Heisenberg model. We find 
\be
A_{\sigma_1\ldots\sigma_N}(k_{P_1},\ldots ,k_{P_N}) = 
\sum_{\pi\in S_{M}} \ A_{\pi} \prod_{t=1}^M\left\lbrace
{1\over \lambda_{\pi_t}-\sin k_{P_{y_t}}+{iU\over 4}}
\prod_{s=1}^{y_t-1}\ {\lambda_{\pi_t}-\sin k_{P_s}-{iU\over 4}
\over \lambda_{\pi_t}-\sin k_{P_s}+{iU\over 4}}\right\rbrace\ ,
%\sum_{\pi\in S_{M}} \ A_{\pi} \prod_{t=1}^M\left\lbrace
%\biggl(\ \prod_{s=1}^{y_t-1}\ {\sin k_{P_s}-\lambda_{\pi_t}-{U\over 4i}
%\over \sin k_{P_s}-\lambda_{\pi_t}+{U\over 4i}}\biggr) 
%{1\over \sin k_{P_{y_t}}-\Lambda_{\pi_t}+{U\over 4i}}\right\rbrace\ ,
\ee
where $1\le y_{_1}< y_{_2}< \ldots < y_{_M}\le N$ are the positions of
the down spins in the sequence $\sigma_1\ldots\sigma_N$ and $A_\pi$
is given by
\be
A_\pi 
= \prod_{1\le l< k\le M} \left({\lambda_{\pi_l} -
\lambda_{\pi_k}-{iU\over 2}\over 
\lambda_{\pi_l} - \lambda_{\pi_k}}\right)\ \ .
\label{AAn}
\ee
The resulting explicit expression for the wave function \r{BAWFN}
coincides with Woynarovich's result \r{wswf}.

\section{Inhomogeneous Heisenberg model}
\label{section:XXX}
\subsection{Algebraic Bethe Ansatz}
Our starting point is a lattice of $N$ spin-1/2's. The corresponding
Hilbert space is $V_1\otimes V_2\otimes\dots\otimes V_N$, where $V_j$
is isomorphic to $C^2$. We define the Pauli matrices ${\vec \tau}= (
\tau^{x}, \tau^{y},\tau^{z}) $ by 
\beq
\tau^x=
\left(
\begin{array}{cc}
0 & 1 \nonumber \\
1 & 0 \nonumber \\
\end{array}
\right) \quad, \qd
\tau^y=
\left(
\begin{array}{cc}
0 & -i  \nonumber \\
i & 0 \nonumber \\
\end{array}
\right) \quad, \qd
\tau^z=
\left(
\begin{array}{cc}
1 & 0 \nonumber \\
0 & -1 \nonumber \\
\end{array}
\right)
\eeq
and $\tau^\pm = \2 (\tau^x \pm i \tau^y)$.

The central object of the Quantum Inverse Scattering Method is the
$R$-matrix, which is a solution of the Yang-Baxter equation. For the
case of the spin-1/2 Heisenberg model it is of the form
\be
{R(\lambda,\mu) =\left(\matrix{f(\mu,\lambda) &0&0&0\cr
0&g(\mu,\lambda)&1&0\cr 0&1&g(\mu,\lambda)&0\cr 0&0&0&f(\mu,\lambda)\cr
}\right) ,}
\label{R}
\ee
where
\be
{f(\mu,\lambda) = 1-{iU\over 2(\mu-\lambda) } \ ,\qquad
g(\mu,\lambda ) = -{iU\over 2(\mu - \lambda) } \ .}
\ee
$R(\lambda,\mu)$ acts on the tensor product space $V_0\otimes V_0$, where
$V_0$ is isomorphic to $C^2$. The Yang-Baxter equation for $R$
is an equation on the space $V_0\otimes V_0\otimes V_0$ and can be
written as
\be
R^{23}(\lambda,\mu)
R^{12}(\lambda,\nu)
R^{23}(\mu,\nu)
=R^{12}(\mu,\nu)
R^{23}(\lambda,\nu)
R^{12}(\lambda,\mu)\ ,
\label{rrr}
\ee
where the superscript indicates in which spaces the $R$ matrix acts
nontrivially. 
We now define an L-operator acting on the tensor product between a
``matrix-space'' $V_0$ and a ``quantum-space'' $V_n$, which is
identified with the Hilbert space over the $n^{\rm th}$ site of our
lattice of spins, by 
\bea
L_n(\lambda) &=&
\frac{\lambda}{\lambda +iU/2} I +\frac{iU/2}{\lambda +iU/2}
\Pi^{(0,n)}\ ,\nn
%\frac{1}{\lambda+iU/2}\left[
%\lambda + iU/4 \, \left(I+{\vec \tau}_0 \cdot {\vec \tau}_n\right)\right] =
&=&\frac{1}{\lambda+iU/2}
\left( 
\begin{array}{cc}
\lambda + (1+\, \tau_n^z) iU/4  &  \tau_n^{-}\ iU/2 \\
\tau_n^{+}\ iU/2 & \lambda + (1-\tau_n^z) iU/4
\end{array} 
\right)\ .
\label{lop}
\eea

The Yang-Baxter equation \r{rrr} implies the following intertwining
relations for the $L$-operator
\be
{R(\lambda,\mu) \left(L_n(\lambda)\otimes L_n(\mu)\right)
=  \left(L_n(\mu)\otimes L_n(\lambda)\right) R(\lambda ,\mu) \ ,}
\label{intL}
\ee
where the tensor product is between matrix spaces, i.e. \r{intL} is a
relation over the space $V_0\otimes V_0\otimes V_n$.

Next we note, that the intertwiner for the $L$-operator
(\ref{intL}) still holds, if we shift both spectral parameters $\lambda$
and $\mu$ by an arbitrary amount $\nu_n$, {\sl i.e.}
\be
{R(\lambda ,\mu) \left(L_n(\lambda -\nu_n)\otimes
L_n(\mu -\nu_n)\right) =  \left(L_n(\mu -\nu_n)\otimes L_n(\lambda
-\nu_n)\right) R(\lambda ,\mu) \ .}
\label{intLinh}
\ee
We now construct an {\sl inhomogeneous monodromy matrix} in the
following way
\be
{T(\mu|\{a_j\}) = L_N(\mu-a_N)L_{N-1}(\mu-a_{N-1})\ldots L_1(\mu-a_1) =
\left(\matrix{A(\mu)&B(\mu)\cr C(\mu)&D(\mu)\cr}\right) \ .}
\label{intT}
\ee
Here $a_1,\ldots ,a_N$ are $N$ arbitrary complex constants.
The intertwiner \r{intLinh} can be lifted to the level of the monodromy
matrix
\be
{R(\lambda ,\mu) \left(T(\lambda|\{a_j\})\otimes T(\mu|\{a_j\})\right)
=  \left(T(\mu|\{a_j\})\otimes T(\lambda|\{a_j\})\right) R(\lambda ,\mu) \ .}
\label{rtt}
\ee

By tracing (\ref{intT}) over the matrix space $V_0$ one then finds
that the {\sl transfer matrices}
\be
\tau(\mu|\{a_j\})={\rm Tr}_0 (T(\mu|\{a_j\}))=A(\mu)+D(\mu)
\label{transferm}
\ee
commute for any values of spectral parameter $\mu$, {\sl i.e.} 
$[\tau(\mu|\{a_j\}),\tau(\nu|\{a_j\})]=0$. That implies that
the transfer matrix is the generating functional of an infinite number
of mutually commuting conserved quantum operators (via expansion in
powers of $\mu$). 

Eigenstates of the transfer matrix are constructed by means of the
Algebraic Bethe Ansatz. Starting point is the choice of a {\sl
reference state}, which is a trivial eigenstate of $\tau(\mu|\{
a_j\})$. 
We choose the saturated ferromagnetic state
\be
\vac = |\up\up\up\ldots\up\rangle = \otimes_{n=1}^N |\up\rangle_n\ .
\ee
The action of the $L$-operator (\ref{lop}) on $|\up\rangle_n$ can be
easily calculated and implies the following actions of the matrix
elements of the monodromy matrix 
\bea
A(\mu)\vac &=& a(\mu)\vac\ ,\quad a(\mu) = 1\ ,\nn
D(\mu)\vac &=& d(\mu)\vac\ ,\quad d(\mu) =
\prod_{j=1}^N \frac{\mu-a_{j} }{\mu-a_{j} + iU/2  } \ ,\nn
C(\mu)\vac &=&0\ ,\qquad B(\mu)\vac \neq 0\ .
\label{AD}
\eea

{From} (\ref{AD}) we see that $B(\lambda)$ play the role of creation
operators. Acting with $B$ corresponds to flipping a spin.
States with $M$ down spins can be constructed as 
\be
F(\lambda_1,\ldots ,\lambda_M)=\prod_{j=1}^M (-2i/U)
B(\lambda_j -iU/4)|0\rangle \ ,
\label{states}
\ee
where we have shifted the spectral parameters and introduced a
particular normalization for later convenience.
The requirement that the states (\ref{states}) ought to be eigenstates
of the transfer matrix puts constraints on the values $\lambda_n$: the
set $\{\lambda_j\}$ must be a solution of the following system of
Bethe Ansatz equations
\be
\label{bae}
\prod_{k=1}^N \frac{\lambda_j-a_{k}+iU/4 }{\lambda_j-a_{k} - iU/4  }
=\prod_{\scriptstyle l=1\atop \scriptstyle l\neq j}^M
\frac{\lambda_j-\lambda_l+iU/2}{\lambda_j-\lambda_l-iU/2}\ ,
\qd j=1,\ldots ,M\ .
\ee
The corresponding eigenvalues of the transfer matrix are
\be
\tau(\mu|\{a_j\})\ F(\lambda_1,\ldots ,\lambda_M)=
\left[a(\mu)\prod_{j=1}^M f(\mu,\lambda_j-iU/4)
+d(\mu)\prod_{j=1}^M f(\lambda_j-iU/4,\mu)\right]
F(\lambda_1,\ldots ,\lambda_M)\ .
\label{eigen}
\ee
For our present purposes we need to consider the transfer matrix
evaluated at the first inhomogeneity. We find
\bea
\tau(a_1|\{a_j\}) &=& {\rm Tr}_0\left[
 L_N(a_1-a_N)L_{N-1}(a_1-a_{N-1})\ldots L_2(a_1-a_2)\Pi^{(0,1)}\right]\nn
&=&
{\rm Tr}_0\left[\Pi^{(0,1)}\Pi^{(0,1)}
L_N(a_1-a_N)\Pi^{(0,1)}\Pi^{(0,1)}L_{N-1}(a_1-a_{N-1})
\Pi^{(0,1)}\Pi^{(0,1)}\ldots \Pi^{(0,1)}L_2(a_1-a_2)\Pi^{(0,1)}\right]\nn
&=&
{\rm Tr}_0\left[\Pi^{(0,1)}
X^{1,N}(a_1,a_N)X^{1,N-1}(a_1,a_{N-1})\ldots X^{1,2}(a_1,a_2)\right]\nn
&=&
X^{1,N}(a_1,a_N)X^{1,N-1}(a_1,a_{N-1})\ldots X^{1,2}(a_1,a_2)\ ,
\label{xtau}
\eea
where $X^{j,k}$ is defined in \r{xjk}. Here we have first used
the explicit form of the L-operator \r{lop} and then the identities
\be
\Pi^{(j,k)}\Pi^{(j,n)}\Pi^{(j,k)}=\Pi^{(k,n)}\ ,\qquad
\Pi^{(j,k)}\Pi^{(j,k)}=I\ .
\label{ppp}
\ee

\subsection{Explicit expressions for the eigenstates}

In this subsection we derive the following expression for the
eigenstates \r{states}
\be
F(\Lambda_1,\ldots ,\Lambda_M)=
\sum_{\{y_{_i}\}} 
\sum_{\pi\in S_{M}} \ A_{\pi} \prod_{t=1}^M\left\lbrace
{1\over \Lambda_{\pi_t}-a_{P_{y_t}}+{iU\over 4}}
\prod_{s=1}^{y_t-1}\ {\Lambda_{\pi_t}-a_{P_s}-{iU\over 4}
\over \Lambda_{\pi_t}-a_{P_s}+{iU\over 4}}\right\rbrace
\prod_{j=1}^M\tau^-_{y_j} \vac\ ,
\label{stateswoy}
\ee
where the summation extends over $1\le y_{_1}< y_{_2}< y_{_3}< ... <
y_{_M}\le N$, $P$ is a permutation of $N$ elements, and the $A_\pi$
is given by
\be
A_\pi 
= \prod_{1\le l< k\le M} {\Lambda_{\pi_l} -
\Lambda_{\pi_k}-{iU\over 2}\over 
\Lambda_{\pi_l} - \Lambda_{\pi_k}}\ \ .
\label{AA}
\ee

Our discussion closely parallels \cite{EKS92a}. The basic tool
for proving \r{stateswoy} is the ``$n$-site generalized model''
\cite{IK84,IKR87} (see also \cite{KBIBo} p.151f and p.171). 

\vskip .5cm
\underline{Generalised two-site model.}
\vskip .5cm

Let us first divide the product of $L$-operators in \r{intT} into two
parts
\bea
T_{_I}(\Lambda) &=& L_n(\Lambda - a_{n})\ldots
L_2(\Lambda -a_{2})\ L_1(\Lambda - a_{1})\ ,\nn
T_{_{II}}(\Lambda) &=& L_N(\Lambda - a_{N})\ldots
L_{n+2}(\Lambda -a_{n+2})\ L_{n+1}(\Lambda - a_{n+1})\ .
\label{intT2}
\eea
Clearly we have
\be
T(\Lambda) = T_{_{II}}(\Lambda)\ \  T_{_{I}}(\Lambda)\ \ .
\ee

Both $T_{_{II}}(\Lambda) $ and $T_{_{I}}(\Lambda)$ are $2\times 2 $
matrices

\be
T_{_{I}}(\Lambda) =
\left(\matrix{A_{_I}(\Lambda)&B_{_I}(\Lambda)\cr
C_{_I}(\Lambda)&D_{_I}(\Lambda)\cr} \right)\ \ ,\ \ 
T_{_{II}}(\Lambda) =
\left(\matrix{A_{_{II}}(\Lambda)&B_{_{II}}(\Lambda)\cr
C_{_{II}}(\Lambda)&D_{_{II}}(\Lambda)\cr} \right)\ \ .
\ee
By construction the matrix elements of $T_{_{I}}(\Lambda)$ commute with
the matrix elements of $T_{_{II}}(\Lambda)$. The commutation relations
of the matrix elements of the same $T$-operator are as in \r{rtt}
\be
R\left(\Lambda_1,\Lambda_2\right) \bigl(T_{\alpha}(\Lambda_1)\otimes
T_{\alpha}(\Lambda_2)\bigr) = \bigl(T_{\alpha}(\Lambda_2)\otimes
T_{\alpha}(\Lambda_1)\bigr) R\left(\Lambda_1,\Lambda_2\right)\ ,\quad
\alpha=I,II\ .
\ee
The matrix elements of $T$ can be expressed in terms of the matrix
elements of $T_{_I}$ and $T_{_{II}}$, {\sl e.g.,}
\be
B(\Lambda) =  A_{_{II}}(\Lambda)\ B_{_{I}}(\Lambda) +
B_{_{II}}(\Lambda) \ D_{_{I}}(\Lambda)\ \ .
\ee
It is also possible to express the vectors $\prod_{j=1}^M B(\Lambda_j)
\ \vac$ in terms of $B_{_I}$ and $B_{_{II}}$. In order to do this we
will use that
\be
C_{\alpha}(\Lambda)\vac = 0\ \ ,\ \ A_{\alpha}(\Lambda)\vac =
a_{\alpha}(\Lambda)\vac\ \ ,\ \ D_{\alpha}(\Lambda)\vac =
d_{\alpha}(\Lambda)\vac\ ,\quad \alpha=I,II.
\ee
Here
\be
     a_{_{I}}(\Lambda) = 1\ ,\ \ d_{_{I}}(\Lambda) =
     \prod_{j=1}^{n}\frac{\Lambda - a_j}{\Lambda - a_j + iU/2}\ ,
\ee
and 
\be
     a_{_{II}}(\Lambda) = 1\ ,\ \ d_{_{II}}(\Lambda) =
     \prod_{j=n+1}^N \frac{\Lambda - a_j}{\Lambda - a_j + iU/2} \ \ .
\ee

In \cite{IK84,IKR87} it was proved that 

\bea
\prod_{j=1}^{M} B(\Lambda_j)\ \vac =
\sum_{{\cal S}_I, \, {\cal S}_{II}}\ 
\prod_{\Lambda_k^I \in {\cal S}_I}\ 
\prod_{\Lambda_m^{II} \in {\cal S}_{II}}
a_{_{II}}(\Lambda_k^I)\ 
d_{_{I}}(\Lambda_m^{II})\ f(\Lambda_k^{I},\Lambda_m^{II})\ 
B_{_{II}}(\Lambda_m^{II})\ B_{_{I}}(\Lambda_k^I)\vac \ .
\label{genmod}
\eea

On the right hand side of \r{genmod} we have summations with respect
to partitions of the set of all $\Lambda_j$'s into two subsets ${\cal
S}_I=\{\Lambda^{I}_k\}$ and ${\cal S}_{II}=\{\Lambda^{II}_m\}$. 
Here $k$ labels different $\Lambda$ in the subset
${\cal S}_I$ and $m$ labels different $\Lambda$ in the subset
${\cal S}_{II}$.
The above model is called generalised two-site model because $T$ is
represented as a product of two factors. This is not sufficient for
our purposes however. Let us therefore now consider the so-called

\vskip .5cm
\underbar{Generalised $k$-site model.}
\vskip .5cm

Let us represent $T$ as a product of $k$ factors
\be
T(\Lambda) = T_{k}(\Lambda)\cdot\cdot\cdot T_{2}(\Lambda)\ 
T_{1}(\Lambda)\ \ .
\label{445}
\ee
Here each $T_{\alpha}(\Lambda)$ is a string of $L$-operators like in
\r{intT2}. The commutation relations of the matrix elements of each of
the $T_{\alpha}(\Lambda)$ is given by the same intertwiner as in \r{rtt}.

The matrix elements of $T_{\beta}(\Lambda)$ commute with the matrix
elements of $T_{\alpha}(\Lambda)$ if $\alpha\ne\beta$ and like for the
two-site model we have
\be
T_{\alpha}(\Lambda) = \left(\matrix{A_\alpha(\Lambda)&
B_\alpha(\Lambda)\cr 
C_\alpha(\Lambda)&D_\alpha(\Lambda)\cr}\right) \ \ ,
\ee
\be 
C_\alpha(\Lambda) \vac = 0\ \ \ ,\ \ \ A_\alpha(\Lambda) \vac =
a_\alpha(\Lambda) \vac 
\ \ \ ,\ \ \ D_\alpha(\Lambda) \vac = d_\alpha(\Lambda) \vac \ \ .
\label{abcd}
\ee

By explicitly multiplying the matrices in \r{445} we can express
$B(\Lambda)$ in terms of matrix elements of the $T_\alpha(\Lambda)$.
Iteration of \r{genmod} leads to the following expression for the
eigenfunctions of the transfer matrix $T(\Lambda)$
\bea
\prod_{j=1}^{M} B(\Lambda_j)\ \vac &=& 
\sum_{{\cal S}_1, \dots, \, {\cal S}_k}
\left(\prod_{\alpha =1}^k\ 
\prod_{\Lambda_{m_\alpha} \in {\cal S}_\alpha}
B_{_{\alpha}}(\Lambda_{m_\alpha}^\alpha)
\vac\right) \ \times\nn
&& \qqd \times \left(\prod_{1\le\alpha < \beta\le k}\ 
\prod_{\Lambda_{m_\alpha} \in {\cal S}_\alpha}\ 
\prod_{\Lambda_{k_\beta} \in {\cal S}_\beta}
a_{_{\beta}}(\Lambda_{m_\alpha}^\alpha)\  
d_{_{\alpha}}(\Lambda_{k_\beta}^\beta)\
f(\Lambda_{m_\alpha}^\alpha,\Lambda_{k_\beta}^{\beta}) \right) .
\label{gm2}
\eea

Here the summation is with respect to the partitions of the set of all
$\Lambda_j$'s into $k$ disjoint subsets ${\cal S}_\beta, \beta =
1,\dots, k$. The index $m_\alpha$ enumerates different $\Lambda$
in the subset ${\cal S}_\alpha$ and the index $k_\beta$ enumerates
different $\Lambda$ in the subset ${\cal S}_\beta$. Equation \r{gm2}
was first proved in \cite{IKR87}.

We now consider the special case of the generalized $N$-site model,
where $N$ is the length of the underlying lattice. This means that
each factor $T_\alpha(\Lambda)$ in our generalised $N$-site model is
identified with an individual $L$-operator in \r{intT}. 
The eigenvalues in \r{abcd} are given by
\be
a_{\alpha}(\Lambda) = 1\ ,\qquad
d_{\alpha}(\Lambda) = \frac{\Lambda - a_\alpha}
{\Lambda - a_\alpha + iU/2}\ .
\ee
The $B$-operators are given by
\be
     B_\alpha(\Lambda) = \frac{iU/2}{\Lambda - a_\alpha + iU/2}
			 \, \tau_\alpha^- \ ,
\ee
and have the important feature that $B_\alpha (\Lambda_1) B_\alpha
(\Lambda_2) = 0$. This implies that each set ${\cal S}_\alpha$ in
\r{gm2} consists of maximally one element. We are now in the position
to write down explicit expressions for the eigenstates \r{states}.

\vskip .5cm
a) \underline{One overturned spin.}
\vskip .5cm
Let us first consider the eigenfunctions in the sector with one
overturned spin. Application of \r{gm2} yields
\be
F( \Lambda ) = (- 2i/U)
B(\Lambda -iU/4)\vac = \sum_{y=1}^{N}\sigma_y^-\vac
{1 \over \Lambda - a_y + {iU\over 4}}
\prod_{i=1}^{y-1}
{\Lambda - a_i - {iU\over 4}\over {\Lambda - a_i + {iU\over 4}}}\ .
\label{1spin}
\ee

Here all sets ${\cal S}_\alpha$ in \r{gm2} except one are empty. This
one (one-element) set is ${\cal S}_{y}= \{\Lambda-iU/4\}$. 

\vskip .5cm
a) \underline{Two overturned spins.}
\vskip .5cm
Now application of \r{gm2} leads to
\bea
F(\Lambda_1, \Lambda_2) &=& \sum_{1\le y_{_1}< y_{_2}\le
N}\sum_{\pi\in S_2} \sigma_{y_1}^-\
\sigma_{y_2}^-\vac \frac{1}{\Lambda_{\pi_1}-a_{y_1}+iU/4}
\frac{1}{\Lambda_{\pi_2}-a_{y_2}+iU/4}\times\nn
&& \qqd \times\left( \prod_{j=1}^{y_1-1}
\frac{\Lambda_{\pi_1} - a_{j} - {iU\over 4}}{\Lambda_{\pi_1} -
a_{j} + {iU\over 4}}\right)\left(\prod_{l=1}^{y_2-1}
\frac{\Lambda_{\pi_2} - a_{l} - {iU\over 4}}
{\Lambda_{\pi_2} - a_{l} + {iU\over 4}}\right)
f(\Lambda_{\pi_1},\Lambda_{\pi_2})\ .
\label{bb}
\eea
Here $\pi$ is a permutation of two elements $1,2$ and
\be
f(\Lambda_{\pi_1},\Lambda_{\pi_2}) = 
{\Lambda_{\pi_1}-\Lambda_{\pi_2} - {iU\over 2}\over
\Lambda_{\pi_1}-\Lambda_{\pi_2}}\ .
\ee
In this case only two subsets ${\cal S}_\alpha$ are nonempty. Each
of them consists of one element (${\cal S}_{y_1} =
\{\Lambda_{\pi_1}-iU/4\}$ and ${\cal S}_{y_2} =
\{\Lambda_{\pi_2}-iU/4\}$). 
Equation \r{bb} is of the desired form \r{stateswoy} if we identify
\be
A_\pi = {{\Lambda_{\pi_1} - \Lambda_{\pi_2}-{iU\over 2}\over
\Lambda_{\pi_1} - \Lambda_{\pi_2}}}\ \ ,
\ee
which is in complete agreement with \r{AA}.

\vskip.5cm
c) \underline{M overturned spins.}
\vskip .5cm
The result for 2 overturned spins generalizes straightforwardly to $M$
overturned spins. The nonempty subsets ${\cal S}_\alpha$ in \r{gm2}
(each of which consists of exactly one element) are ${\cal S}_{y_1} =
\Lambda_{\pi_1}-iU/4$,  ${\cal S}_{y_2} = \Lambda_{\pi_2}-iU/4$, \dots,
${\cal S}_{y_M} = \Lambda_{\pi_M}-iU/4$, where $\pi$ is some
permutation of $M$ elements. A straightforward calculation then yields
\r{stateswoy} and \r{AA}.

\clearpage

%%%%%%%%%%%%%%%%%%%%%%%%%%%%%%%%%%%%%%%%%%%%%%%%%%%%%%%%%%%%%%%%%%%%%%%%
%     scdappb.tex                                                      %
%%%%%%%%%%%%%%%%%%%%%%%%%%%%%%%%%%%%%%%%%%%%%%%%%%%%%%%%%%%%%%%%%%%%%%%%

%\addtolength{\evensidemargin}{-2.5cm}
%\addtolength{\oddsidemargin}{-2.5cm}
%\addtolength{\textwidth}{4.6cm}
\tabcolsep3mm

\section{\boldmath The spectrum of three electrons with one-down spin 
for $L=6$}
We present a complete list of eigenstates of the Hubbard Hamiltonian 
(\ref{hamiltonian}) in section V for the case $N=3$ and $M=1$ for a
6-site system ($L=6$) and $U=5$. The energy levels are listed in 
increasing order. 

The energy eigenvalues obtained by direct numerical 
diagonalization of the Hamiltonian (the Householder-QR method) and by
the Bethe ansatz method coincide within  an error of $\CO(10^{-15})$. 
In the table only the energy eigenvalues obtained by Bethe ansatz
are listed. The last digit for each numerical value has a rounding
error. The symbols $S$ and $P$ denote the spin and momentum of the
eigenstate, respectively.  

There are 90 eigenstates for $N=3$ and $M=1$, as we enumerated in
section V.C.3. The numerical results shown in the table confirm the 
completeness of the Bethe ansatz. 

\bigskip

\begin{tabular}{|c|c|c|r|r|}
\hline
No. & Energy & $S$ & $P/(\pi/3)$ &  types \\ \hline 

 1  & $-4.19862084914891$ & 1/2 & 5 & real $I_j=  0.5,-0.5,-1.5,
       J=  0.5 $ \\ \cline{2-5}
 & \multicolumn{4}{|c|}{$ k_j=  0.542662224387082,
    - 0.315837723840216, - 1.27402205174346,
    \La_\alpha= 0.587983554411128$} \\ \hline
 2  & $-4.19862084914891$ & 1/2 & 1 & real $I_j=  1.5, 0.5,-0.5,
       J= -0.5$ \\ \cline{2-5}
 & \multicolumn{4}{|c|}{$k_j =  1.27402205174346, 0.315837723840216,
    - 0.542662224387082, \La_\alpha= - 0.587983554411128$} \\ \hline
 3  & $-4.00000000000000$ & 3/2 & 0 & quartet $I_j= 1,0,-1$ \\ \hline
 4  & $-3.35402752146807$ & 1/2 & 4 & real $I_j=  0.5,- 0.5,- 1.5,
       J= - 0.5 $ \\ \cline{2-5}
 & \multicolumn{4}{|c|}{$ k_j = 0.231206350487730, - 0.682137186480516,
    -1.64346426640041, \La_\alpha = - 1.27426628748753$} \\ \hline
 5  & $- 3.35402752146807$ & 1/2 & 2 & real $I_j=  1.5, 0.5,-0.5,
       J=  0.5 $ \\ \cline{2-5}
 & \multicolumn{4}{|c|}{$ k_j= 1.64346426640041, 0.682137186480516,
    -0.231206350487730, \La_\alpha= 1.27426628748753$} \\ \hline
 6  & $- 2.63449090541439$ & 1/2 & 0 & real $I_j=  1.5,-0.5,-1.5,
       J=  0.5 $ \\ \cline{2-5}
 & \multicolumn{4}{|c|}{$ k_j= 1.52991315867874, -0.279871354037999,
    -1.25004180464074, \La_\alpha= 0.845079113394529$} \\ \hline
 7  & $-2.63449090541439$ & 1/2 & 0 & real $I_j=  1.5, 0.5,-1.5,
       J= -0.5 $ \\ \cline{2-5}
 & \multicolumn{4}{|c|}{$ k_j= 1.25004180464074, 0.279871354037999,
    -1.52991315867874, \La_\alpha= -0.845079113394529$} \\ \hline
 8  & $-2.21221243379665$ & 1/2 & 4 & real $I_j=  0.5,-0.5,-2.5,
    J=  0.5 $ \\ \cline{2-5}
 & \multicolumn{4}{|c|}{$ k_j= 0.554943548673639, -0.307417851034944,
    -2.34192080003189, \La_\alpha= 0.644785906059544$} \\ \hline
 9  & $-2.21221243379665$ & 1/2 & 2 & real $I_j=  2.5, 0.5,-0.5,
    J= -0.5 $ \\ \cline{2-5}
 & \multicolumn{4}{|c|}{$ k_j= 2.34192080003189, 0.307417851034944,
    -0.554943548673640, \La_\alpha= -0.644785906059544$} \\ \hline
10  & $-2.17072407234717$ & 1/2 & 5 & real $I_j=  1.5,-0.5,-1.5,
      J= -0.5 $ \\ \cline{2-5}
 & \multicolumn{4}{|c|}{$ k_j= 1.22622874924835, -0.660511074050219,
    -1.61291522639472, \La_\alpha= -1.15790499577381$} \\ \hline
11  & $-2.17072407234717$ & 1/2 & 1 & real $I_j=  1.5, 0.5,-1.5,
   J=  0.5 $ \\ \cline{2-5}
 & \multicolumn{4}{|c|}{$ k_j= 1.61291522639472, 0.660511074050219,
    -1.22622874924835, \La_\alpha= 1.15790499577381$} \\ \hline
12 & $-2.00000000000000$ & 3/2 & 1 & quartet $I_j= 2,0,-1$ \\ \hline
13 & $-2.00000000000000$ & 3/2 & 5 & quartet $I_j= 1,0,-2$ \\ \hline
14 & $-2.00000000000000$ & 3/2 & 3 & quartet $I_j= 2,1,0$ \\ \hline
15 & $-2.00000000000000$ & 3/2 & 3 & quartet $I_j= 0,-1,-2$ \\ \hline
\end{tabular} 
\clearpage
\begin{tabular}{|c|c|c|r|r|}
\hline
No. & Energy & $S$ & $P/(\pi/3)$ &  types \\ \hline

16  & $-1.68312858421806$ & 1/2 & 3 & real $I_j=  0.5,-0.5,-2.5,
      J= -0.5 $ \\ \cline{2-5}
 & \multicolumn{4}{|c|}{$ k_j= 0.261468569600159, -0.628328469155396,
    -2.77473275403456, \La_\alpha= -0.993984117066770$} \\ \hline
17  & $-1.68312858421806$ & 1/2 & 3 & real $I_j=  2.5, 0.5,-0.5,
      J=  0.5 $ \\ \cline{2-5}
 & \multicolumn{4}{|c|}{$ k_j= 2.77473275403456, 0.628328469155396,
    -0.261468569600159, \La_\alpha= 0.993984117066770$} \\ \hline
18 & $-1.00000000000000$ & 3/2 & 2 & quartet $I_j= 2,1,-1$ \\ \hline
19 & $-1.00000000000000$ & 3/2 & 4 & quartet $I_j= 1,-1,-2$ \\ \hline
20 & $-1.00000000000000$ & 3/2 & 2 & quartet $I_j= 3,0,-1$ \\ \hline
21 & $-1.00000000000000$ & 3/2 & 4 & quartet $I_j= 3,1,0$ \\ \hline
22  & $-0.867143471169408$ & 1/2 & 3 & real $I_j=  0.5,-1.5,-2.5,
      J=  0.5 $ \\ \cline{2-5}
 & \multicolumn{4}{|c|}{$ k_j= 0.515010200239397, -1.28773972583561,
    -2.36886312799358, \La_\alpha= 0.460329439565352$} \\ \hline
23  & $-0.867143471169409$ & 1/2 & 3 & real $I_j=  2.5, 1.5,-0.5,
      J= -0.5 $ \\ \cline{2-5}
 & \multicolumn{4}{|c|}{$ k_j= 2.36886312799358, 1.28773972583561,
    -0.515010200239397, \La_\alpha= -0.460329439565352$} \\ \hline
24  & $-0.763936521983257$ & 1/2 & 2 & real $I_j= -0.5,-1.5,-2.5,
      J=  0.5 $ \\ \cline{2-5}
 & \multicolumn{4}{|c|}{$ k_j= -0.404815536059732, -1.33874801493405,
    -2.44522665379261, \La_\alpha= 0.071452164223840$} \\ \hline
25  & $-0.763936521983257$ & 1/2 & 4 & real $I_j=  2.5, 1.5, 0.5,
      J= -0.5 $ \\ \cline{2-5}
 & \multicolumn{4}{|c|}{$ k_j= 2.44522665379261, 1.33874801493405,
    0.404815536059732, \La_\alpha= -0.07.1452164223840$} \\ \hline
26  & $-0.724935621196484$ & 1/2 & 5 & real $I_j=  1.5, 0.5,-2.5,
      J= -0.5 $ \\ \cline{2-5}
 & \multicolumn{4}{|c|}{$ k_j= 1.27598401405461, 0.318718889893781,
    -2.64190045514499, \La_\alpha= -0.568959086784317$} \\ \hline
27  & $-0.724935621196484$ & 1/2 & 1 & real $I_j=  2.5,-0.5,-1.5,
      J=  0.5 $ \\ \cline{2-5}
 & \multicolumn{4}{|c|}{$ k_j= 2.64190045514499, -0.318718889893781,
    -1.27598401405461, \La_\alpha= 0.568959086784317$} \\ \hline
28  & $-0.633335023683704$ & 1/2 & 5 & real $I_j=  1.5,-0.5,-2.5,
      J=  0.5 $ \\ \cline{2-5}
 & \multicolumn{4}{|c|}{$ k_j= 1.54060965024643, -0.274638622892597,
    -2.31316857855043, \La_\alpha= 0.886033905628400$} \\ \hline
29  & $-0.633335023683705$ & 1/2 & 1 & real $I_j=  2.5, 0.5,-1.5,
      J= -0.5 $ \\ \cline{2-5}
 & \multicolumn{4}{|c|}{$k_j= 2.31316857855043, 0.274638622892597,
    -1.54060965024643, \La_\alpha= -0.886033905628400$} \\ \hline
30  & $-0.441363635441783$ & 1/2 & 4 & real $I_j=  1.5,-0.5,-2.5,
      J= -0.5 $ \\ \cline{2-5}
 & \multicolumn{4}{|c|}{$k_j= 1.24802157741378, -0.602680979199234,
    -2.73973570060774, \La_\alpha= -0.869103625350605$} \\ \hline
\end{tabular}
\clearpage
\begin{tabular}{|c|c|c|r|r|}
\hline
No. & Energy & $S$ & $P/(\pi/3)$ &  types \\ \hline 

31  & $-0.441363635441782$ & 1/2 & 2 & real $I_j=  2.5, 0.5,-1.5,
      J=  0.5 $ \\ \cline{2-5}
 & \multicolumn{4}{|c|}{$k_j= 2.73973570060774, 0.602680979199234,
    -1.24802157741378, \La_\alpha= 0.869103625350605$} \\ \hline
32  & $-0.164932552352930$ & 1/2 & 0 & real $I_j=  1.5, 0.5,-2.5,
      J=  0.5 $ \\ \cline{2-5}
 & \multicolumn{4}{|c|}{$k_j= 1.61980144017864, 0.665436431311280,
    -2.28523787148992, \La_\alpha= 1.18390417778921$} \\ \hline
33  & $-0.164932552352929$ & 1/2 & 0 & real $I_j=  2.5,-0.5,-1.5,
      J= -0.5 $ \\ \cline{2-5}
 & \multicolumn{4}{|c|}{$k_j= 2.28523787148992, -0.665436431311280,
    -1.61980144017864, \La_\alpha= -1.18390417778921$} \\ \hline
34 & 0.00000000000000 & 3/2 & 0 & quartet $I_j= 2,0,-2$ \\ \hline
35 & 0.00000000000000 & 3/2 & 1 & quartet $I_j= 3,1,-1$ \\ \hline
36  &  0.041639350031250 & 1/2 & 2 & real $I_j=  0.5,-1.5,-2.5,
      J= -0.5 $ \\ \cline{2-5}
 & \multicolumn{4}{|c|}{$k_j= 0.242333631055016, -1.61486765220468,
    -2.81625618363673, \La_\alpha= -1.16526625596158$} \\ \hline
37  &  0.041639350031250 & 1/2 & 4 & real $I_j=  2.5, 1.5,-0.5,
      J=  0.5 $ \\ \cline{2-5}
 & \multicolumn{4}{|c|}{$k_j= 2.81625618363673, 1.61486765220468,
    -0.242333631055016, \La_\alpha= 1.16526625596158$} \\ \hline
38  &  0.599066446144297 & 1/2 & 1 & real $I_j= -0.5,-1.5,-2.5,
      J= -0.5 $ \\ \cline{2-5}
 & \multicolumn{4}{|c|}{$k_j= -0.702579425111547, -1.67294123942593,
    -2.86046709144551, \La_\alpha= -1.39028884293989$} \\ \hline
39  &  0.599066446144296 & 1/2 & 5 & real $I_j=  2.5, 1.5, 0.5,
   J=  0.5 $ \\ \cline{2-5}
 & \multicolumn{4}{|c|}{$k_j= 2.86046709144551, 1.67294123942593,
    0.702579425111547, \La_\alpha= 1.39028884293989$} \\ \hline
40  &  0.625362070788712 & 1/2 & 4 & real $I_j=  1.5,-1.5,-2.5,
      J=  0.5 $ \\ \cline{2-5}
 & \multicolumn{4}{|c|}{$k_j= 1.49854259024225, -1.26094870462919,
    -2.33198898800626, \La_\alpha= 0.722114430777064$} \\ \hline
41  &  0.625362070788713 & 1/2 & 2 & real $I_j=  2.5, 1.5,-1.5,
      J= -0.5 $ \\ \cline{2-5}
 & \multicolumn{4}{|c|}{$k_j= 2.33198898800626, 1.26094870462919,
    -1.49854259024225, \La_\alpha= -0.722114430777064$} \\ \hline
42 & 1.00000000000000 & 3/2 & 1 & quartet $I_j= 3,1,0$ \\ \hline
43 & 1.00000000000000 & 3/2 & 5 & quartet $I_j= 3,0,-1$ \\ \hline
44 & 1.00000000000000 & 3/2 & 1 & quartet $I_j= 2,1,-1$ \\ \hline
45 & 1.00000000000000 & 3/2 & 5 & quartet $I_j= 1,-1,-2$ \\ \hline
\end{tabular}
\clearpage
\begin{tabular}{|c|c|c|r|r|}
\hline
No. & Energy & $S$ & $P/(\pi/3)$ &  types \\ \hline 
46  & 1.24466526322010 & 1/2 & 3 & real $I_j=  1.5,-1.5,-2.5,
      J= -0.5 $ \\ \cline{2-5}
 & \multicolumn{4}{|c|}{$k_j= 1.23359323035324, -1.58513579306981,
    -2.79005009087323, \La_\alpha= -1.05370338517746$} \\ \hline
47  &  1.24466526322010 & 1/2 & 3 & real $I_j=  2.5, 1.5,-1.5,
      J=  0.5 $ \\ \cline{2-5}
 & \multicolumn{4}{|c|}{$k_j= 2.79005009087323, 1.58513579306981,
    -1.23359323035324, \La_\alpha= 1.05370338517746$} \\ \hline
48  &  1.26651528335317 & 1/2 & 0 & real $I_j=  2.5,-0.5,-2.5,
      J=  0.5 $ \\ \cline{2-5}
 & \multicolumn{4}{|c|}{$k_j= 2.65775114820892, -0.311801389555702,
    -2.34594975865322, \La_\alpha= 0.614983937128585$} \\ \hline
49  &  1.26651528335317 & 1/2 & 0 & real $I_j=  2.5, 0.5,-2.5,
      J= -0.5 $ \\ \cline{2-5}
 & \multicolumn{4}{|c|}{$k_j= 2.34594975865322, 0.311801389555702,
    -2.65775114820892, \La_\alpha= -0.614983937128585$} \\ \hline
50  &  1.55774931070832 & 1/2 & 5 & real $I_j=  2.5,-0.5,-2.5,
      J= -0.5 $ \\ \cline{2-5}
 & \multicolumn{4}{|c|}{$k_j= 2.31151260511968, -0.609456221165668,
    -2.74925393515061, \La_\alpha= -0.901701357480205$} \\ \hline
51  &  1.55774931070832 & 1/2 & 1 & real $I_j=  2.5, 0.5,-2.5,
      J=  0.5 $ \\ \cline{2-5}
 & \multicolumn{4}{|c|}{$k_j= 2.74925393515061, 0.609456221165668,
    -2.31151260511968, \La_\alpha= 0.901701357480206$} \\ \hline
52 & 2.00000000000000 & 3/2 & 0 & quartet $I_j= 3,2,1$ \\ \hline
53 & 2.00000000000000 & 3/2 & 0 & quartet $I_j= 3,-1,-2$ \\ \hline
54 & 2.00000000000000 & 3/2 & 2 & quartet $I_j= 3,1,-2$ \\ \hline
55 & 2.00000000000000 & 3/2 & 4 & quartet $I_j= 3,2,-1$ \\ \hline
56  &  2.59087887855288 & 1/2 & 4 & real $I_j=  2.5,-1.5,-2.5,
      J=  0.5 $ \\ \cline{2-5}
 & \multicolumn{4}{|c|}{$k_j= 2.60714264000049, -1.28680359174393,
    -2.36753659945316, \La_\alpha= 0.468661303303178$} \\ \hline
57  &  2.59087887855288 & 1/2 & 2 & real $I_j=  2.5, 1.5,-2.5,
      J= -0.5 $ \\ \cline{2-5}
 & \multicolumn{4}{|c|}{$k_j= 2.36753659945316, 1.28680359174393,
    -2.60714264000049, \La_\alpha= -0.468661303303178$} \\ \hline
58 & 3.00000000000000 & 1/2 & 3 & $\eta$-pairing $\otimes I_j= 0$
   \\ \hline
59  &  3.24859982950560 & 1/2 & 4 & real $I_j=  2.5,-1.5,-2.5,
      J= -0.5 $ \\ \cline{2-5}
 & \multicolumn{4}{|c|}{$ k_j= 2.29420005680492, -1.59212356246732,
    -2.79647159673079, \La_\alpha= -1.07985902893246$} \\ \hline
60  &  3.24859982950560 & 1/2 &  2 & real $I_j=  2.5,1.5,-2.5,
      J= 0.5 $ \\ \cline{2-5}
 & \multicolumn{4}{|c|}{$ k_j= 2.79647159673079, 1.59212356246732,
    -2.29420005680492, \La_\alpha= 1.07985902893246$} \\ \hline
\end{tabular}
\clearpage
\begin{tabular}{|c|c|c|r|r|}
\hline
No. & Energy & $S$ & $P/(\pi/3)$ &  types \\ \hline 
61  &  3.53450070903252 & 1/2 & 4 & complex $m=10.0, \ell= 6.0,
      J^{'}= 1.5, \sigma= -1.0 $ \\ \cline{2-5}
 & \multicolumn{4}{|c|}{$q= 2.18262982829371, \xi= 1.51998108466021,
    k_3= 6.10671585537857, \Lambda = 1.96074971503998$} \\ \cline{2-5}
 & \multicolumn{4}{|c|}{$\Re \delta =  0.138234567206297\times 10^{-3}$,
    $\Im \delta = 0.236218958183487\times 10^{-3}$} \\ \hline
62  &  3.53450070903252 & 1/2 &  2 & complex $m= 8.0, \ell= 1.0,
      J^{'}=-1.5, \sigma=  1.0 $ \\ \cline{2-5}
 & \multicolumn{4}{|c|}{$ q= 4.10055547888588, \xi= 1.51998108466021,
    k_3= 0.176469451801019, \Lambda = -1.96074971503998$} \\ \cline{2-5}
 & \multicolumn{4}{|c|}{$\Re \delta = -0.138234567211404\times 10^{-3}$,
    $\Im \delta = 0.236218958176826\times 10^{-3}$ } \\ \hline
63 & 4.00000000000000 & 3/2 & 3 & quartet $I_j= 3,2,-2$ \\ \hline
64 & 4.00000000000000 & 1/2 & 4 & $\eta$-pairing $\otimes I_j= 1$ \\
      \hline
65 & 4.00000000000000 & 1/2 & 2 & $\eta$-pairing $\otimes I_j=-1$ \\
      \hline
66  &  4.13147838656489 & 1/2 & 5 & complex $m= 5.0, \ell= 1.0,
      J^{'}= 1.5, \sigma= -1.0 $ \\ \cline{2-5}
 & \multicolumn{4}{|c|}{$ q= 2.23719190707960, \xi= 1.45368482839940,
    k_3= 0.761603941823783, \Lambda = 1.77325189299191$} \\ \cline{2-5}
 & \multicolumn{4}{|c|}{$\Re \delta =  0.307921735414718\times 10^{-3}$,
    $\Im \delta = 0.266758568093550\times 10^{-3}$} \\ \hline
67  &  4.13147838656489 & 1/2 & 1 & complex $m=13.0, \ell= 6.0,
      J^{'}=-1.5, \sigma=  1.0 $ \\ \cline{2-5}
 & \multicolumn{4}{|c|}{$ q= 4.04599340009998, \xi= 1.45368482839940,
    k_3= 5.52158136535580, \Lambda = -1.77325189299191$} \\ \cline{2-5}
 & \multicolumn{4}{|c|}{$\Re \delta = -0.307921735405836\times 10^{-3}$,
    $\Im \delta = 0.266758568103764\times 10^{-3}$ } \\ \hline
68  &  4.33177213111493 & 1/2 & 5 & complex $m=11.0, \ell= 6.0,
      J^{'}= 0.5, \sigma= -1.0 $ \\ \cline{2-5}
 & \multicolumn{4}{|c|}{$ q= 2.77491534511596, \xi= 1.10073629130283,
    k_3= 5.96934237293065, \Lambda = 0.601275696875128$} \\ \cline{2-5}
 & \multicolumn{4}{|c|}{$\Re \delta = -0.273302194108371\times 10^{-2}$,
    $\Im \delta = -0.199420060157429\times 10^{-2}$ } \\ \hline
69  &  4.33177213111493 & 1/2 &  2 & complex $m= 7.0, \ell= 1.0,
      J^{'}=-0.5, \sigma=  1.0 $ \\ \cline{2-5}
 & \multicolumn{4}{|c|}{$ q= 3.50826996206362, \xi= 1.10073629130283,
    k_3= 0.313842934248937, \Lambda = 0.601275696875128$} \\ \cline{2-5}
 & \multicolumn{4}{|c|}{$\Re \delta = -0.119981837180916\times 10^{1}$,
    $\Im \delta = -0.199420060157118\times 10^{-2}$ } \\ \hline
70  &  4.57467498439465 & 1/2 &  0 & complex $m= 6.0, \ell= 1.0,
      J^{'}= 0.5, \sigma=  1.0 $ \\ \cline{2-5}
 & \multicolumn{4}{|c|}{$ q= 2.88935479499487, \xi= 1.07299907022393,
    k_3= 0.504475717189853, \Lambda = 0.411558289383258$} \\ \cline{2-5}
 & \multicolumn{4}{|c|}{$\Re \delta = -0.399317226400675\times 10^{-2}$,
    $\Im \delta = 0.222938603299694\times 10^{-3}$ } \\ \hline
\end{tabular}
\clearpage
\begin{tabular}{|c|c|c|r|r|}
\hline
No. & Energy & $S$ & $P/(\pi/3)$ &  types \\ \hline 
71  &  4.57467498439465 & 1/2 & 0 & complex $m=12.0, \ell= 6.0,
      J^{'}=-0.5, \sigma= -1.0 $ \\ \cline{2-5}
 & \multicolumn{4}{|c|}{$ q= 3.39383051218472, \xi= 1.07299907022393,
    k_3= 5.778709589989730, \Lambda = 0.411558289383258$} \\ \cline{2-5}
 & \multicolumn{4}{|c|}{$\Re \delta = -0.819123406502514\times 10^{0}$,
    $\Im \delta = 0.222938603298584\times 10^{-3}$ } \\ \hline
72  &  4.71197559752276 & 1/2 &  3 & complex $m= 9.0, \ell= 5.0,
      J^{'}= 1.5, \sigma= -1.0 $ \\ \cline{2-5}
 & \multicolumn{4}{|c|}{$q= 2.16082673394224, \xi= 1.54894763250639,
    k_3= 5.10312449288489, \Lambda = 2.04356190829635$} \\ \cline{2-5}
 & \multicolumn{4}{|c|}{$\Re \delta = 0.892857558656424\times 10^{-4}$,
    $\Im \delta = 0.211992582481946\times 10^{-3}$ } \\ \hline
73  &  4.71197559752276 & 1/2 &  3 & complex $m= 9.0, \ell= 2.0,
      J^{'}=-1.5, \sigma=  1.0 $ \\ \cline{2-5}
 & \multicolumn{4}{|c|}{$q= 4.12235857323734, \xi= 1.54894763250639,
    k_3= 1.18006081429470, \Lambda = -2.04356190829635$} \\ \cline{2-5}
 & \multicolumn{4}{|c|}{$\Re \delta = -0.892857558563165\times 10^{-4}$,
    $\Im \delta = 0.211992582494824\times 10^{-3}$ } \\ \hline
74  &  5.58600548103459 & 1/2 & 4 & complex $m=10.0, \ell= 5.0,
      J^{'}= 0.5, \sigma= -1.0 $ \\ \cline{2-5}
 & \multicolumn{4}{|c|}{$ q= 2.72669142318074, \xi= 1.11641164129461,
    k_3= 5.01859266560450, \Lambda = 0.683372926842314$} \\ \cline{2-5}
 & \multicolumn{4}{|c|}{$\Re \delta = -0.186687577849687\times 10^{-2}$,
    $\Im \delta = -0.244844007284084\times 10^{-2}$ } \\ \hline
75  &  5.58600548103459 & 1/2 &  2 & complex $m= 8.0, \ell= 2.0,
      J^{'}=-0.5, \sigma=  1.0 $ \\ \cline{2-5}
 & \multicolumn{4}{|c|}{$ q= 3.55649388399885, \xi= 1.11641164129461,
    k_3= 1.26459264157509, \Lambda = 0.683372926842313$} \\ \cline{2-5}
 & \multicolumn{4}{|c|}{$\Re \delta = -0.136487897790614\times 10^{1}$,
    $\Im \delta = -0.244844007284262\times 10^{-2}$ } \\ \hline
76  &  5.95823319001949 & 1/2 &  0 & complex $m= 6.0, \ell= 2.0,
      J^{'}= 1.5, \sigma= -1.0 $ \\ \cline{2-5}
 & \multicolumn{4}{|c|}{$q= 2.27376322866332, \xi= 1.41368501399761,
    k_3= 1.73565884985295, \Lambda = 1.66056016294642$} \\ \cline{2-5}
 & \multicolumn{4}{|c|}{$\Re \delta =  0.455885386176691\times 10^{-3}$,
    $\Im \delta = 0.245749596857303\times 10^{-3}$ } \\ \hline
77  &  5.95823319001949 & 1/2 & 0 & complex $m=12.0, \ell= 5.0,
      J^{'}=-1.5, \sigma=  1.0 $ \\ \cline{2-5}
 & \multicolumn{4}{|c|}{$ q= 4.00942207851627, \xi= 1.41368501399761,
    k_3= 4.54752645732663, \Lambda = -1.660560162946420$} \\ \cline{2-5}
 & \multicolumn{4}{|c|}{$\Re \delta = -0.455885386181354\times 10^{-3}$,
    $\Im \delta = 0.245749596852640\times 10^{-3}$ } \\ \hline
78 & 6.00000000000000 & 1/2 & 5 & $\eta$-pairing $\otimes I_j= 2$ \\
      \hline
79 & 6.00000000000000 & 1/2 & 1 & $\eta$-pairing $\otimes I_j=-2$ \\
      \hline
80  &  6.03334376215914 & 1/2 & 1 & complex $m= 7.0, \ell= 2.0,
      J^{'}= 0.5, \sigma=  1.0 $ \\ \cline{2-5}
 & \multicolumn{4}{|c|}{$ q= 2.96456730023144, \xi= 1.06124089913747,
    k_3= 1.40124825791330, \Lambda = 0.288686041886590$} \\ \cline{2-5}
 & \multicolumn{4}{|c|}{$\Re \delta = -0.375435197170576\times 10^{-2}$,
    $\Im \delta = 0.208597324485038\times 10^{-2}$ } \\ \hline
\end{tabular}
\clearpage
\begin{tabular}{|c|c|c|r|r|}
\hline
No. & Energy & $S$ & $P/(\pi/3)$ &  types \\ \hline 
81  &  6.03334376215914 & 1/2 & 5 & complex $m=11.0, \ell= 5.0,
      J^{'}=-0.5, \sigma= -1.0 $ \\ \cline{2-5}
 & \multicolumn{4}{|c|}{$q= 3.31861800694814, \xi= 1.06124089913747,
    k_3= 4.88193704926629, \Lambda = 0.288686041886588$} \\ \cline{2-5}
 & \multicolumn{4}{|c|}{$\Re \delta = -0.573617731801462$,
    $\Im \delta = 0.208597324485194\times 10^{-2}$ } \\ \hline
82  &  6.70841301401077 & 1/2 &  2 & complex $m= 8.0, \ell= 4.0,
      J^{'}= 1.5, \sigma= -1.0 $ \\ \cline{2-5}
 & \multicolumn{4}{|c|}{$q= 2.16393969303817, \xi= 1.54471884698899,
    k_3= 4.04970102349645, \Lambda = 2.03142493862260$} \\ \cline{2-5}
 & \multicolumn{4}{|c|}{$\Re \delta =  0.956259360065381\times 10^{-4}$,
    $\Im \delta = 0.215691965483877\times 10^{-3}$ } \\ \hline
83  &  6.70841301401077 & 1/2 & 4 & complex $m=10.0, \ell= 3.0,
      J^{'}=-1.5, \sigma=  1.0 $ \\ \cline{2-5}
 & \multicolumn{4}{|c|}{$q= 4.11924561414142, \xi= 1.54471884698899,
    k_3= 2.23348428368314, \Lambda = -2.031424938622590$} \\ \cline{2-5}
 & \multicolumn{4}{|c|}{$\Re \delta = -0.956259360207490\times 10^{-4}$,
    $\Im \delta = 0.215691965477882\times 10^{-3}$ } \\ \hline
84 & 7.00000000000000 & 1/2 & 0 & $\eta$-pairing $\otimes I_j=3$ \\
      \hline
85  &  7.48332665113181 & 1/2 & 1 & complex $m= 7.0, \ell= 3.0,
      J^{'}= 1.5, \sigma= -1.0 $ \\ \cline{2-5}
 & \multicolumn{4}{|c|}{$ = 2.20081321450656, \xi= 1.49694330556893,
    k_3= 2.92875642936307, \Lambda = 1.89535079698079$} \\ \cline{2-5}
 & \multicolumn{4}{|c|}{$\Re \delta =  0.187321491183834\times 10^{-3}$,
    $\Im \delta = 0.252337664304214\times 10^{-3}$ } \\ \hline
86  &  7.48332665113181 & 1/2 & 5 & complex $m=11.0, \ell= 4.0,
      J^{'}=-1.5, \sigma=  1.0 $ \\ \cline{2-5}
 & \multicolumn{4}{|c|}{$q= 4.08237209267303, \xi= 1.49694330556893,
    k_3= 3.354428877816510, \Lambda = -1.89535079698079$} \\ \cline{2-5}
 & \multicolumn{4}{|c|}{$\Re \delta = -0.187321491188275\times 10^{-3}$,
    $\Im \delta = 0.252337664298441\times 10^{-3}$ } \\ \hline
87  &  7.59363119464461 & 1/2 &  3 & complex $m= 9.0, \ell= 4.0,
      J^{'}= 0.5, \sigma= -1.0 $ \\ \cline{2-5}
 & \multicolumn{4}{|c|}{$q= 2.74080575567842, \xi= 1.11156342668968,
    k_3= 3.94316644941255, \Lambda = 0.659158265098385$} \\ \cline{2-5}
 & \multicolumn{4}{|c|}{$\Re \delta = -0.212814421543628\times 10^{-2}$,
    $\Im \delta = -0.234942161399343\times 10^{-2}$ } \\ \hline
88  &  7.59363119464461 & 1/2 &  3 & complex $m= 9.0, \ell= 3.0,
      J^{'}=-0.5, \sigma=  1.0 $ \\ \cline{2-5}
 & \multicolumn{4}{|c|}{$q= 3.54237955150117, \xi= 1.11156342668968,
    k_3= 2.34001885776704, \Lambda = 0.659158265098384$} \\ \cline{2-5}
 & \multicolumn{4}{|c|}{$\Re \delta = -0.131618838598134\times 10^{1}$,
    $\Im \delta = -0.234942161399543\times 10^{-2}$ } \\ \hline
89  &  8.02701965828632 & 1/2 &  2 & complex $m= 8.0, \ell= 3.0,
   J^{'}= 0.5, \sigma=  1.0 $ \\ \cline{2-5}
 & \multicolumn{4}{|c|}{$q= 2.89722498813356, \xi= 1.07154335884761,
    k_3= 2.58313043330567, \Lambda = 0.398665945793610$} \\
         \cline{2-5}
 & \multicolumn{4}{|c|}{$\Re \delta = -0.401340764685176\times 10^{-2}$,
    $\Im \delta = 0.414789194807641\times 10^{-3}$ } \\ \hline
90  &  8.02701965828632 & 1/2 & 4 & complex $m=10.0, \ell= 4.0,
      J^{'}=-0.5, \sigma= -1.0 $ \\ \cline{2-5}
 & \multicolumn{4}{|c|}{$q= 3.38596031904603, \xi= 1.07154335884761,
    k_3= 3.70005487387392, \Lambda = 0.398665945793609$} \\ \cline{2-5}
 & \multicolumn{4}{|c|}{$\Re \delta = -0.793318483940372$,
    $\Im \delta = 0.414789194806753\times 10^{-3}$ } \\ \hline
\end{tabular}

%\bibliographystyle{phys}
%\bibliography{hub,thermo}

\end{document}